\documentclass[twocolumn]{aastex61}

\usepackage{graphicx}  
\usepackage{float}     
\usepackage{amsmath}
\usepackage{longtable}
\usepackage{soul}
\usepackage{CJK}

\shorttitle{Millimeter line variation in IRC\,+10216}
\shortauthors{He, Dinh-V-Trung \& Hasegawa}

\begin{document}

\title{Monitor variability of millimeter lines in IRC\,+10216 
      \footnote{Thanks to the support of observations by Academia Sinica, 
      Institute of Astronomy and Astrophysics (ASIAA), Taiwan.}}


\correspondingauthor{J.H. He}
\email{jinhuahe@ynao.ac.cn}

\author[0000-0002-3938-4393]{J.H. He}
\affiliation{Yunnan Observatories, Chinese Academy of Sciences, 396 Yangfangwang, Guandu District, Kunming, 650216, P. R. China}
\affiliation{Chinese Academy of Sciences South America Center for Astronomy, China-Chile Joint Center for Astronomy, Camino El Observatorio \#1515, Las Condes, Santiago, Chile}
\affiliation{Key Laboratory for the Structure and Evolution of Celestial Objects, Chinese Academy of Sciences, 396 Yangfangwang, Guandu District, Kunming, 650216, P. R. China}
\affiliation{Center for Astronomical Mega-Science, Chinese Academy of Sciences, 20A Datun Road, Chaoyang District, Beijing, 100012, P. R. China}

\author{Dinh-V-Trung}
\affiliation{Institute of Physics, Vietnam Academy of Science and Technology, 10 DaoTan, BaDinh, Hanoi, Vietnam}

\author{T.I. Hasegawa}
\affiliation{Academia Sinica, Astronomy and Astrophysics, AS/NTU. No.1, Sec. 4, Roosevelt Rd, Taipei 10617, Taiwan}


\begin{abstract}

A single dish monitoring of millimeter maser lines SiS\,J=14-13 and HCN\,$\nu_2=1^{f}$\,J=3-2 and several other rotational lines is reported for the archetypal carbon star IRC\,+10216. Relative line strength variations of 5\%$\sim$30\% are found for eight molecular line features with respect to selected reference lines. Definite line-shape variation is found in limited velocity intervals of the SiS and HCN line profiles. The asymmetrical line profiles of the two lines are mainly due to the varying components. Their dominant varying components of the line profiles have similar periods and phases as the IR light variation, although both quantities show some degree of velocity dependence; there is also variability asymmetry between the blue and red line wings of both lines. Combining the velocities and amplitudes with a wind velocity model, we suggest that the line profile variations are due to SiS and HCN masing lines emanating from the wind acceleration zone. The possible link of the variabilities to thermal, dynamical and/or chemical processes within or under this region is also discussed.

\end{abstract}


\section{Introduction} \label{sec:intro}

Variability study of molecular emission lines in strongly pulsating carbon rich Asymptotic Giant Branch (AGB) stars (carbon stars) is very rare perhaps because of the lack of strong radio masers in them and the poor absolute millimeter line flux calibration accuracy for most ground based telescopes. Only few works contributed to this topic \citep{Carl1990,Groe1998,Cern2014}.The discovery of weak masers in millimeter wavelength toward some carbon stars \citep[see e.g.,][]{Henk1983,Turn1987,Luca1989,Schi2000,Cern2000,Fonf2006,Gong2017} offers the possibility to monitor the low rotational transitions of circumstellar molecules with existing millimeter telescopes. These maser lines are usually emitted from the inner region of the circumstellar envelop (CSE). This makes them good tools to study the properties of the dynamical atmosphere of the pulsating stars.

IRC\,+10216 (or CW\,Leo) is an archetypal carbon star that is very rich in molecular line emission \citep[see the most recent examples in][among others]{He08,Tene10a,Cern2010,Neuf2011,Pate2011,Deci2015,Gong2015} and serves as a benchmark for molecular line studies in circumstellar envelopes. Most of the weak maser lines in carbon stars were also discovered in this object. For example, the first centimeter maser SiS J=1-0 (18.155\,GHz) was discovered in IRC\,+10216 by
\citet{Henk1983}, confirmed by \citet{Nguy1984} and newly explored by \citet{Gong2017}. Later on, mm masers were found in more ground and vibrational levels of SiS, including J=11-10, 14-13 and 15-14 lines in the $v=0$ state and J=12-11, 13-12 and 14-13 lines in the $v=1$ state in this star \citep{Turn1987,Fonf2006}. Another
molecule also versatile in maser emission is HCN. At least four maser lines have been found in the low lying vibrational states in IRC\,+10216: ($01^{1c}0$), J=2-1 \citep{Luca1989}; ($02^00$), J=1-0 \citep{Luca1992}; ($04^00$), J=9-8 \citep{Schi2000} and ($01^10$)-($04^00$), J=10-9 \citep{Schi2003}. Maser
phenomenon has also been reported for CS\,v=1\, J=3-2 \citep{High2000,Cern2000} and possibly for the oxygen bearing molecular lines OH 1665/1667 MHz \citep{Ford2003} and even H$_2$O $6_{1,6}-5_{2,3}$ \citep{Han1998} in this star.

However, millimeter line variability study is relatively lacking for IRC\,+10216. Beside the earlier attempt of \citet{Carl1990}, the only work that successfully revealed the millimeter line variability in this star is \citet{Cern2014} who used Herschel HIFI spectra. Combining several sparsely sampled epochs, they found that some lines of spatially extended species such as the SiCC and low-J CO lines showed little variation, while some other molecular lines such as high-J CO, CS, SiO, SiS, HCN, CCH\,N=6-5, and H$_2$O lines all show evident variability. Particularly, the varying CCH\,N=6-5 line was argued to be emitted from an extended region in the CSE which, together with the far infrared (IR) variability of dust emission found by \citet{Groe2012}, demonstrated the importance of IR excitation/heating of dust and gas in the outer parts of the CSE. 

In the inner CSE of IRC\,+10216, some molecular emission lines are also expected to vary with time because of the variation of IR light and gas temperature, the rise and fall of pulsation shocks and the ensuing varying shock chemistry \citep[see particularly the chemical simulations for the case of IRC\,+10216 by][]{Cher2011,Cher2012}, and possibly dust formation instability. 

Masers are interesting phenomena in the space and are one of the best tracers of variability due to their sensitive dependence upon the variation of pumping agents \citep{Elit1992}. However, the lack of strong radio masers makes them almost useless to carbon stars. Weaker maser lines in millimeter wavelengths could regain their usefulness to carbon stars with large interferometers like ALMA. Identifying these maser lines and investigate their properties through variability studies is a good approach for both single dishes and interferometers.

The large flux calibration uncertainty that prevented us from detecting the line variation with ground based telescopes can be greatly reduced by studying the relative variation of lines in the same side band. The relative line variations could bring us important information of these processes.

We report in this work the discovery of relative line strength and line shape variation for eight line features around 1.1\,mm wavelength, including two candidate millimeter maser lines, SiS\,$v=0$, J=14-13 and HCN\,$\nu_2=1^{1f}$, J=3-2, toward IRC\,+10216. The design of the monitoring, the observations and data processing are described in Sect.~\ref{sec:obs}, the results and the characteristics of line intensity and line profile variabilities are presented in Sect.~\ref{sec:results}. We discuss the remaining uncertainties and how to understand the results in Sect.~\ref{sec:discuss} and summarize the new findings in Sect.~\ref{sec:summary}. The formulation of the relative line variability, part of the results and some extended discussions are given in appendices.

\section{Monitoring strategy, observation and data processing} \label{sec:obs}

\subsection{Observations} \label{sec:observation}

The monitoring observations were made during a period of 523 days from Dec. 3, 2007 to May 9, 2009, with a gap of about a half year (169 days, from May 26 to Nov. 11, 2008). Thus the monitoring period covers nearly a whole IR variation period \citep[630 days,][]{Ment12}. In total, 27 epochs are sampled. The first 17 epochs before the gap were sampled roughly once every 10 days, while the rest 10 epochs were sampled roughly once every 20 days. However, there was a trouble of the spectrometer on Apr. 25, 2008 in the LSB, which leads to the LSB data unusable at that epoch. Therefore, useful LSB data are available only for 26 epochs.

The ALMA band-6 receiver was used at the 10\,m Submillimeter Telescope (SMT) of Arizona Radio Observatory on Mt. Graham. Both the USB and LSB were observed simultaneously with Filter Banks that have a 1\,GHz frequency coverage and 1\,MHz resolution each (centered at 266.80 and 254.55\,GHz respectively). The weather was better in the first period than in the second, with an average $\tau(1.3\,{\rm mm})=0.27$ and $0.36$ before and after the time gap respectively. The system temperatures were about 300\,K in LSB and 350\,K in USB on average. The average on-source integration time was about 51\,min, ranging from 30 to 76\,min at various epochs. This results in an average RMS baseline noise of 12\,mK at a spectral resolution of 1\,MHz for both side bands. The observation was done in beam switch mode with a $2\arcmin$ throw at 2 Hz frequency. The image rejection ratio were measured each time for both side bands and it was found to be usually in the range of 10-20\,db. When the image rejection was not very high, the strongest line SiS\,J=14-13 in our LSB can emerge in the USB as a false feature at several epochs (see details in the results below). The beam size is about $29\arcsec$. A main beam efficiency of 0.75 can be adopted for both side bands.  

\subsection{Data processing} \label{sec:reduce}

The observational data were reduced using the GILDAS/CLASS software. The baseline is always straight and thus only a linear baseline was removed from each side band spectrum (after averaging together sub-scans). The spectra at individual epochs are also averaged together to yield very deep overall average spectra with an RMS baseline noise of  1.6\,mK in LSB and 2.0\,mK in USB. The deep overall average spectra allows us to identify many weak line features in the two 1-GHz spectral windows that are usually invisible in individual epoch data. 

Some detected line features are actually not single lines but groups of blended lines from different carriers or different transitions of the same molecules, we treat the blended lines as a whole in this work.

\section{Results} \label{sec:results}

\subsection{Line identification and line profile examples} \label{sec:lineiden}

The very deep cumulative exposure in the spectra averaged over all epochs (overall average spectra hereafter) allows us to detect 47 lines in the two side bands (including blended lines, but hyper-fine splits are merged and counted as a single line). The line carriers are identified using the CDMS\footnote{http://www.astro.uni-koeln.de/cdms/catalog} \citep{Mull01,Mull05} and Splatalogue\footnote{http://www.cv.nrao.edu/php/splat/} line lists. Beside eight unidentified lines (U-lines; three in USB and five in LSB), the rest 39 lines belong to 17 molecules. Rotational lines from excited vibration states are detected for C$_4$H, HCN, H$^{13}$CN and SiS. The transition list is given in Table~\ref{tab_lines} in Appendix~\ref{sec:lineiden_detail} (on-line only). Some lines are partially or totally blended with each other and considered together as a single composite line feature in the variability analysis. In total, 30 line features are defined (18 in USB and 12 in LSB). The eight strongest line features used for light curve analysis are summarized in Table~\ref{tab_linegroups}. The plots of the overall average spectral and more comments are given in Appendix~\ref{sec:lineiden_detail}.
\begin{table*}
 \centering
 \begin{minipage}{180mm}  
  \caption{Strongest line features (lines or line groups) for the variability study and fitted light-curve parameters.}
  \label{tab_linegroups}
  \begin{tabular}{llrrrlrrr}
  \hline
  \hline
Freq      & line feature                                & Va       & Vb       & $P$   & $A$             & $\phi_0$   & $Z$             & $\chi^2_{\rm min}$\\
MHz       &                                           & km/s     & km/s     & day   & K\,km\,s$^{-1}$ &            &     K\,km\,s$^{-1}$ & \\
\hline                                                                   
\multicolumn{9}{l}{LSB:}                                                                        \\
254103.20  & SiS\,14-13                                 & -41.0    & -11.0    &  680(18) & 7.36(0.20)  & $-0.03(0.01)$  & 136.00(0.15) & 4391  \\
254216.14  & $^{30}$SiO\,6-5                            & -41.0    & -11.1    &  759(75) & 0.46(0.04)  & $+0.63(0.06)$  &   4.35(0.03) &  143  \\
254683.00  & Na$^{37}$Cl,CH$_2$NH,HC$_3$N               & -63.1    &  10.4    &  708(50) & 0.64(0.04)  & $+0.56(0.05)$  &   4.55(0.03) &   53  \\
\multicolumn{9}{l}{USB:}                                                                                                                          \\
266500.88  & C$_4$H\,$^2\Pi_{3/2}\,\nu_7=1^{1e}$        & -41.0    & -11.0    &  695(50) & 0.61(0.04)  & $+0.00(0.02)$  &   2.08(0.03) &   42  \\
266771.19  & C$_4$H\,$^2\Pi_{1/2}\,\nu_7=1^{1f}$        & -41.0    & -11.0    &  678(40) & 0.74(0.04)  & $-0.01(0.01)$  &   2.91(0.04) &   65  \\
267120.00  & C$_3$N,$^{13}$CCCN,HCN($\nu_2=2^{2f+2e}$), & -45.0    &   7.7    &  586(31) & 0.62(0.05)  & $+0.01(0.02)$  &   2.13(0.05) &   75  \\
            & C$_4$H($\nu_7=2^2$)                        &          &          &          &             &                &              &       \\
267199.28  & HCN\,$\nu_2=1^{1f}$\,J=3-2                 & -41.0    &  -9.0    &  445( 9) & 1.21(0.11)  & $+0.61(0.01)$  &  14.84(0.09) & 1412  \\
267243.00  & $^{29}$SiS,HCN($\nu_2=2^0$)                & -42.0    & -10.0    & 1182(58) & 3.53(0.25)  & $-0.01(0.05)$  &  11.66(0.25) & 1242  \\
\hline
\end{tabular} \\
Note: Average lab frequency is used to represent a blended line group and define its systemic velocity -26.5\,km\,s$^{-1}$. The other quantities are the velocity range (Va$\sim$Vb) for the integrated line strength, the four light curve parameters ($P$, $A$, $\phi_0$ and $Z$) defined in Eq.~(\ref{eq_cos}) and the minimum $\chi^2$ value for the best fit. The maximum phase is expressed in IR variation phase and zero IR phase corresponds to the IR maximum on JD\,2454554.0. For the four line features in the right panel of Fig.~\ref{fig_IintLightCurve}, the phase corresponds to the maximum of the fitted light curve after the time gap of our monitoring.\\
\end{minipage}
\end{table*}

\begin{figure*}
 \centering
\includegraphics[angle=0,scale=0.38]{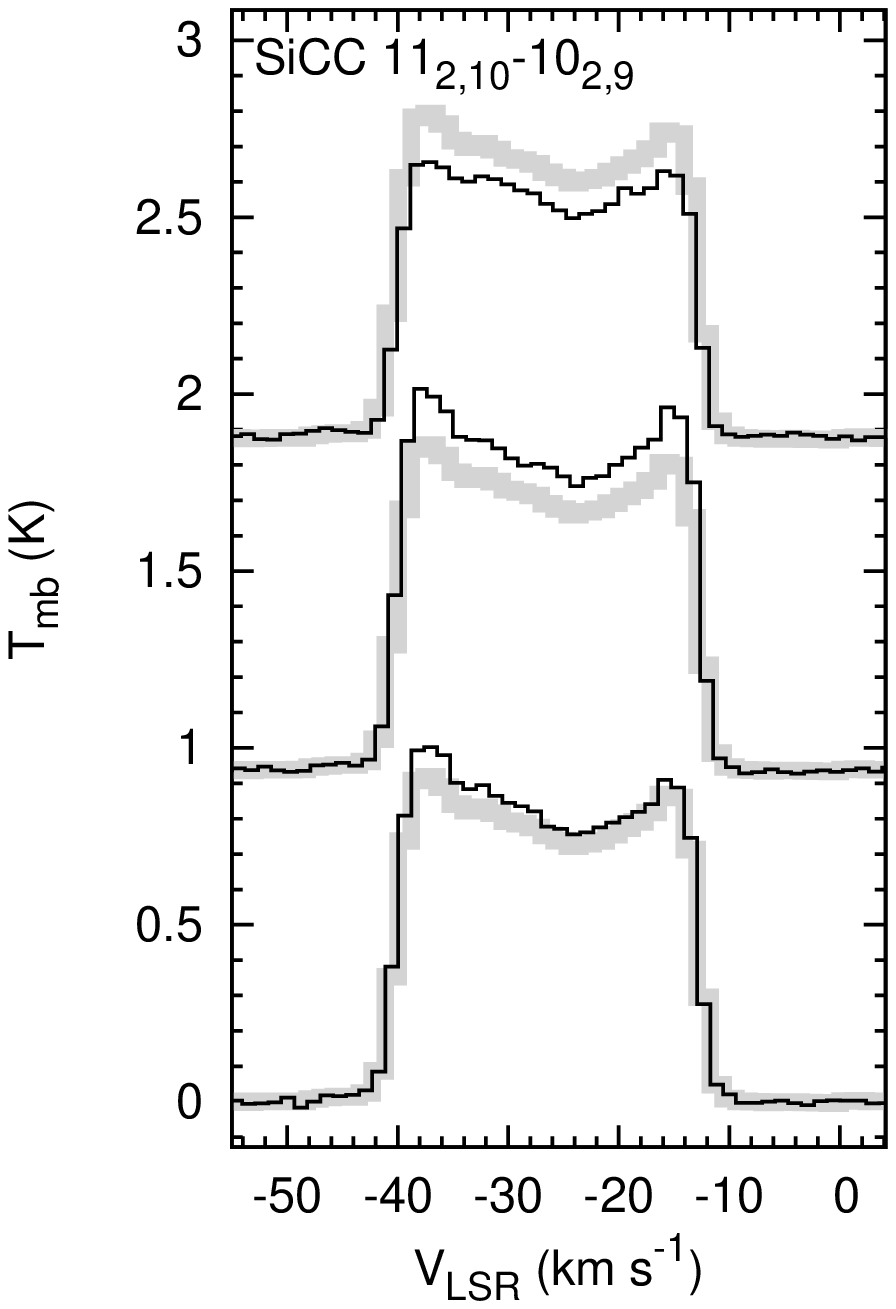}
\includegraphics[angle=0,scale=0.38]{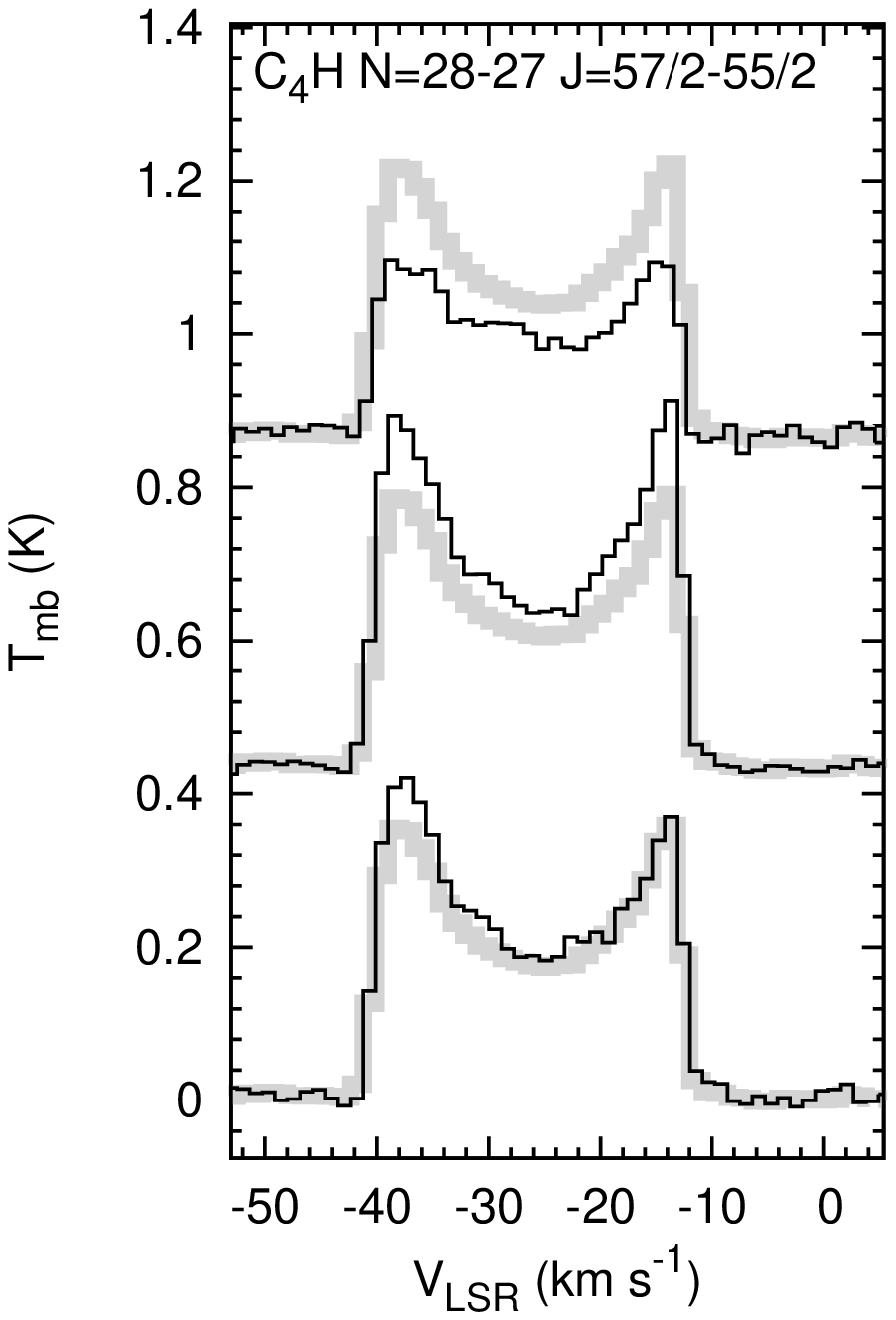}
\includegraphics[angle=0,scale=0.38]{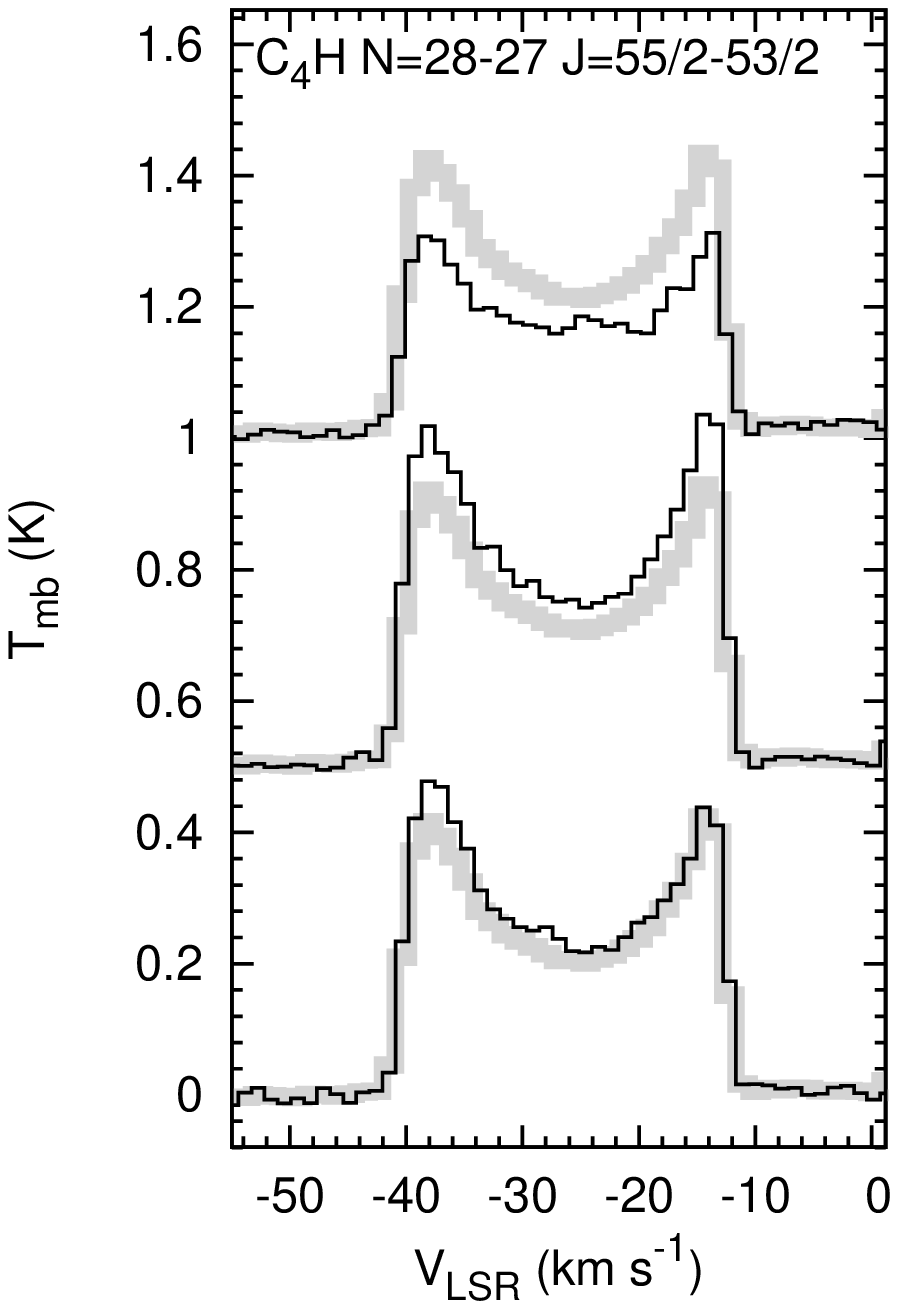}\\
\includegraphics[angle=0,scale=0.38]{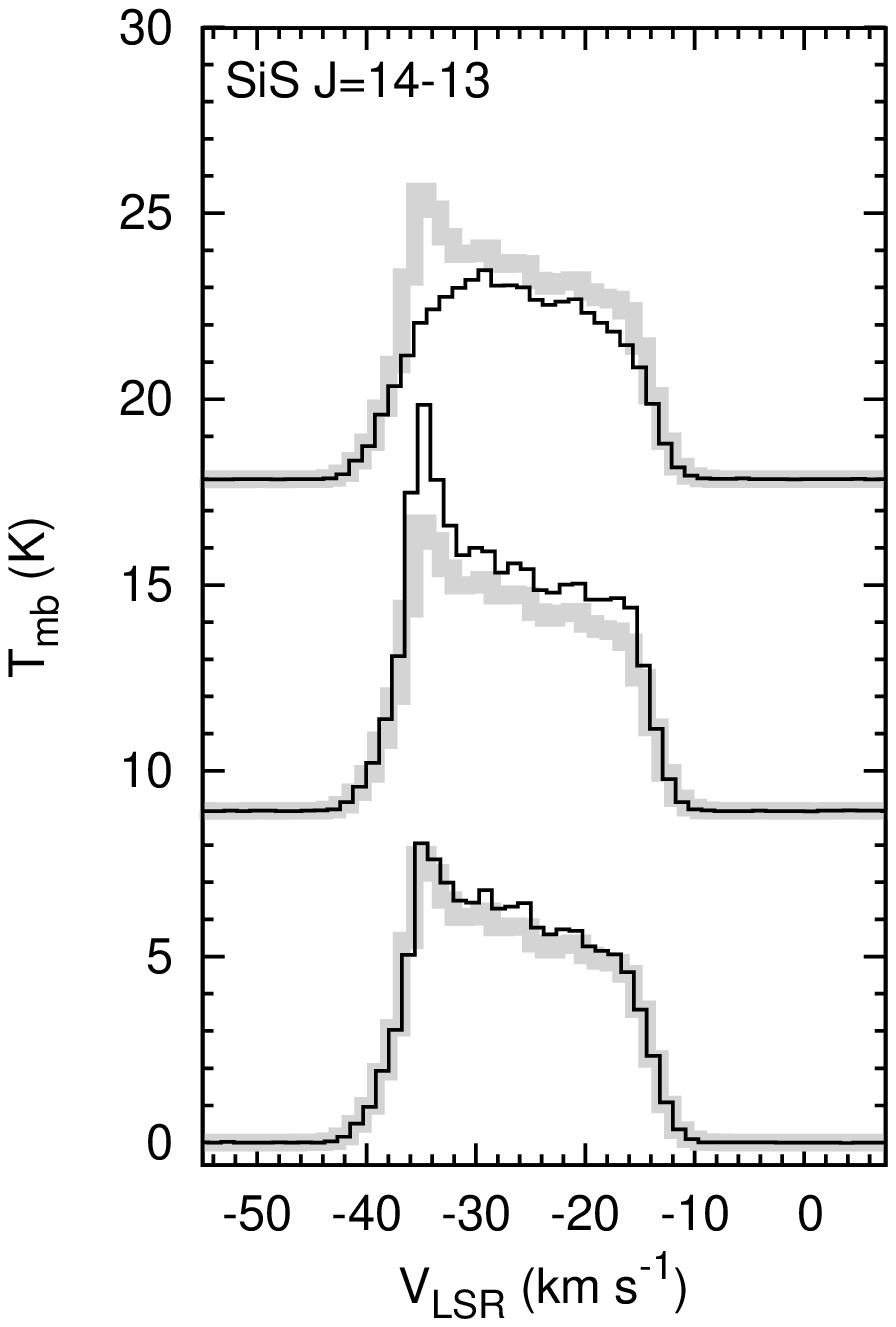}
\includegraphics[angle=0,scale=0.38]{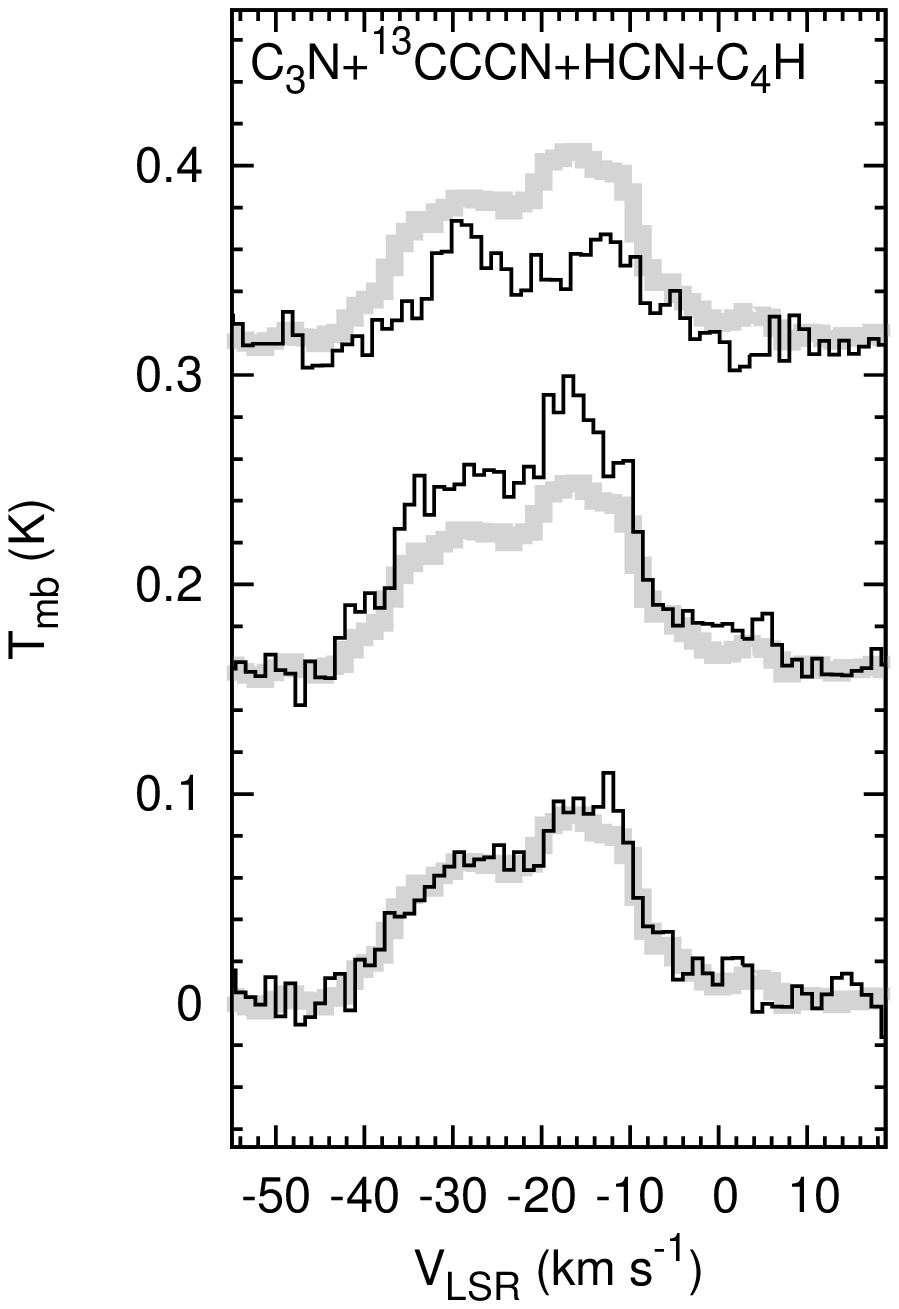}
\includegraphics[angle=0,scale=0.38]{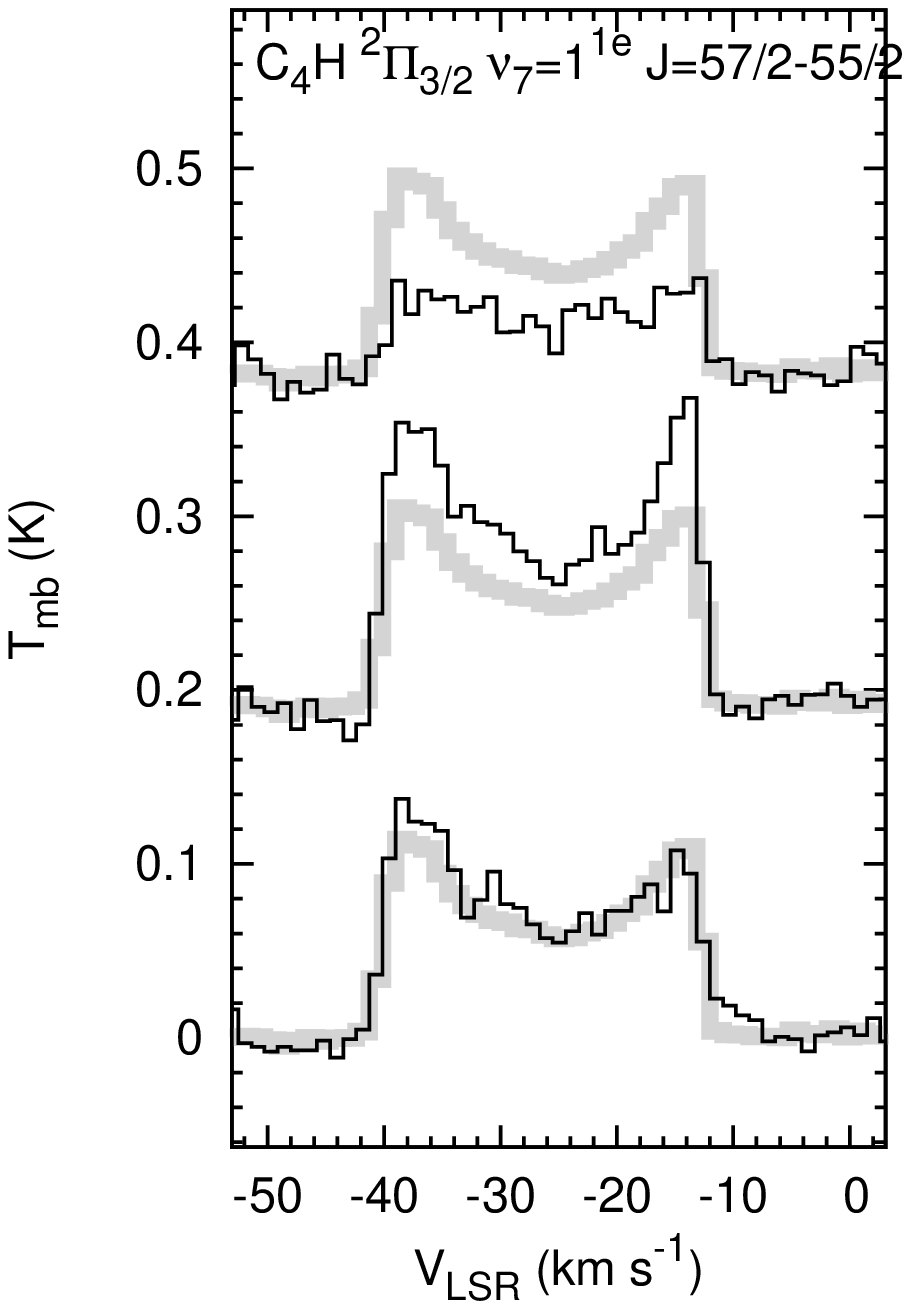}\\
\includegraphics[angle=0,scale=0.38]{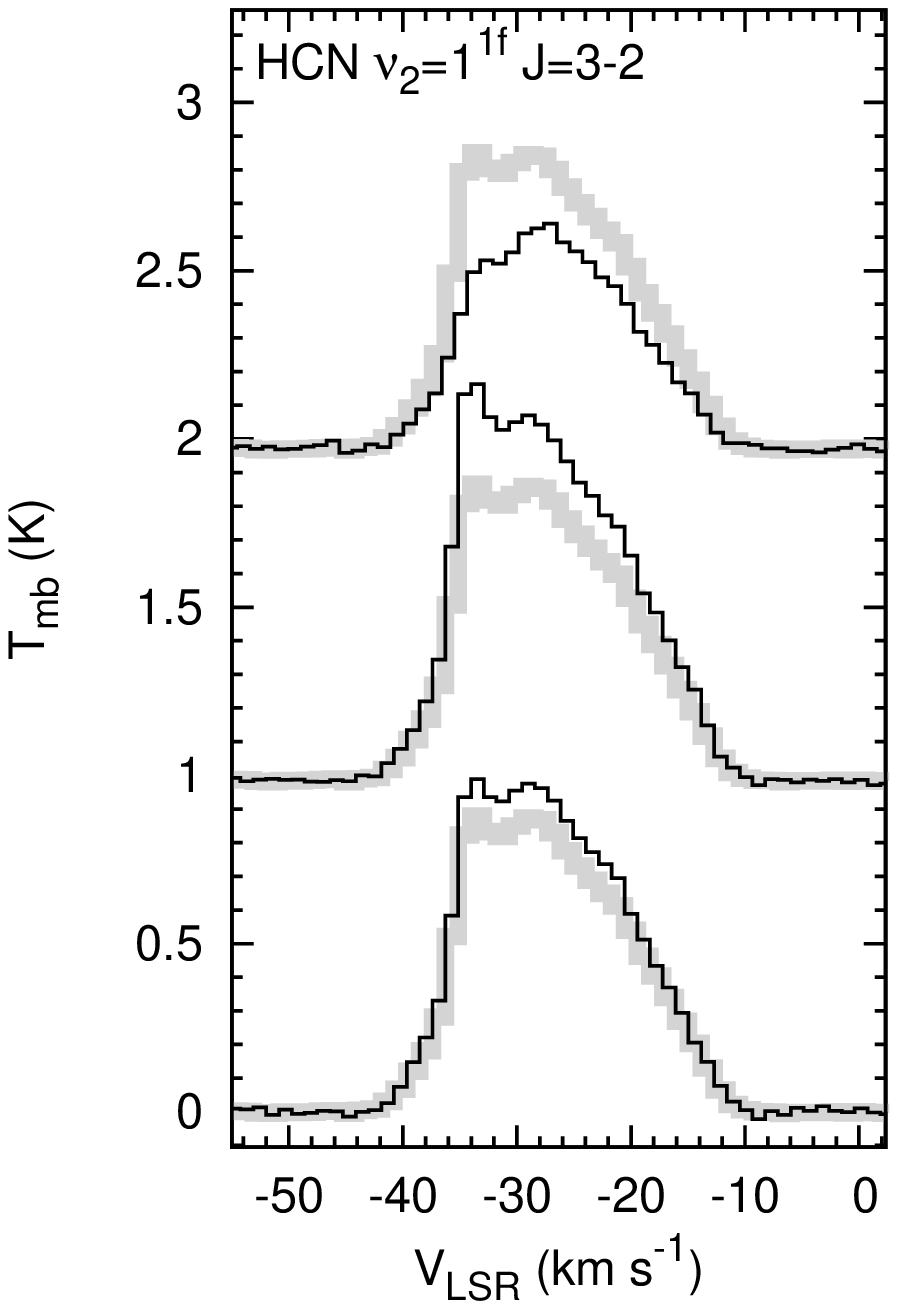}
\includegraphics[angle=0,scale=0.38]{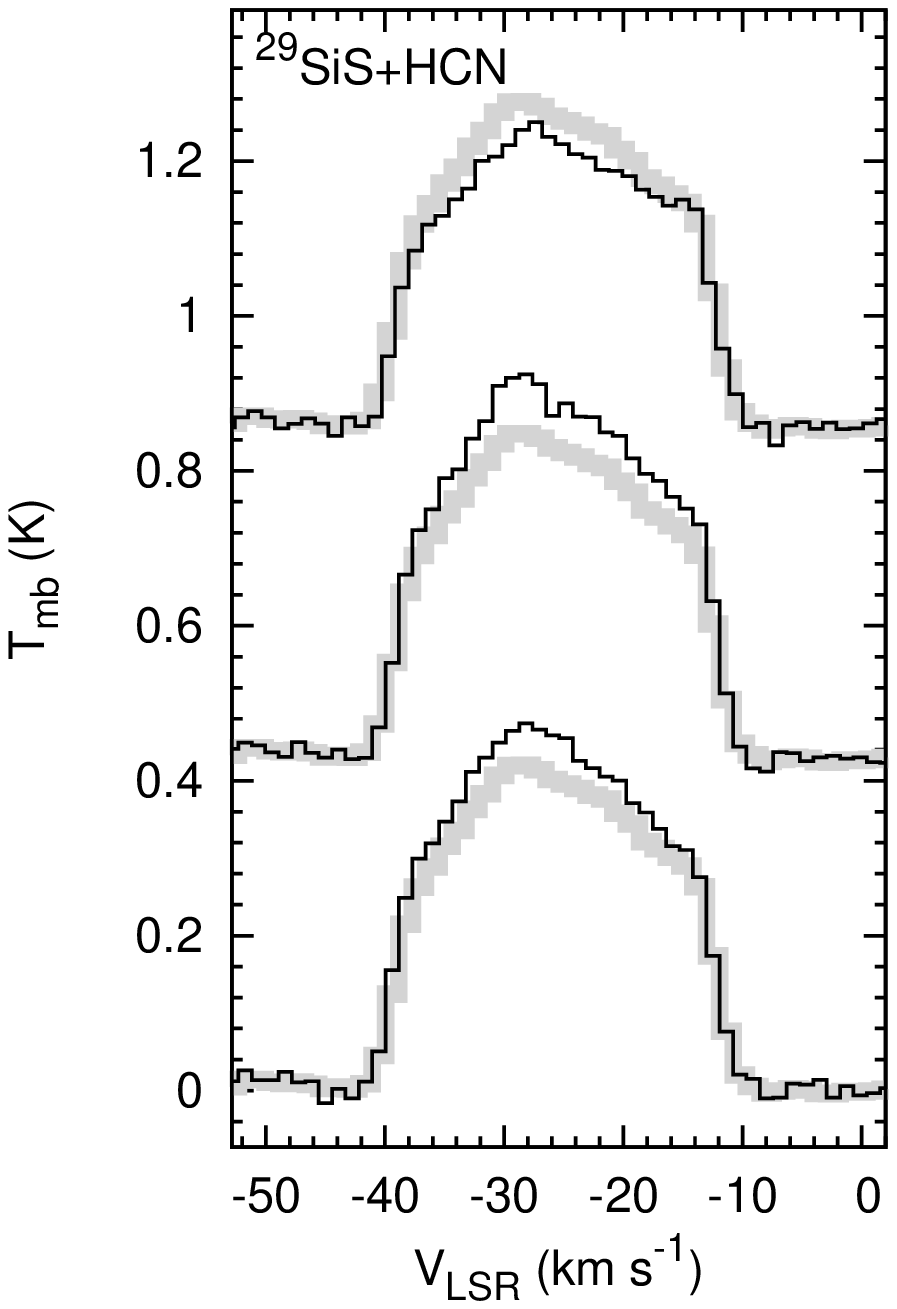}
\includegraphics[angle=0,scale=0.38]{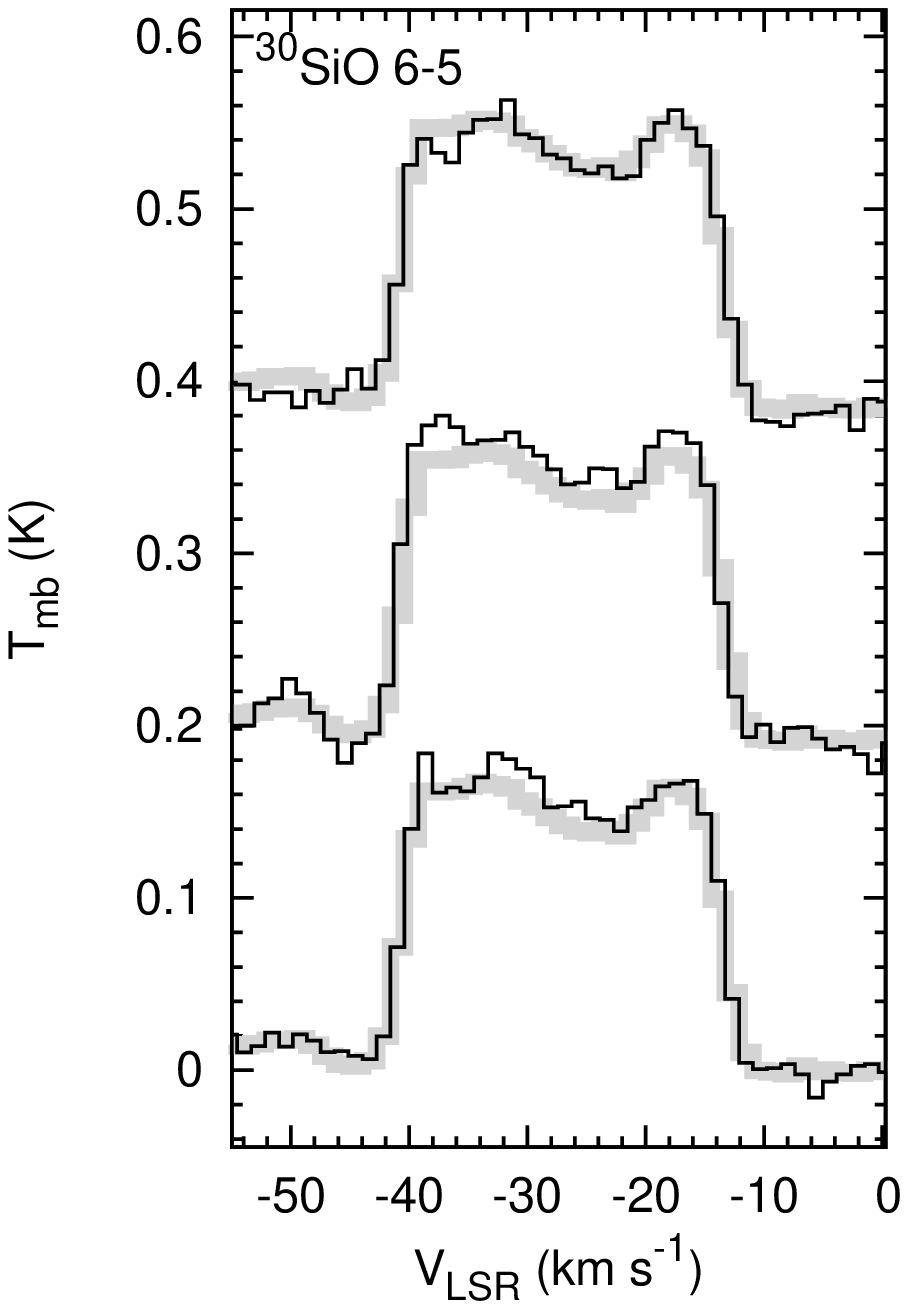}
\caption{Examples of varying spectral line profiles (without in-band calibration) at three selected epochs (from bottom curve to top curve in each panel): 0 day (corresponding to JD\,2454438.0 when most lines look similar as their average spectra --- the thick gray line ), 138 day (close to the IR maximum), 501 day (close to the IR minimum). {\em Top row}: the three in-band calibrator lines; {\em middle row}: three line features whose in-band calibrated strengths will be varying nearly in-phase with the IR light; {\em bottom row}: three line features whose in-band calibrated strengths will be varying differently from the IR light. }
 \label{fig_spec}
\end{figure*}

As an example, we show in Fig.~\ref{fig_spec} the spectral line profiles of selected stronger line features at three selected epochs to illustrate how they change with time. In particular they include the three in-band calibrator lines  in the top row and the  two candidate maser lines of SiS and HCN (the left panels in the middle and bottom rows).

\subsection{Relative variation of line strength} \label{sec:var-Iint}
 
The lines strengths are calibrated using relatively stable lines in each side band as calibrators: SiCC\,$11_{2,10}-10_{2,9}$ (LSB), and the average strength of C$_4$H\,N=28-27, J=57/2-55/2 and J=55/2-53/2 (USB). They are expected to be stable because they are from the vibrationally ground state and have extended emission regions in IRC\,+10216. Particularly,  the {\em Herschel} observations by \citet{Cern2014} have nicely confirmed the stability of the SiCC rotational lines in large beams.

An `in-band flux calibration factor' for each side band at each epoch is defined in two steps: first we compute the average strength of the calibrator lines over all epochs; then the calibration factor is set to be equal to the average to individual flux ratio of the calibrator-line. Each target line feature is in-band calibrated by multiplying with this factor. The calibration can be done either to integrated line strengths or to line profiles. 
However we note that, if the reference lines are also varying, their variations will be intricately involved in the so calibrated line fluxes. 

The relative line variation will be compared to the well established near infrared (NIR) variations. We adopt the NIR light curve of IRC\,+10216 from \citet{Ment12}. They used the NIR photometry of \citet{Shen11} from Dec. 10, 1999 to Nov. 11, 2008 and found a weighted-mean period of $630.0\pm 2.9$\,days and a maximum epoch at JD\,$2454554.0\pm 7.4$ from the light curves of H, K, L, M bands. Their light curves are slightly different from the NIR light curves of \citet{Mens01} who used photometry data during 1965 to 1998. Thus, the newer one is closer to our observation epochs.  \citet{Mens01} mentioned a K-band amplitude of about $1^{\rm m}$\footnote{They originally mentioned $2^{\rm m}$ which we guess is the full range of variation, because otherwise it is too large compared to the light curves in \citet{Ment12}.}. The amplitudes in the Fig.~A.1 of \citet{Ment12} range from about $0.76^{\rm m}$ in J-band to $0.55^{\rm m}$ in M-band. Thus, the NIR continuum flux variation amplitudes are in the range from 60\% to a factor of 2.5. The variation amplitude of the stellar luminosity can be inferred from Eq.~(1) of  \citet{Mens01} to be about 44\%. 

\begin{figure*}
 \centering
\includegraphics[angle=0,scale=0.45]{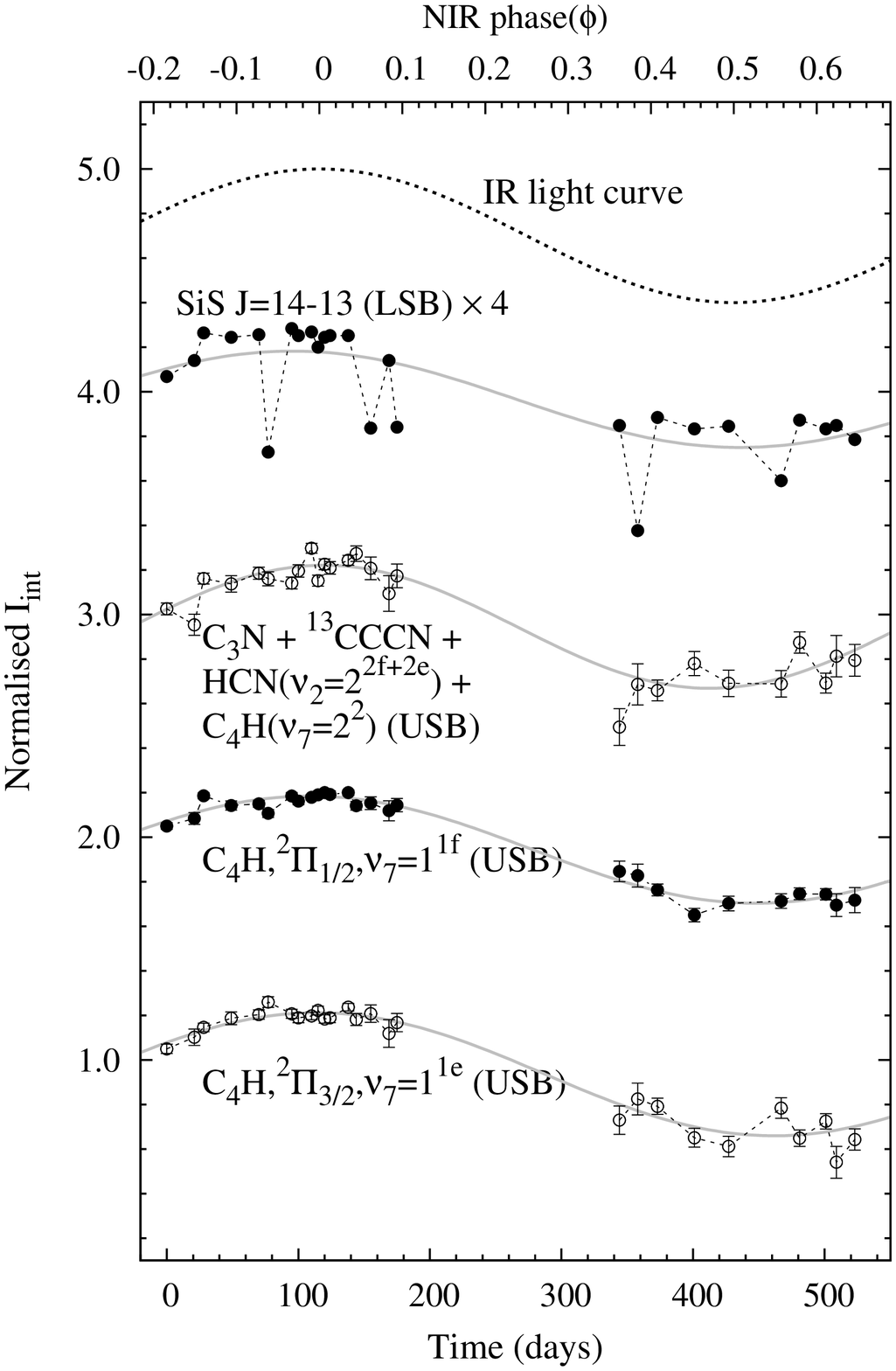}
\includegraphics[angle=0,scale=0.45]{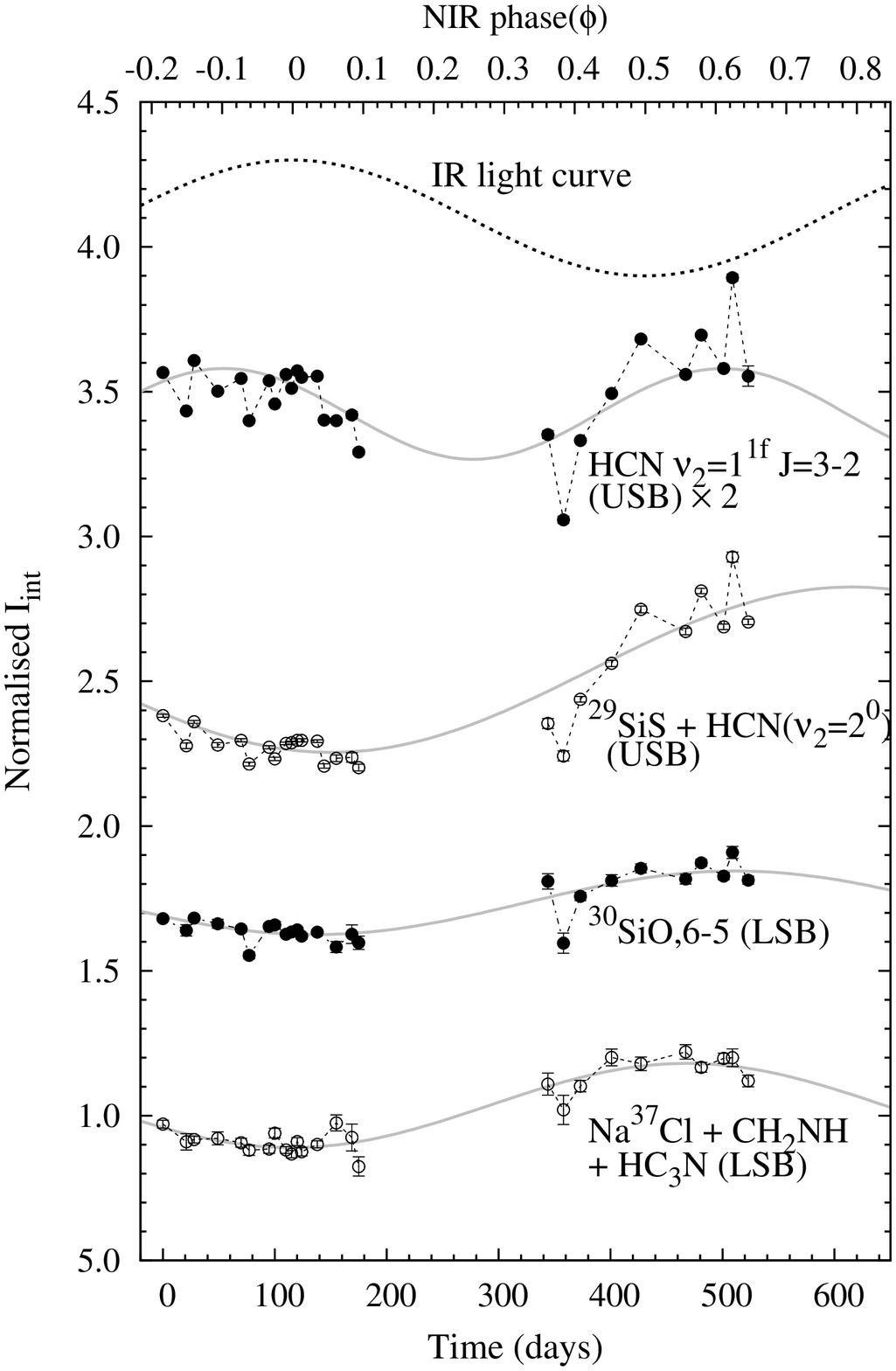}
 \caption{Light curves of integrated line intensity (after in-band calibration): {\em (Left)} line features that are (at least roughly) in-phase with the near infrared light; {\em (Right)} line features that are varying differently than the NIR light. The gray curves are the cosine function fittings (the fitted parameters are in Table~\ref{tab_linegroups}). The light curves are normalized to an average of unity and then shifted upward for clarity. The zero date is our first epoch (JD\,2454438.0) while the nearest zero NIR phase (at maximum, see the upper axis and the top dotted curves) is at JD\,2454554.0. The SiS and HCN line light curve are scaled up by a factor of four and two respectively for clarity.}
 \label{fig_IintLightCurve}
\end{figure*}

The relative light curves of the eight strongest line features are compared with the extrapolated NIR light curve in Fig.~\ref{fig_IintLightCurve}. The fluctuation of data points is larger in the second half of each light curves due to worse weather. The four light curves that closely follow the period and phase of the NIR variation are gathered in the left panel, while the other four that vary differently are in the right panel.

We can characterize the main properties of the relative light curves by fitting a sinusoidal light curve model to each of them:
\begin{equation}
\label{eq_cos}
I_{\rm int}(t) = A\cos[2\pi (t/P-\phi_0)]+Z.
\end{equation}
Here $t$ is the time, $A$, $P$ and $\phi_0$ are the amplitude, period and phase of maximum of the fitted sinusoidal component and $Z$ is the  the average line strength. All four parameters are free parameters determined through $\chi^2$ minimization weighted by observation errors. For convenience, the phase $\phi_0$ is expressed in terms of the NIR phase. 

The fitting results are shown in Table~\ref{tab_linegroups}. The last column of the table gives the minimum $\chi^2$ values of the model fitting. They range from about $40$ to $4390$ and are always larger than the degree of freedom (22 or 23, the number of epochs minus the number of free independent parameters), indicating that either additional statistic errors in addition to baseline RMS noise exist or the observed relative light curves can not be adequately represented by a single sinusoidal light curve. Thus the $1\,\sigma$ statistical errors of the parameters in the parentheses in Table~\ref{tab_linegroups} could be underestimated to some extent.

Despite of the possible underestimate of the parameter uncertainties, the fitting results in Fig.~\ref{fig_IintLightCurve} and Table~\ref{tab_linegroups} still reveal several important facts: 
(1) Most of the line features have their fitted periods not far from the NIR period of 630 days, although longer (e.g., 1182\,days for $^{29}$SiS + HCN\,$\nu_2=2^0$) and shorter (e.g., 444\,days for HCN\,$\nu_2=1^{\rm 1f}$\,J=3-2) periods exist;
(2) The relative amplitude ($A/Z$ from Table~\ref{tab_linegroups}) can reach about 30\% for the weaker line features, but remains at only 5\%-8\% for the two strongest lines, SiS\,J=14-13 and HCN\,$\nu_2=1^{\rm 1f}$\,J=3-2. Therefore, the relative line variations in this work are no larger than the NIR or luminosity variation (higher than 40\%).

\subsection{Line shape variation} \label{sec:var-profile}

We mainly discuss the channel by channel line shape changes of the two strongest lines, HCN\,$\nu_2$=$1^{\rm 1f}$\,J=3-2 and SiS\,J=14-13, in this subsection. The possible line shape variation of several other weaker lines is discussed in the on-line only Appendix~\ref{sec:App_subvrange} and is argued to be mainly due to pointing errors.

The variation of line shape can be investigated in two ways. One way is to directly normalize the line profile with respect to a specified reference velocity channel. The other way is to first apply the in-band calibration and then straightforwardly decompose the light curves of each velocity channel into three components: a `constant component' that does not change with time, a `co-varying component' that is entirely in-phase with the variation in the reference channel, and a `differential variation component' that is independent of the variation in the reference channel (see more details in the next paragraphs). The two ways involve the flux uncertainties differently, so that the comparison of their results can help us assess the successfulness of the in-band calibration procedure. 

Here we explain the above decomposition approach in more details. In a velocity channel, the minimum of the light curve (determined after a smoothing among five neighboring epochs) is taken as the level of constant component and removed from the light curve, leaving the purely varying component (with a minimum of zero). The purely varying component in the reference channel serves as the template of the co-varying component. It is scaled to other channels according to the strength of the constant component, and subtracted from the purely varying component to yield the differential variation component. The reference channel thus has no differential variation component by definition. 

The decomposition approach, as compared to the normalization, has better links to the various contributing physical processes in the CSE: the constant component mainly reflects the contribution from the lines of some abundant species in the outer cooler part of the CSE that are mainly collisionally excited and the minimum emission level of some varying lines that do not entirely extinguish at their minimum epochs; the co-varying component roughly represents those globally varying processes that can be represented by the variation in the reference channel (e.g., those driven by the varying IR light); the differential variation component then wraps all the rest variation processes occurring locally in individual velocity (and also spatial) components (e.g., radially beamed masers or local density enhancements).

\subsubsection{Results of normalization approach}
\label{sec:normlisation}

\begin{figure*}
 \centering
 \includegraphics[angle=0,scale=0.45]{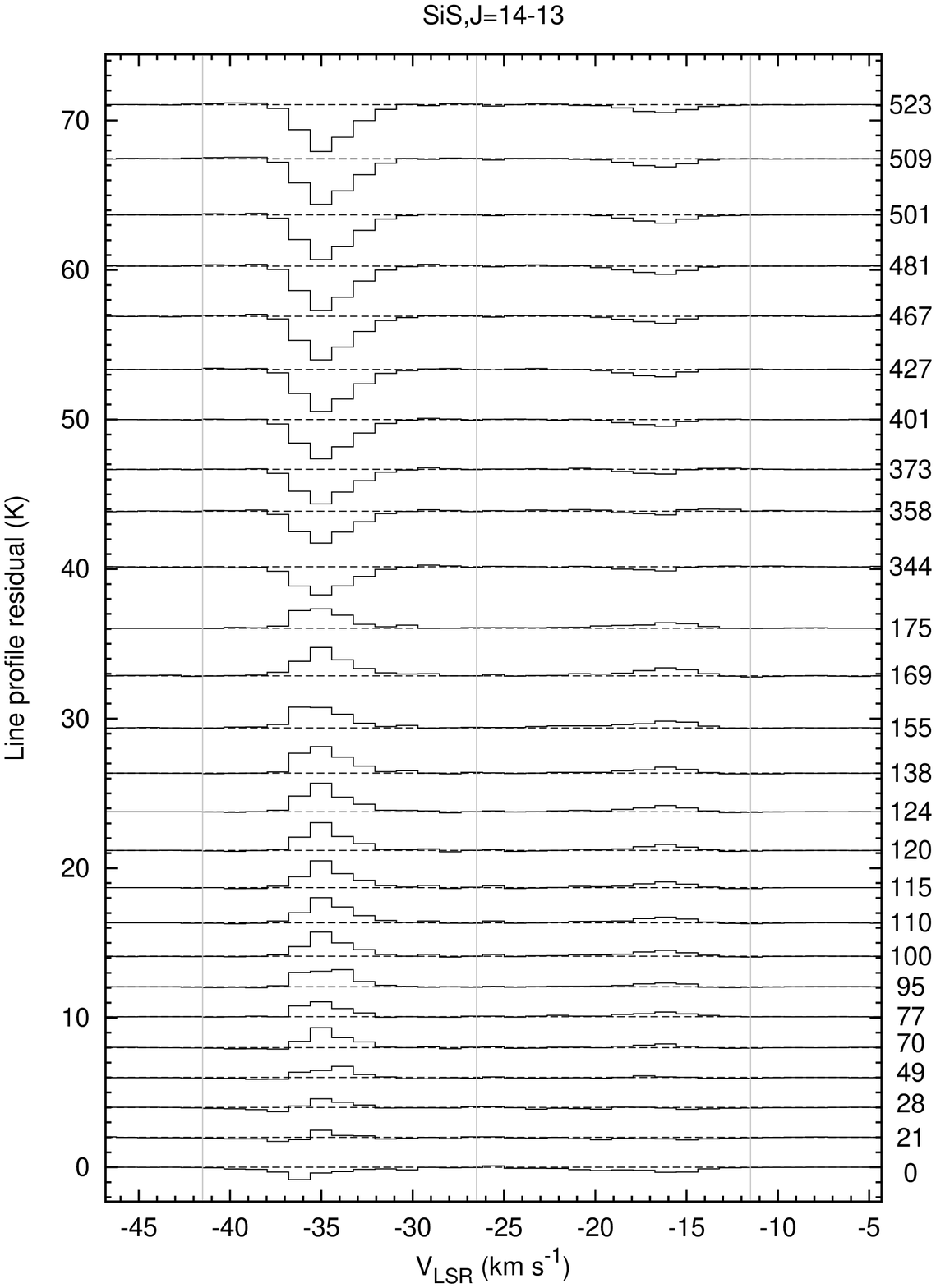}
 \includegraphics[angle=0,scale=0.45]{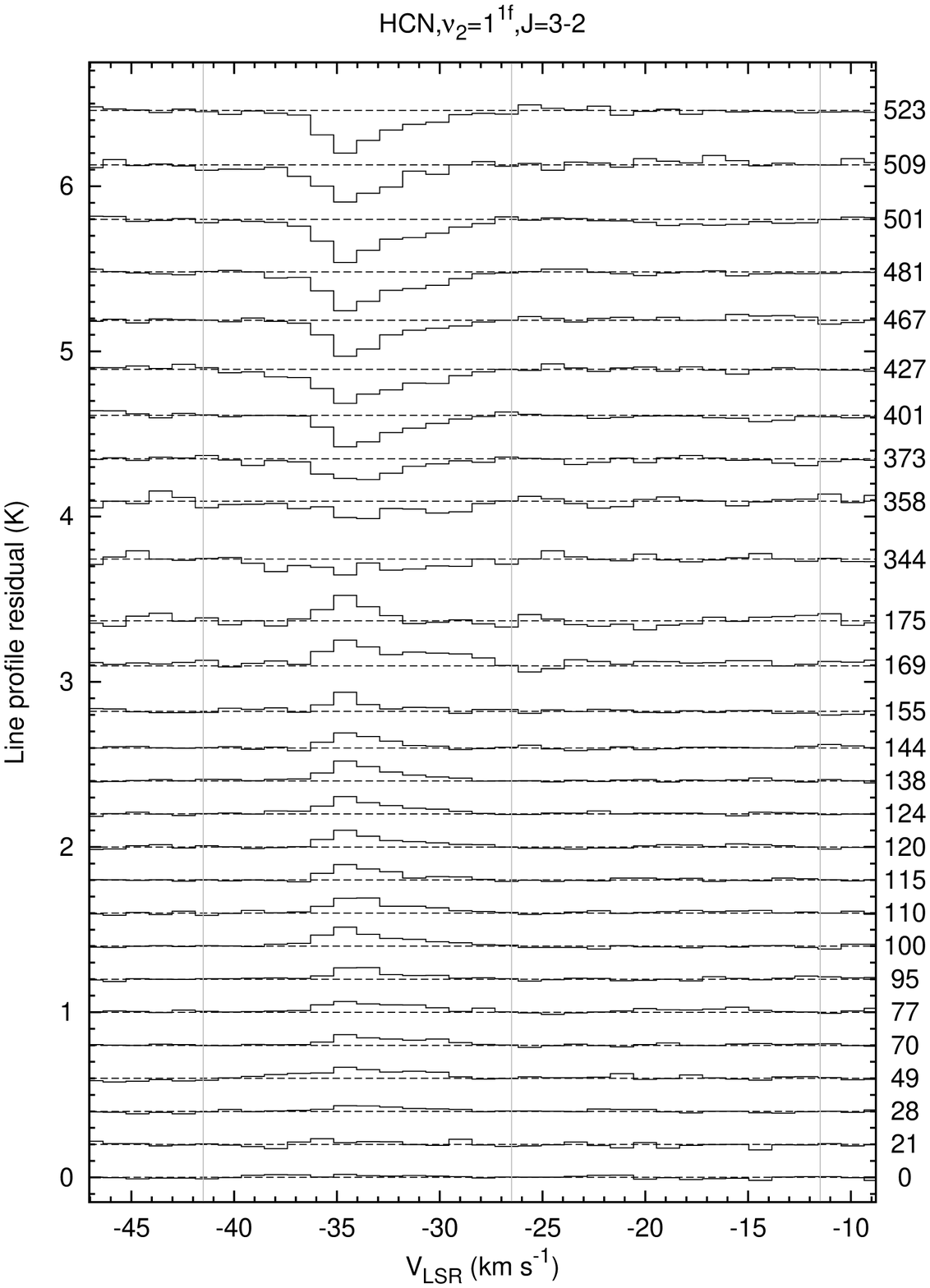}
 \caption{The variation of line profiles expressed in terms of line profile residual (after the normalization and average profile subtraction). The epochs are marked at the right edge of the plots, with the zero epoch being on JD\,2454438.0. The horizontal dashed lines mark the zero value for each epoch, while the vertical gray lines mark the systemic velocity at -26.5\,km\,s$^{-1}$ and the two expansion velocity extremes at $\pm 14.5$\,km\,s$^{-1}$. }
 \label{fig_profile}
\end{figure*}

After the normalization with respect to the systemic velocity channel (at -26.5\,km\,s$^{-1}$) and subtraction of the observed average line profile, we obtain the `line profile residual' as shown in Fig.~\ref{fig_profile}. 
In the left panel, the SiS\,J=14-13 line shows variability in two velocity intervals, with one in the blue wing and the other in the red wing. The variation in the blue wing is much stronger than in the red wing. The central part (-30 to -18\,km\,s$^{-1}$) of the spectrum shows little variation with respect to the systemic velocity channel, which means these velocity channels must be varying roughly in-phase with each other. 
In the right panel of the figure, the HCN\,$\nu_2=1^{\rm 1f}$\,J=3-2 line only shows variability in the blue wing of the line profile. However, the variation is visible in a broader range of velocity than the SiS\,J=14-13 line. 

\begin{figure*}
 \centering
 \includegraphics[angle=0,scale=0.39]{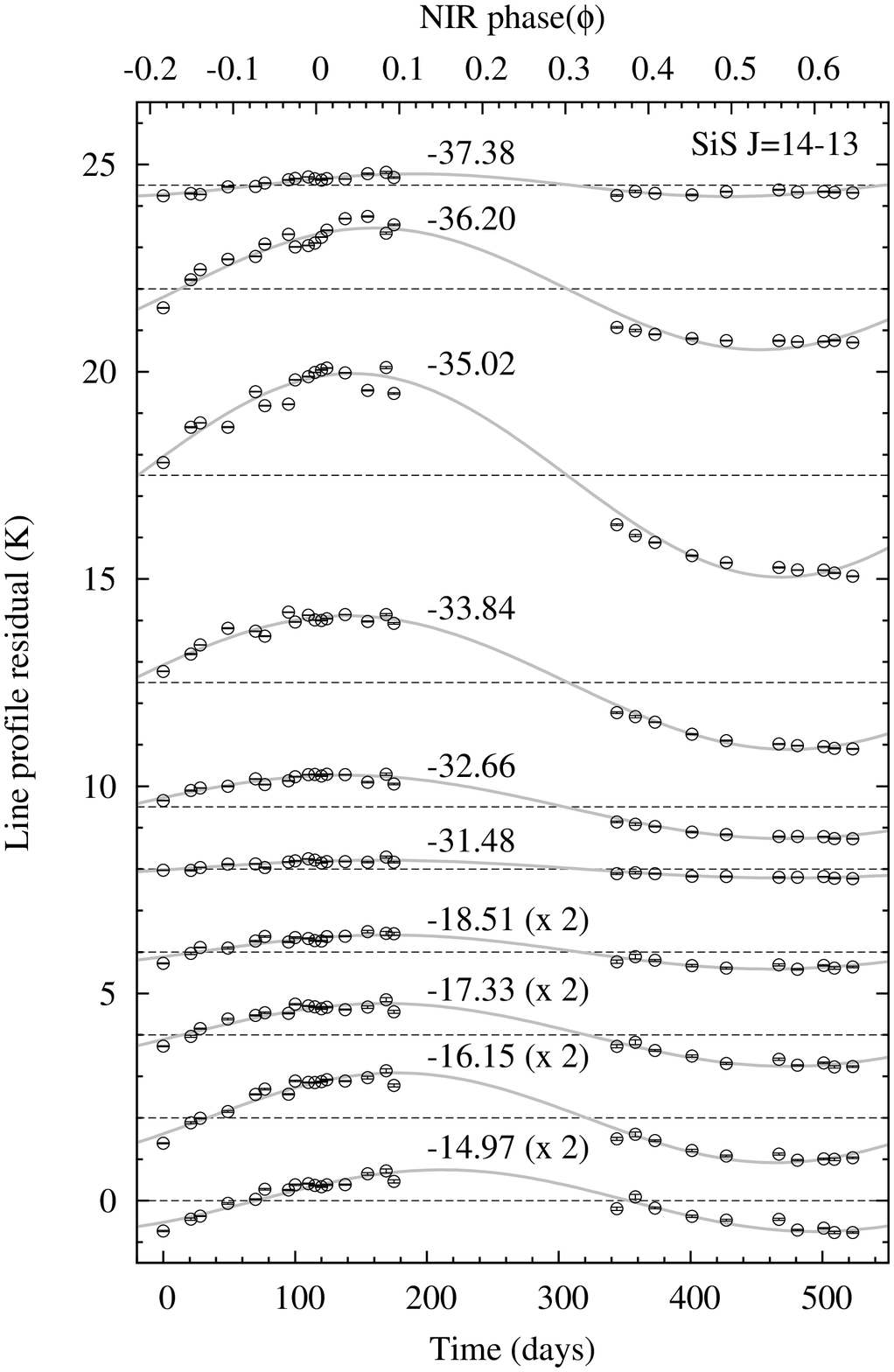}
 \includegraphics[angle=0,scale=0.39]{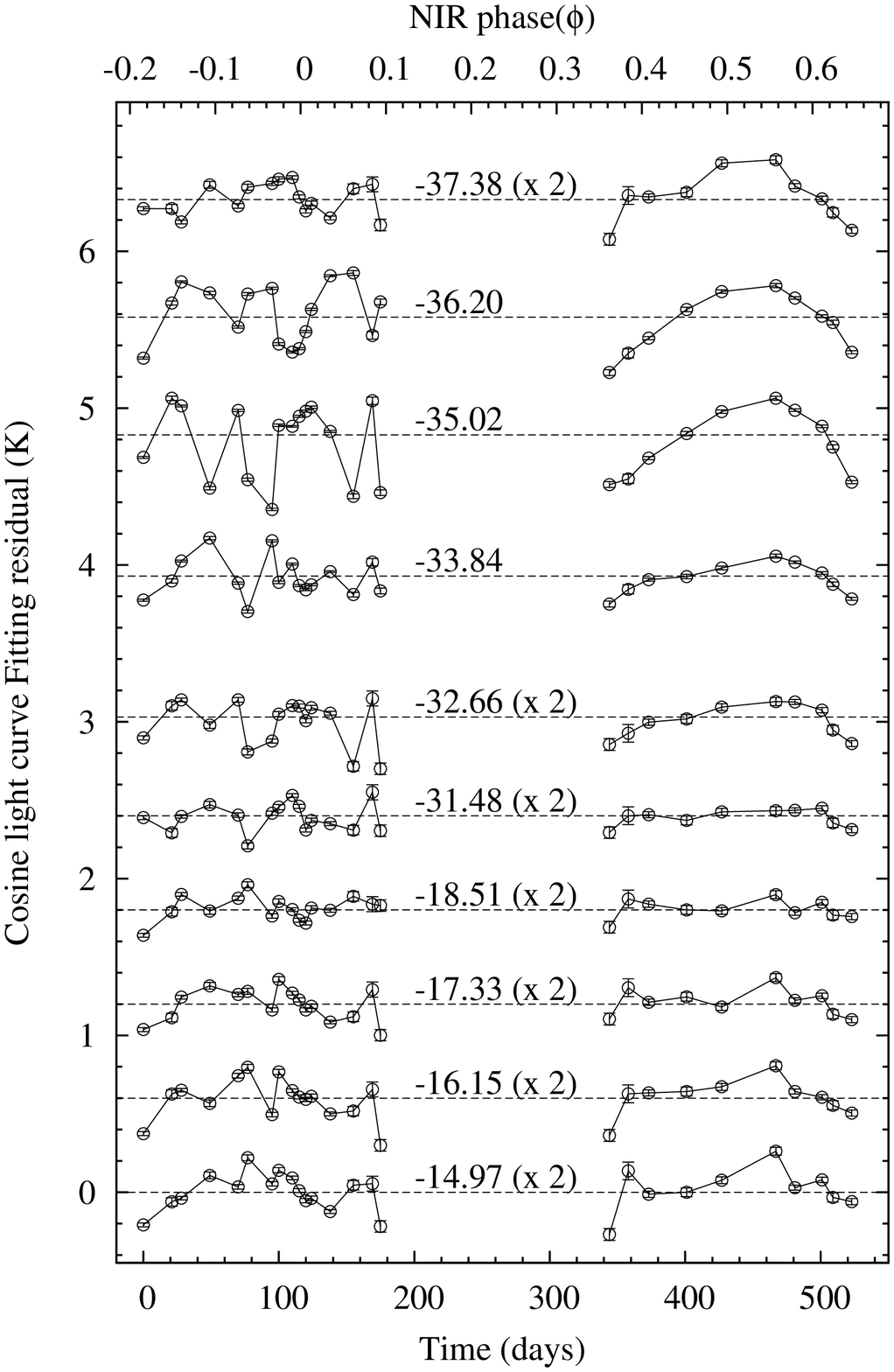}
 \caption{Channel by channel light curves ({\em left}) and the residual of cosine light curve fitting ({\em right}) of SiS J=14-13 line. The line profile residual is from Fig.~\ref{fig_profile}. The different velocity channels are offset for clarity, while the horizontal dashed lines mark the zero axis of the fitted cosine light curves or fitting residual. The velocity channel (km\,s$^{-1}$) is marked by the side of each zero line. The light curves of some weaker velocity channels are scaled up by a factor of two (including their error bars) to show details. }
 \label{fig_chan_lc_SiS}
\end{figure*}

\begin{figure*}
 \centering
 \includegraphics[angle=0,scale=0.39]{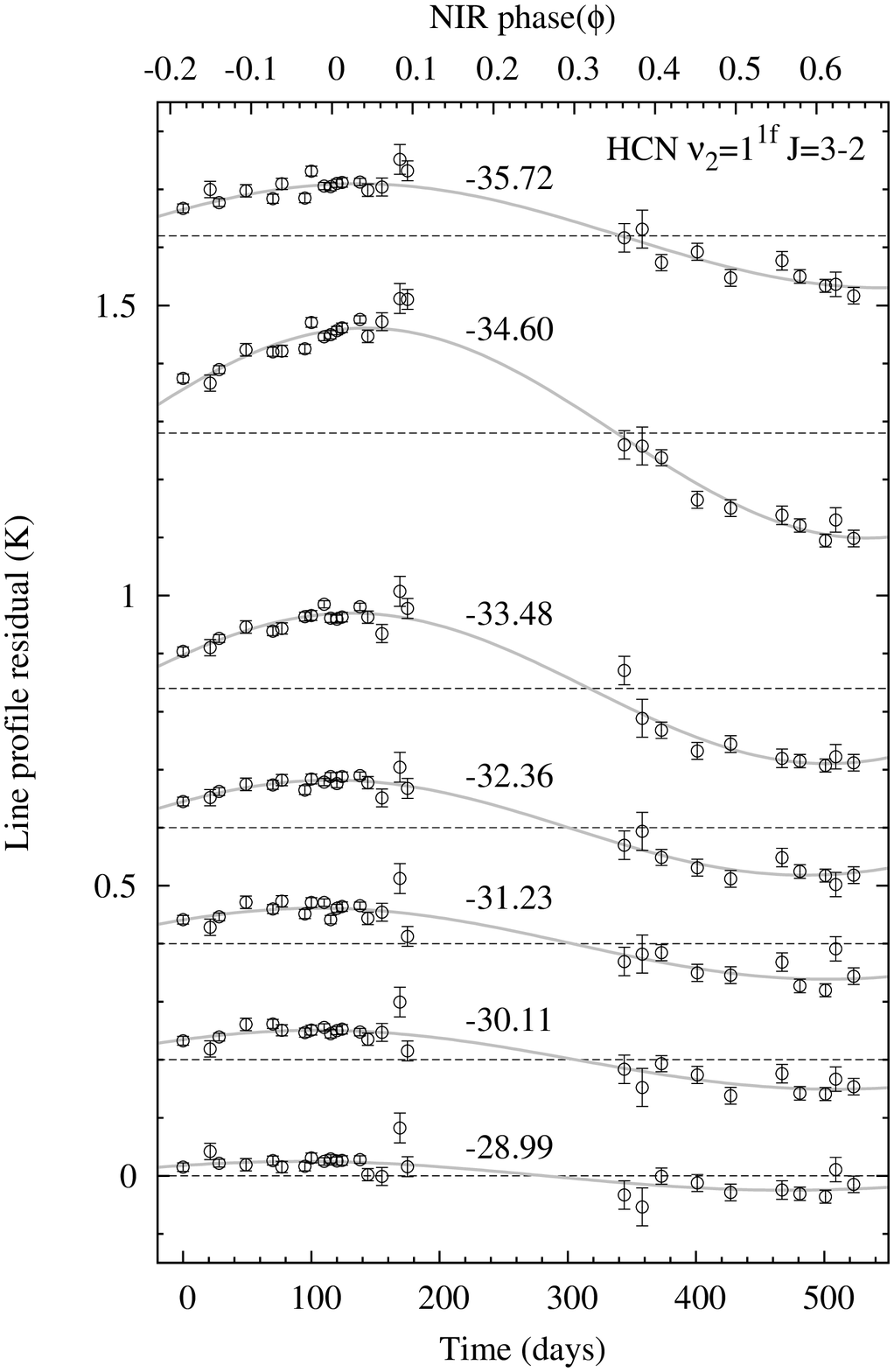}
 \includegraphics[angle=0,scale=0.39]{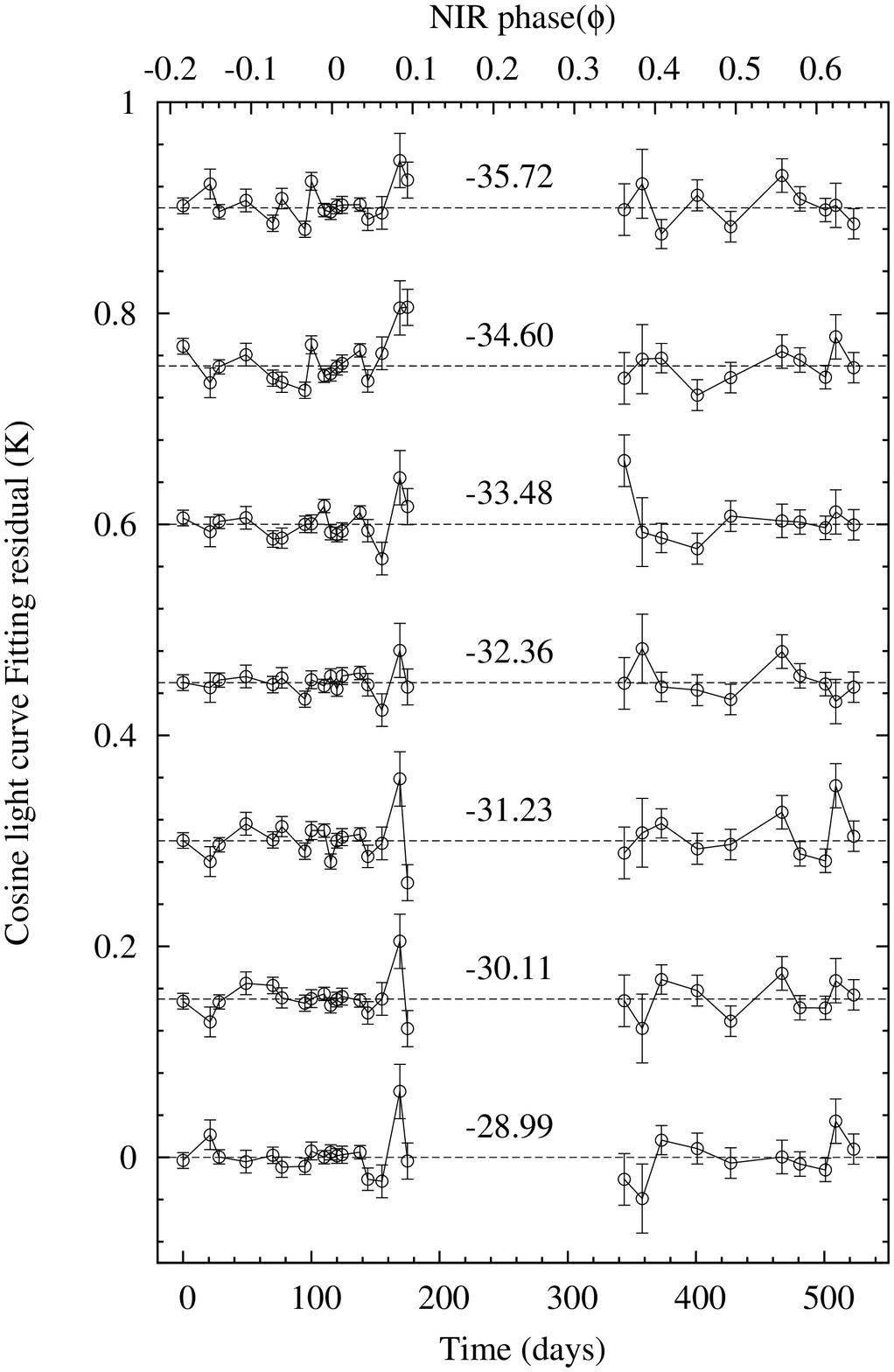}
 \caption{The same as Fig.~\ref{fig_chan_lc_SiS} but for the HCN $\nu_2=1^{\rm 1f}$ J=3-2 line. }
 \label{fig_chan_lc_HCN}
\end{figure*}

To quantify the channel by channel variability, we plot in Figs.~\ref{fig_chan_lc_SiS} and \ref{fig_chan_lc_HCN} the light curves of all velocity channels that show definite variability, together with their cosine function fits and fitting residual. The fittings are generally good.
The minimum $\chi^2$ values are about 20-60 for the HCN line. They are not too large compared to the number of freedom of 23 (= 27 data points - 4 free parameters), which means the cosine light curve is not a too bad model. 
For the SiS line, however, the minimum $\chi^2$ values are still larger than several hundreds ($\gg 22$, the degree of freedom). This means that the offsets of data point from the cosine fits still can not be explained by the statistical errors from baseline noise. In the strongest channels (the top 6 curves in the right panel of Fig.~\ref{fig_chan_lc_SiS}), smaller variability on shorter time scales of 2-4 weeks (earlier than the time gap) and large curvy shape (later than the time gap) can be seen. These additional variabilities and non-sinusoidal features are weaker by factors of 4-7 than the major sinusoidal components.

\begin{figure*}
 \centering
 \includegraphics[angle=-90,scale=0.7,trim=0 0 0 2cm,clip]{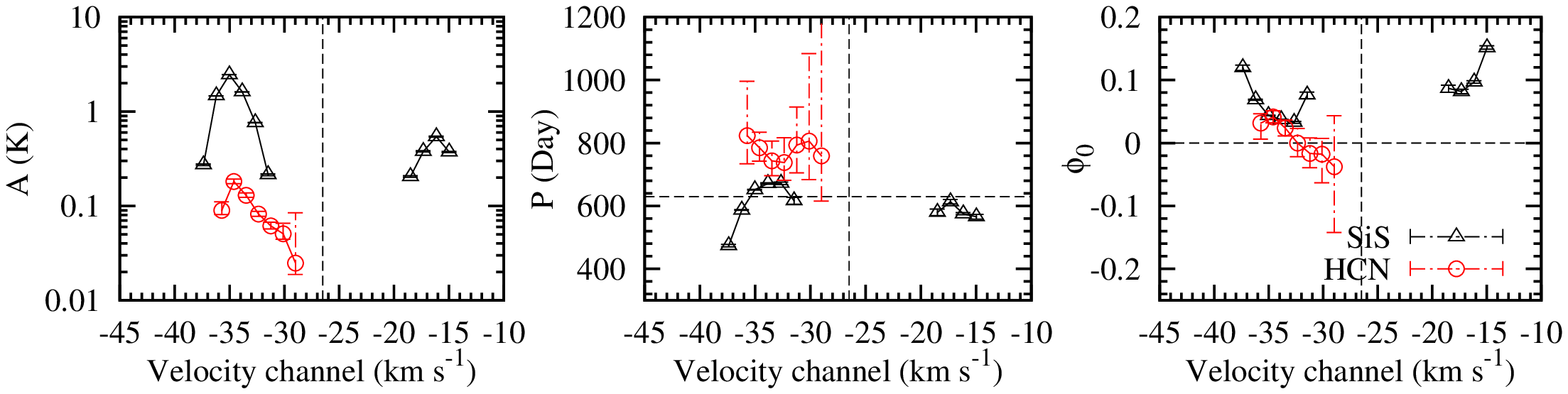}
 \caption{The light curve parameters of SiS J=14-13 and HCN $\nu_2=1^{\rm 1f}$ J=3-2 lines shown as functions of velocity channels: amplitude ($A$), period ($P$) and maximum time expressed in NIR phase ($\phi_0$). See the definition of the quantities in Eq.~(\ref{eq_cos}). The horizontal dashed lines mark the NIR period of $630$ days in the middle panel and the zero value in the right panel, while the vertical dashed lines mark the systemic velocity -26.5\,km\,s$^{-1}$ in all panels. (A color version is available on-line only.)}
 \label{fig_chan_lc_fit}
\end{figure*}

The light curve parameters, the amplitudes, periods and phases, are useful to understanding the physical nature of these varying spectral features. They are plotted as functions of velocity channels in Fig.~\ref{fig_chan_lc_fit}. The following properties can be recognized:
\begin{enumerate}
\item{{\em(Top left panel)} The maximum variation amplitude occurs around the velocity channel $-35\pm 0.6$\,km\,s$^{-1}$ in the blue wing of both SiS and HCN line profiles and around $-16.2\pm0.6$\,km\,s$^{-1}$ in the red wing of the SiS line (the uncertainties are simply set to about one half of the channel width). We call these velocity channels the {\em peak-channels} here after. Thus, the blue and red peak-channels are at relative velocities of $V_{\rm sys}-8.5$\,km\,s$^{-1}$ and $V_{\rm sys}+10.3$\,km\,s$^{-1}$, indicating a left-right asymmetry of about 2\,km\,s$^{-1}$.}
\item{{\em(Left panel)} The amplitude distribution of the HCN line is highly asymmetric with respect to the systemic velocity of the object. Variation only appears in blue wing and extends to velocities very close to $V_{\rm sys}$.}
\item{{\em(Left panel)} The full width at half maximum (FWHM, denoted as $\Delta V$) of the amplitude curve can be measured in the blue wing of the line profiles to be about 3\,km\,s$^{-1}$ for both the SiS and HCN lines. This can be roughly converted to a characteristic velocity dispersion of $\sigma_v=\Delta V /\sqrt[]{8\ln 2}=\sim 1.3$\,km\,s$^{-1}$ for the varying components of both lines.} 
\item{{\em(Left panel)} The amplitude in the blue peak-channel of the SiS line is $2.456\pm0.004$\,K, being $\sim 4.5$ times larger than that of its red peak-channel ($0.541\pm 0.005$\,K) and $\sim 13.6$ times larger than that of the blue peak-channel of the HCN line ($0.18\pm 0.01$\,K).} 
\item{{\em (Middle panel)} The SiS line has a variation period of $\sim 610$\,days around its red peak-channel, $\sim 670$\,days around its blue peak-channel, while the HCN line has a longer period of $\sim 730$\,days around its blue peak-channel.}
\item{{\em (Middle panel)} The variation periods show a velocity dependence, with the SiS line periods decreasing in channels away from its blue/red peak-channels in both directions while the HCN line periods increasing away from its blue peak-channel in both directions. The SiS line periods are close to the NIR period of 630\,days, while the HCN line periods are $\sim 100$-$200$\,days longer. }
\item{{\em (Right panel)} The blue peak-channels of the SiS and HCN lines show similar phase delay of $0.04\pm 0.01$ ($\sim 25\pm 6$ days) with respect to the NIR light, while the red peak-channel of the SiS line shows an additional phase delay of $0.053\pm 0.001$ ($\sim 33.4\pm 0.6$ days) with respect to its blue peak-channel. }
\item{{\em (Right panel)} The variation phases also show velocity dependence, with the SiS line phases increasing in channels farther away from its blue/red peak-channels in both directions while the HCN line phases decreasing in channels farther away from its blue peak-channel in both directions.}
\end{enumerate}

In a summary, the variability properties of the blue and red line wings are not symmetrical and there exist velocity dependent period and phase variation trends. The SiS and HCN lines show both similar and different or even opposite behaviors.

\subsubsection{Results of decomposition approach}
\label{sec:decomposition}

\begin{figure}
\centering
\includegraphics[scale=.4]{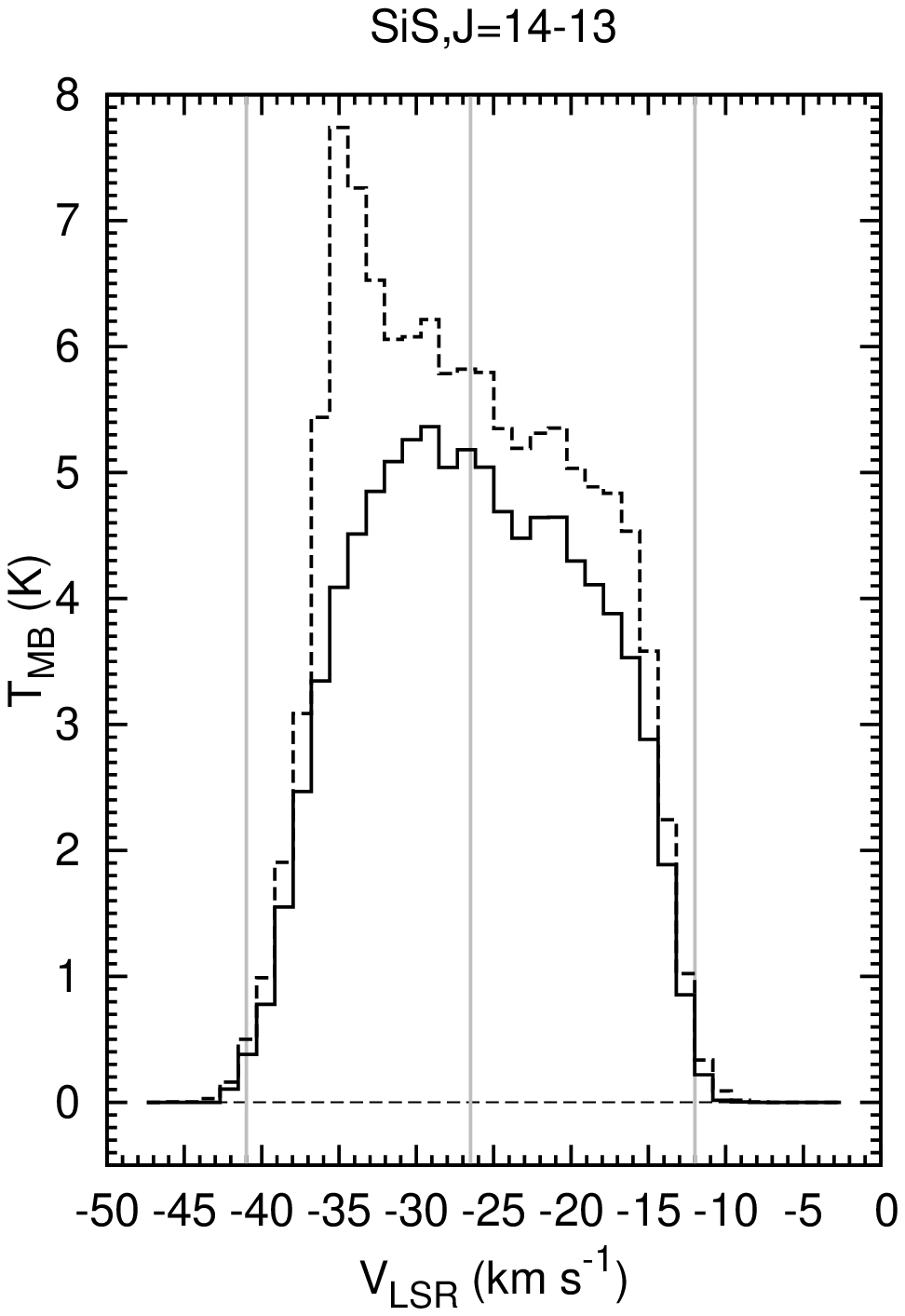}
\includegraphics[scale=.4]{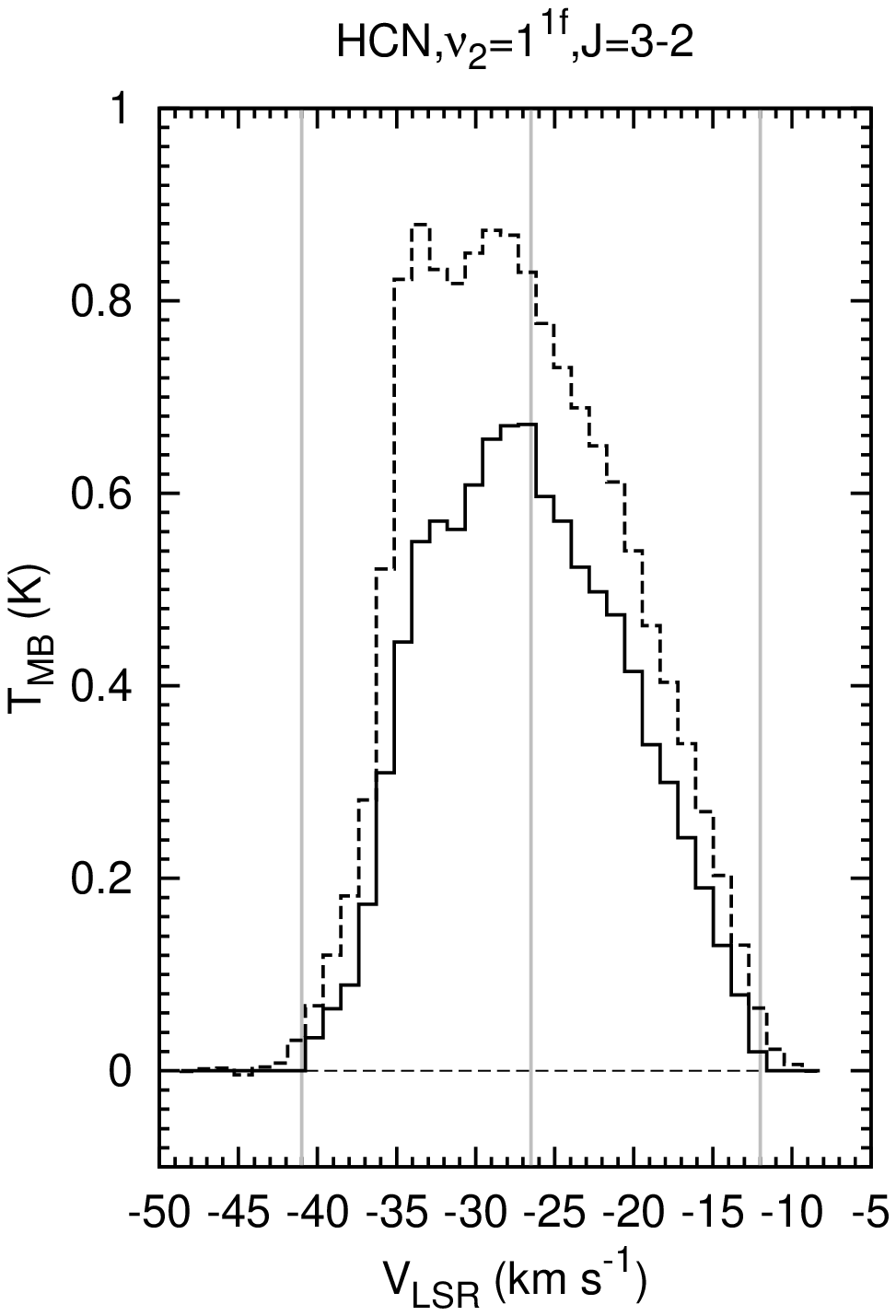}
\caption{Comparison of the constant component (solid histogram) of SiS J=14-13 and HCN $\nu_2=1^{\rm 1f}$ J=3-2 lines with their corresponding observed average line profiles (dashed histogram). The vertical gray lines mark the systemic velocity -26.5\,km\,s$^{-1}$ and the terminal outflow velocity 14.5\,km\,s$^{-1}$ at both sides.
}
\label{fig_const_profile}
\end{figure}

We first show the constant components of SiS J=14-13 and HCN $\nu_2=1^{\rm 1f}$ J=3-2 in Fig.~\ref{fig_const_profile}. Compared with the average line profiles in the same figure, the constant component profiles look more left-right symmetrical, as expected for a typical thermal line in this star. Therefore, the asymmetry in the average line profiles of the two lines mainly originates from the varying components.
\begin{figure*}
\centering
\includegraphics[angle=-90,scale=.4]{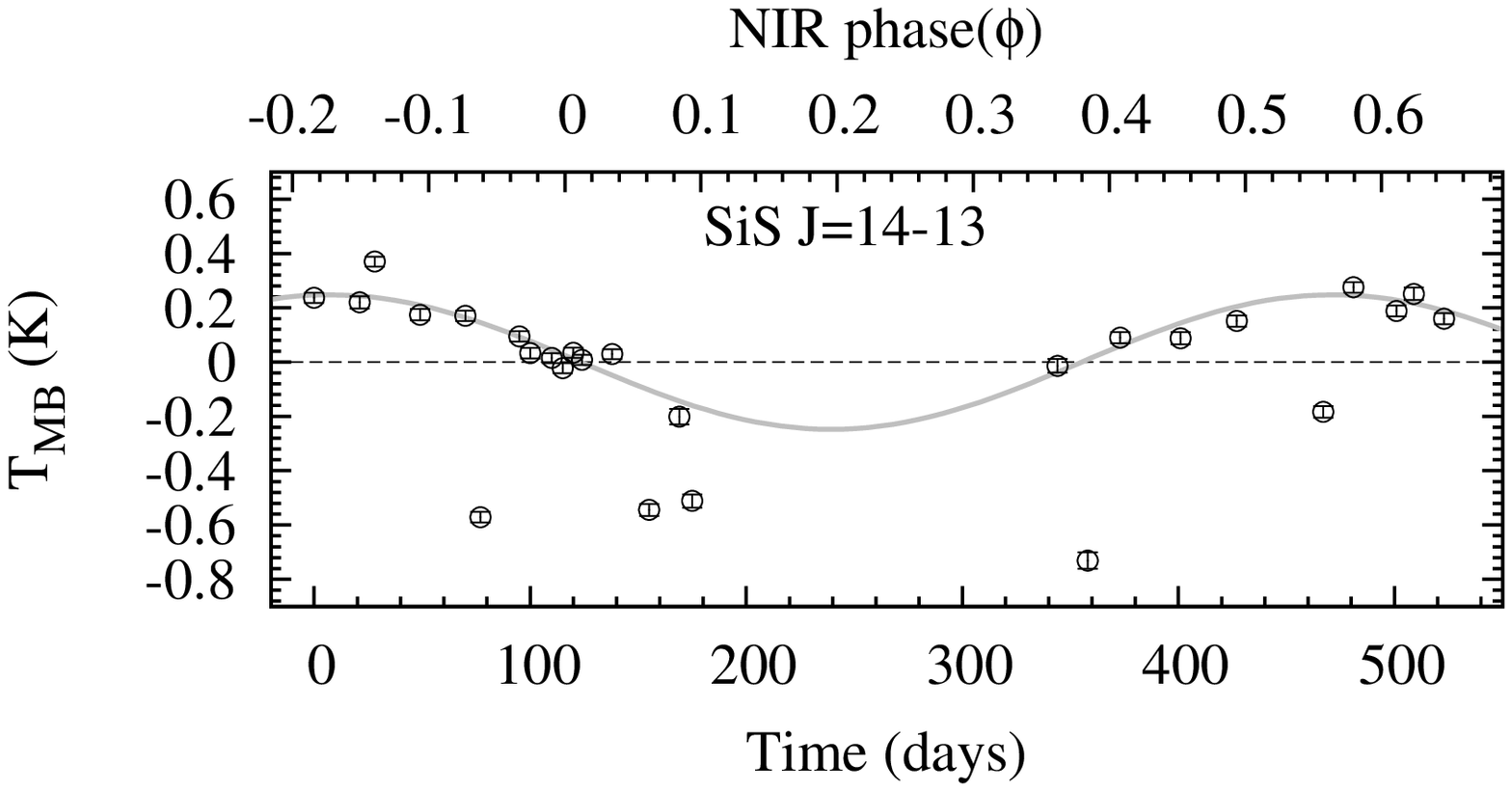}
\includegraphics[angle=-90,scale=.4]{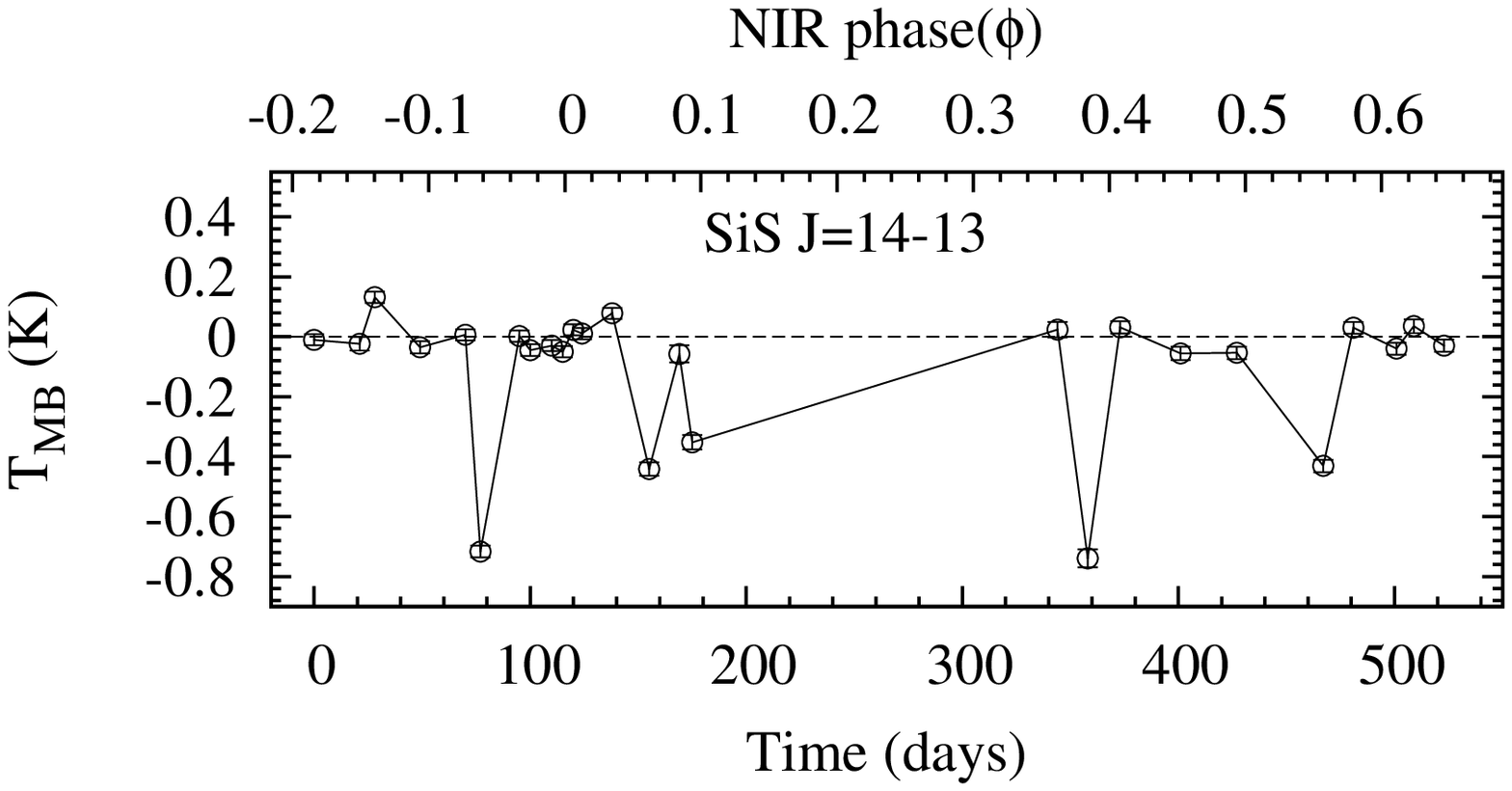}
\includegraphics[angle=-90,scale=.4]{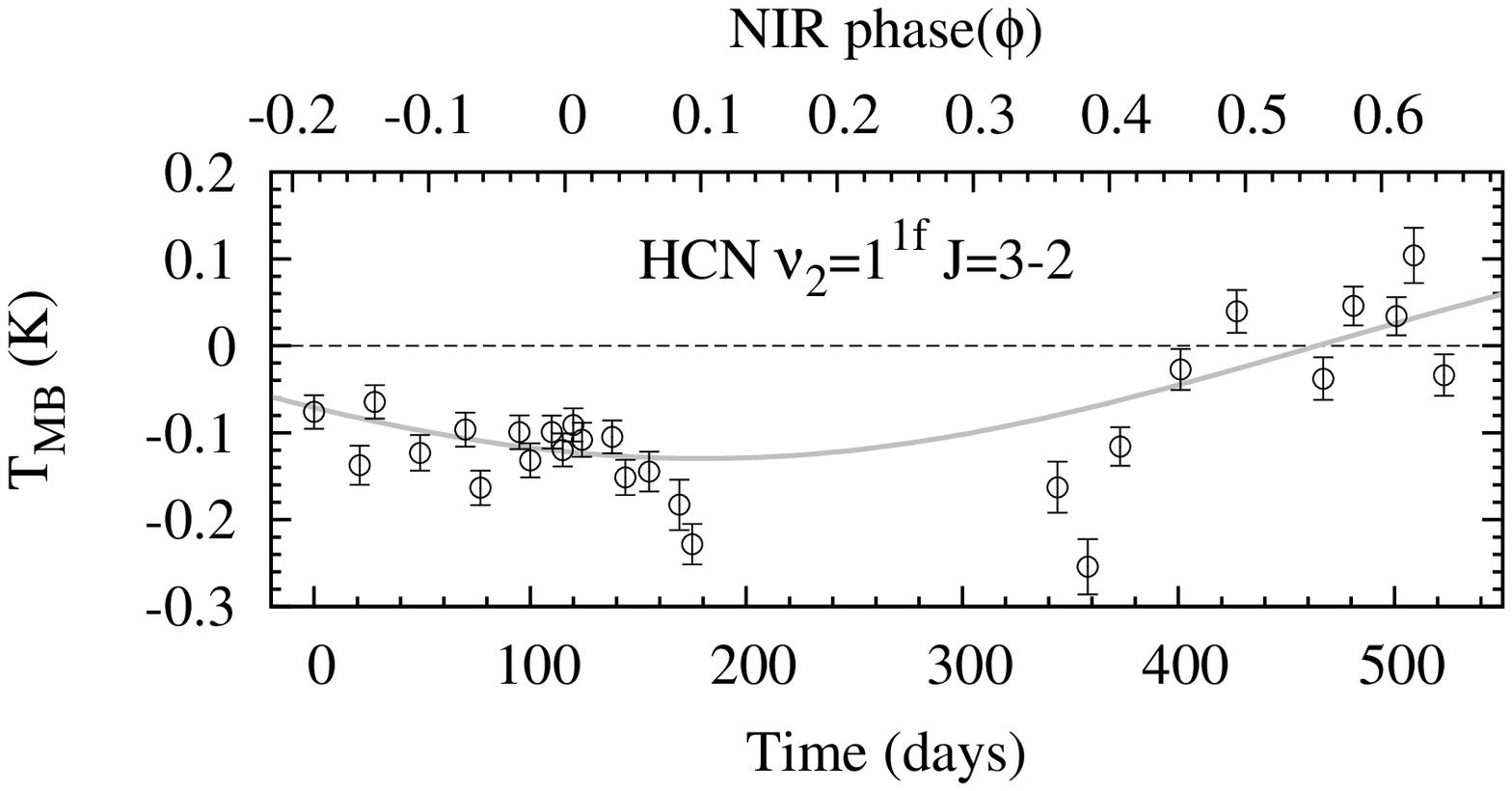}
\includegraphics[angle=-90,scale=.4]{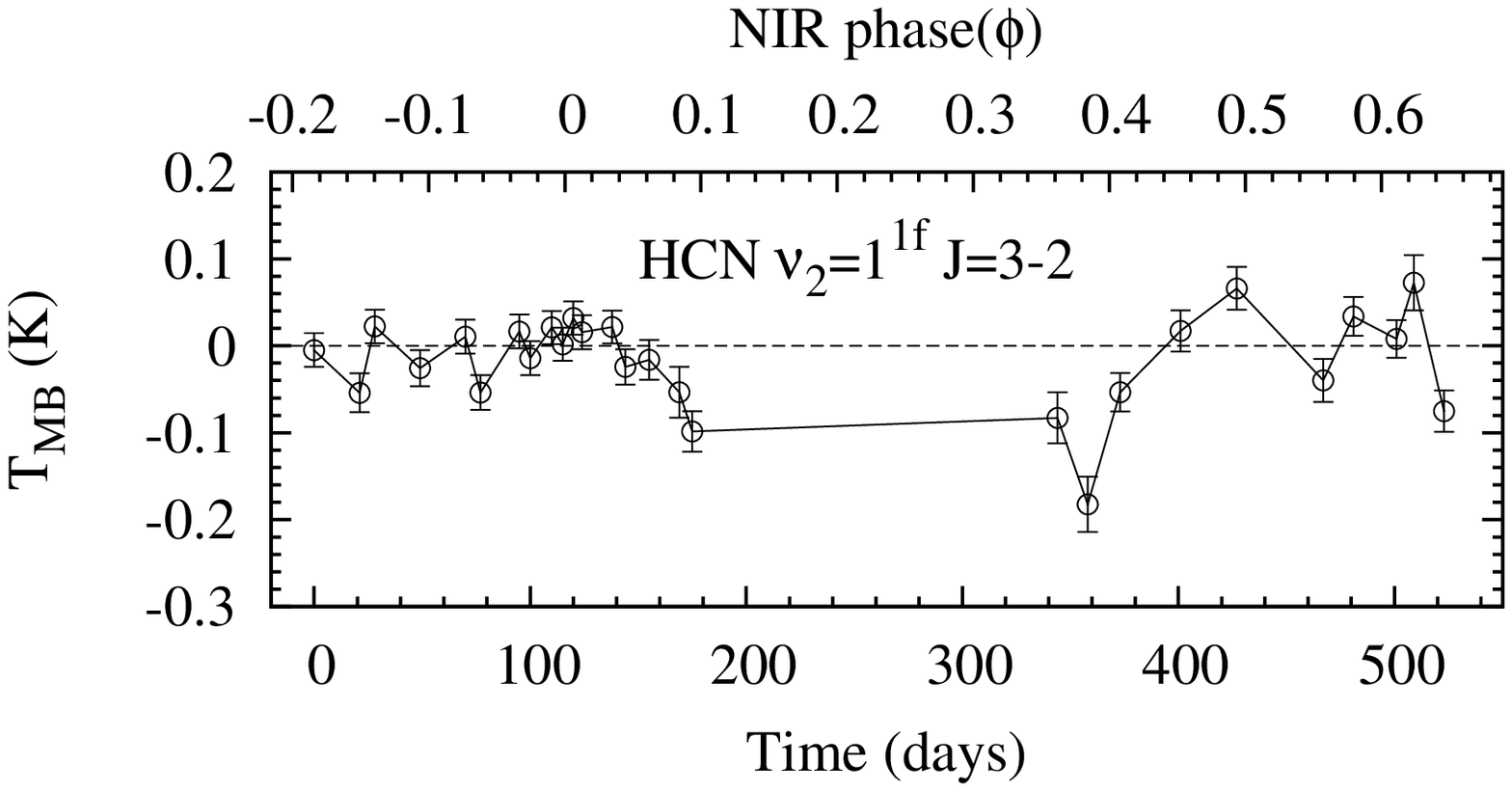}
\caption{The light curves of the co-varying component (in the $V_{\rm sys}$ channel) of the SiS J=14-13 and HCN $\nu_2=1^{\rm 1f}$ J=3-2 lines. {\em Left panels} show the light curve fitting and the {\em right panels} show corresponding fitting residual. The horizontal dashed lines mark the zero axis of the fitted cosine light curves or fitting residual. The five outlier epochs in the SiS light curve (the exceptionally weaker points) are excluded from the light curve fitting of both SiS and HCN lines.
}
\label{fig_covar}
\end{figure*}

Different from the normalization approach, the decomposition approach provides information on the variation in the reference channel at $V_{\rm sys}$ (after in-band calibration): the co-varying component. The period and phase of this component of both the SiS and HCN lines are shown in Fig.~\ref{fig_covar} to be significantly different from the NIR light variation (NIR phase marked on the top axis). 
\begin{figure*}
 \centering
 \includegraphics[angle=-90,scale=0.7,trim=0 0 0 2cm,clip]{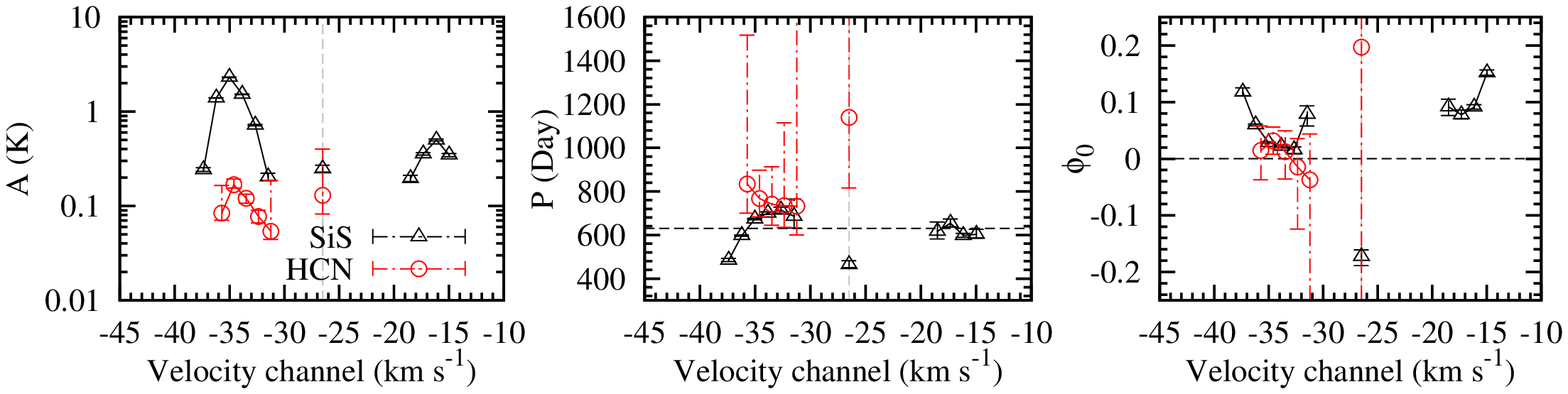}
 \caption{The same as Fig.~\ref{fig_chan_lc_fit} but for the co-varying and differential variation components of the SiS J=14-13 and HCN $\nu_2=1^{\rm 1f}$ J=3-2 lines. The co-varying components are at the systemic velocity -26.5\,km\,s$^{-1}$. (A color version is available on-line only.)
 }
 \label{fig_chan_lc_fit_3parts}
\end{figure*}

The channel-by-channel light curves of the differential variation components look very similar to the light curves in the normalization approach (in Figs.~\ref{fig_chan_lc_SiS} and \ref{fig_chan_lc_HCN}). We do not show the similar light curves, but show in Fig.~\ref{fig_chan_lc_fit_3parts} the parameters from sinusoidal light curve fitting. The light curve parameters of the co-varying component is also plotted at the $V_{\rm sys}$ channel for comparison.

Comparing the parameters of the co-varying components (at $V_{\rm sys}$) with that of the differential variation components (in line wings) in Fig.~\ref{fig_chan_lc_fit_3parts}, we see the following new facts: 
\begin{enumerate}
\item {\em (Left panel)} The peak amplitudes of the differential variation components are significantly larger (SiS line) or slightly larger (HCN line) than that of the co-varying component.
\item {\em (Middle panel)} The period of the co-varying component ($464\pm 16$\,day) of the SiS line is much shorter than that of the differential variation components and the NIR light (630\,day). The period of the co-varying component of the HCN line ($1139^{+1061}_{-323}$\,day) looks much longer than that of the differential variation components and the NIR light, although its uncertainty is quite large.
\item {\em (Right panel)} The phase of the co-varying component of the SiS line is also very different from that of the differential variation components. The phase of the co-varying component of the HCN line is very noisy.
\end{enumerate}

\begin{figure}
 \centering
 \includegraphics[angle=-90,scale=0.6]{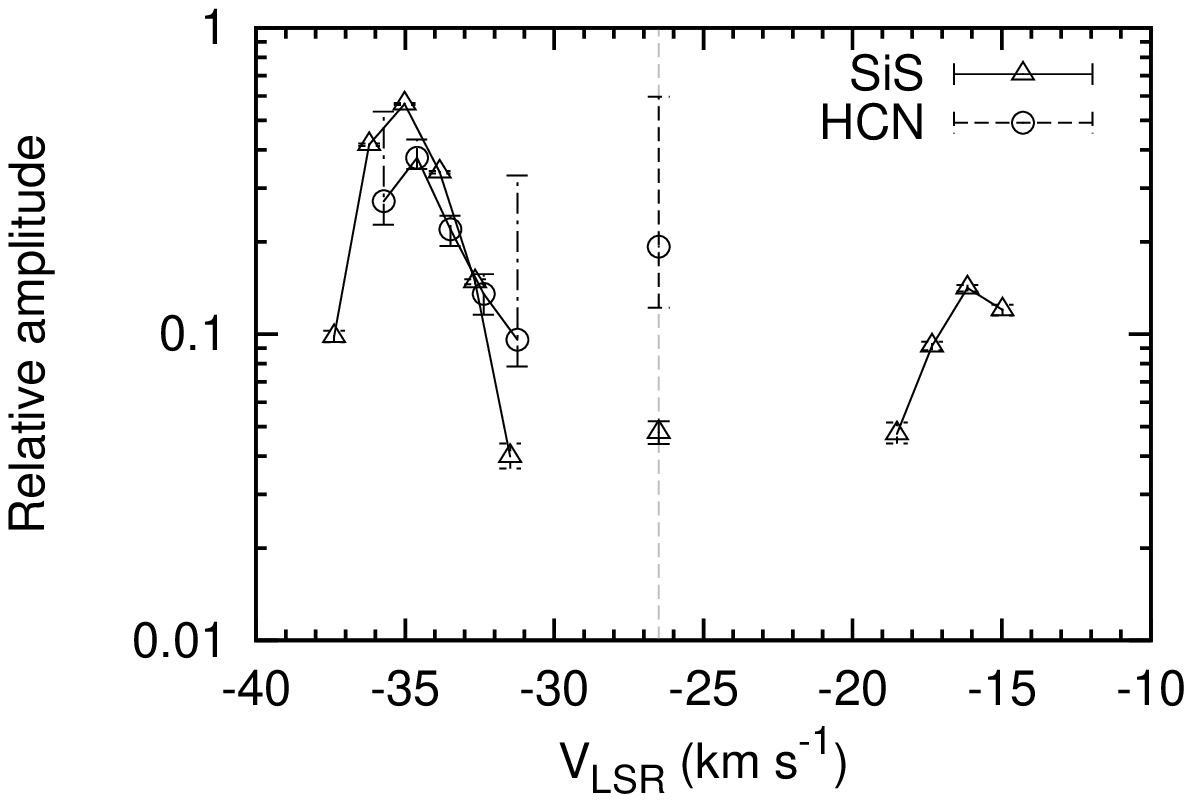}
 \caption{The channel by channel distribution of relative amplitudes (the absolute amplitude to constant component ratio) of the co-varying and differential variation components of the SiS J=14-13 and HCN $\nu_2=1^{\rm 1f}$ J=3-2 lines. The co-varying components are at the systemic velocity (the vertical line).
 }
 \label{fig_chan_relative_amp}
\end{figure}

The varying component to constant component ratios (relative amplitudes) are found to be smaller than unity in all velocity channels, as shown in Fig.~\ref{fig_chan_relative_amp}. The SiS line has a peak relative amplitude of $56.4\pm 0.5$\% in the blue wing and $14.1\pm 0.4$\% in the red wing for its differential variation components, while its co-varying component has a much smaller relative amplitude of only $4.8\pm 0.4$\%. The HCN line has a peak relative amplitude of $38\pm 4$\% for its differential variation components in its blue wing, while its co-varying component has an noisy relative amplitude of $19^{+40}_{-7}$\%.

\section{Discussion} \label{sec:discuss}
\subsection{Uncertainties in the relative light curves} \label{sec: discuss_uncertainty}

Although the traditional flux calibration uncertainty (typically $\sim20$\%) has been greatly reduced in the in-band calibrated relative light curves and the line shape light curves, there could be remaining effects of the uncertain factors that are neither completely removed nor fully reflected by the baseline noise of the spectral line data. They may prevent us from understanding the astrophysical origin of the relative light curves. 

The most important additional source of uncertainties is the telescope pointing fluctuation.  Different molecular lines can be excited in different regions of the CSE (e.g., compact, extended or shell like regions) and the gas density and chemical abundances are not necessarily homogeneously or smoothly distributed. Thus, the different lines and the different velocity channels of the same line can respond to the random telescope pointing offsets differently. The pointing uncertainty of the telescope is usually better than $3\arcsec$ ($3\sigma$), being small compared to the beam size of $29\arcsec$. The intensity variation of a point source is only about 2\% at such a beam offset, being much smaller than the intensity variation (usually 10\%-20\%) found in this work. An extended smooth emission region would be even less sensitive to such small pointing offsets. For those molecules with a shell like emission structure that has a comparable size as our telescope beam (e.g., our in-band calibrator lines of SiCC and C$_4$H belong to this case), both their strengths and line shapes may be more sensitive to the telescope pointing offsets. However, as we discuss  in detail in the on-line only Appendix~\ref{sec:App_subvrange}, the effect of this to their line strengths is no larger than 3\%. In addition, the pointing error should be random in nature and thus can not explain the large regular long-period variation found in this work. 

The other additional uncertainty sources are less important. There is no strong telluric line in the bandpass. The side band leakage is not an issue in our data, because the strong lines in the image side bands were all known before hand and were carefully arranged to avoid overlapping with lines in the signal side band. The image rejections were also high (10-20\,db). The spectral baselines are also very straight and thus are largely free of baseline problems. 

Therefore, we conclude that the observed regular long-period variations of both line strengths and line shapes are not due to observational error, but are physical in the object.

\subsection{Interpretation of the line strength variation} 
\label{sec: discuss_rel_lc}

We have no reliable evidence to prove how and where the observed in-band calibrated line variations (Sect.~\ref{sec:var-Iint}) are produced. However, we can discuss one possibility: they very possibly originate from the inner part of the CSE. The line emission in the outer cooler part of the CSE is usually dominated by collisional excitation by H$_2$ which is not very sensitive to the central star pulsation. However, observable line emission variation could arise through various mechanisms in the inner hot part of the CSE through larger changes of dust and gas temperatures due to the varying heating by IR light, pulsation shock propagation in the extended atmosphere, variation of chemical abundances due to periodically varying shock chemistry, dynamical instability of the dust formation processes \citep{Flei1995} and maser effects. Particularly, the shock propagation can result in diverse variation periods and phases \citep[see e.g.][]{Bess1996,Hoef1998,Loid1999,Hoef2003,Gaut2004,Nowo2005,Nowo2010}. The fluctuating chemistry in the shock regions \citep[e.g.,][]{Cher2011,Cher2012,Mari2016,Gobr2016} can produce temporal variation of abundances in the shock regions and at larger downstream radii to produce observable effects in line emission and this process could have periods and phases different from the IR light. 

The feasibility of attributing the observed line variations to small regions within or under the wind acceleration zone can be tested by checking the possible brightness temperatures of the varying components of the line features at various assumed size scales. From Table~\ref{tab_linegroups}, the amplitudes of all eight line features range from 0.46-7.36\,K\,km\,s$^{-1}$. Assuming a line width of 29\,km\,s$^{-1}$ (for those blended lines, we assume one line dominates), the absolute amplitudes in main beam temperature in our beam of $29\arcsec$ is about 16-254\,mK. If the varying line emission all comes from the wind acceleration region \citep[diameter of $22\,R_*$ or $0.44\arcsec$;][]{Deci2015}, the brightness temperatures are 70-1103\,K, which is in a reasonable range. If however, some of the varying line emission would directly come from the small shock region \citep[diameter of $6\,R_*$ or $0.12\arcsec$;][]{Nowo2010,Mari2016}, the line width would be much narrower \citep[e.g., two times the sound speed: 4\,km\,s$^{-1}$;][]{Mari2016}, the similarly scaled line brightness temperatures become $6700-1\times 10^5$\,K. This is too high for thermal line emission. Thus, the observed line variations can not solely come from the small shock region, unless some of them are masers.

Furthermore, literature data and excitation considerations also support that the line variation can originate from the inner CSE. Five of the eight varying line features are rotational transitions in vibrationally excited states and thus are mainly excited in the inner hot dense regions of the CSE. For the remaining three line features, SiS\,J=14-13 ($E_{\rm up}=91$\,K), $^{30}$SiO\,J=6-5 ($E_{\rm up}=43$\,K), and the blended line feature Na$^{37}$Cl+CH$_2$NH+HC$_3$N ($E_{\rm up}=$116, 18 and 165\,K respectively), all of the involved molecules have a strong compact emission component near to the central star, as revealed by the sub-millimeter imaging spectral line survey of \citet{Pate2011} or recent ALMA observations \citep[e.g.,][]{Veli2015}. 

Therefore, the in-band calibrated millimeter-line light-curves very possibly originate from the small region before the terminal wind velocity is reached. They have the potential to be used as tracers of the dynamical processes in this region around all AGB stars.

\subsection{Interpretation of line shape variation.} \label{sec: discuss_rel_subrange}

The good agreement of the channel-by-channel line-shape light-curves between the normalization and decomposition approaches has demonstrated that the in-band line flux calibration is successful and thus the light curve parameters from the decomposition approach (Figs.~\ref{fig_chan_lc_fit_3parts} and \ref{fig_chan_relative_amp}) can be trusted. Fig.~\ref{fig_chan_relative_amp} shows that the relative variation amplitudes in the strongest varying channels in the line wings are larger than that in the $V_{\rm sys}$ channel, which explains why we see the variation of line shape of the SiS J=14-13 and HCN $\nu_2=1^{\rm 1f}$ J=3-2 lines. Because the variation in the $V_{\rm sys}$ channel has different period and phase than in the line wings (Figs.~\ref{fig_covar} and \ref{fig_chan_lc_fit_3parts}), the light curves in the normalization approach in Sect.~\ref{sec:normlisation} should be the combined effect of the intrinsic variations in both the line wings and the $V_{\rm sys}$ channel. 

Below we combine the velocities of the peak channels, the absolute peak amplitudes, and a wind velocity model from literature to demonstrate that only masers can produce the observed SiS and HCN line profile variations. We will discuss this problem for three cases: the emission region of the varying components of the two lines are at the outer edge of the wind acceleration zone ($\sim 11\,R_*$), or in the outer stead wind zone  ($> 11\,R_*$), or inside or under the wind acceleration zone  ($< 11\,R_*$), according to the wind velocity model of \citet{Deci2015}.

If the varying region is located at the outer edge of the wind acceleration zone, its radial direction must make an angle of $\sim 54^\circ$ to the sight line to project the terminal wind speed of $V_{\rm exp}=14.5$\,km\,s$^{-1}$ to the blue-peak channel at $V_{\rm p}=-8.5$\,km\,s$^{-1}$. This  corresponds to a linear shift of $b=8.9\,R_*$ from the central star position on the sky plane. As discussed in Sect.~\ref{sec:normlisation}, the FWHM of the variation amplitude around the peak channels is $\Delta V=\sim 3$\,km\,s$^{-1}$, corresponding to a length $\Delta l$ of the varying region as
\begin{equation}
\Delta l = \frac{\Delta V}{\sqrt{V_{\rm exp}^2-V_{\rm p}^2}}b \approx 0.255b\approx 2.27\,R_* \approx 0.045''.
\label{eq:thickness}
\end{equation}
For a round region, the blue peak amplitudes of 2.46\,K and 0.18\,K of the SiS and HCN lines in  our telescope beam of $29''$ (Sect.~\ref{sec:normlisation}) can be scaled into brightness temperatures of about $1.0\times 10^6$\,K and $7.5\times 10^4$\,K, respectively. They are far higher than the thermal temperature of less than 2000\,K in the CSE. The varying components must be masers. The weaker varying maser emission in the red-shifted line-wing could be due to occultation of the radio photosphere of the central star.

Is it possible that the varying regions are at larger radii so that their emission is still thermal? Lets assume a thermal line brightness temperature of 2000\,K. This requires a varying region size of $\Delta l = \sim 50\,R_*$ ($\sim 1''$) for the blue-peak of the SiS line or $\Delta l = \sim 14\,R_*$ ($\sim 0.28''$) for the blue-peak of the HCN line. Reversing Eq.~(\ref{eq:thickness}) yields offsets of the varying regions from the central star of $b=\sim 196\,R_*$ ($3.9''$ for SiS) and $b=\sim 55\,R_*$ ($1.1''$ for HCN) on the sky plane. The larger thermal line variation amplitudes in these regions perhaps mean higher local densities. The much smaller variation amplitudes in the red line wings of both the SiS and HCN lines mean that the corresponding density enhancements at the far side of the CSE should be much weaker.  However, such strongly asymmetric density enhancements at $>1.1''-3.9''$ offsets were never observed in high resolution mappings of this star \citep[e.g., the ALMA observations by][]{Deci2015,Veli2015,Quin2016}. In addition, it is also lack of driving mechanisms for the larger line variation amplitudes in so far regions from the central star. Therefore, thermal emission is not able to interpret the observed line profile variations.

Because the varying SiS and HCN masers peak in line wings, they are most probably located along the direction of the central star so that they are beaming radially, just like the case of the OH 1612\,GHz masers around oxygen-rich AGB stars. The fact that the peak channel velocity is smaller than steady wind velocity indicates that the maser regions must be located within the wind acceleration zone. Taking the wind velocity curve in the one step wind acceleration model from \citet[][the wind velocity linearly increases from 2\,km\,s$^{-1}$ at $\sim 5\,R_*$ to the terminal wind velocity of 14.5\,km\,s$^{-1}$ at $\sim 11\,R_*$]{Deci2015}, we find the maser radius to be 8.1\,$R_*$ for the blue peak channel of both SiS and HCN lines\footnote{Interestingly, \citet{Shin2009} used eSMA to find that the emission of the $\lambda$-doubling conjugate line, HCN\,$(01^{1e}0)$\,J=3-2, is distributed in a very small shell like structure that has a radius of about $23\,R_*$ ($0.38\arcsec$) and a thickness of about $15\,R_*$ ($0.25\arcsec$). The size of the shell is about 2-3 times larger than the estimated radius of the HCN\,$(01^{1f}0)$\,J=3-2 maser region in this work. However, it is not clear if that shell structure has anything to do with the varying masers.} and 9.0\,$R_*$ for the red-peak channel of the SiS line. The velocity range of the varying channels ($\Delta V=\sim 3$\,km\,s$^{-1}$) corresponds to a radial maser length of only $\sim 1.44R_*$ ($7.2\times 10^{13}$\,cm or $0.029\arcsec$), which is typical for a maser \citep{Elit1992}. The maser region is usually elongated along the radial beaming direction. Conservatively assuming spherical maser regions, the peak variation amplitudes in our $29\arcsec$ beam can be scaled to brightness temperatures of $>2.5\times 10^6$\,K, $>5.4\times 10^5$\,K, and $>1.8\times 10^5$\,K for the blue and red peak channels of the SiS line and the blue peak channel of the HCN line, respectively. 

We stress that the central region of the CSE of IRC\,+10216 has complex structures \citep[see e.g.,][]{Tuth2005,Mura2005,Leao2006,Menu2007,Fonf2014,Jeff2014,Deci2015} and even binary system has been proposed \citep{Deci2015,Cern2015,Kim2015}. It is higher desirable to do high resolution observations to confirm the masers directly.

\subsection{Maser properties} \label{sec:discuss_maser_properties}

The variation amplitude (and thus maser line strength) is much weaker in the red-shifted line wing than in the blue shifted line wing for both the SiS and HCN lines. This can be naturally explained by central-star occultation, because the estimated maser-region size ($1.44R_*$) is comparable to the measured diameter ($4\,R_*$) of the radio photosphere of the star \citep{Ment12}. On the other hand, the amplification of the radiation from the central star by the maser system could also contribute.

The blue peak-channel velocity of the SiS line is closer to the systemic velocity than the red peak-channel by $\sim 2$\,km\,s$^{-1}$. This could be also the consequence of the occultation of the red shifted maser component by the star. The elongated maser region should be smaller at smaller radius (thus smaller velocity shift). The star has blocked away more maser emission from the inner smaller maser sub-regions, resulting in a redder maser velocity to be seen in the red line wing. On the other hand, the self-absorption to the blue shifted maser line emission by the SiS molecules in the outer CSE is also able to shift the blue maser peak to a velocity closer to $V_{\rm sys}$ \citep{Hugg1986}. 

In the strongest blue shifted velocity channels of the SiS\,J=14-13 line profile, additional variation with shorter periods of about 2-4 weeks and flat light curves are also found in different parts of the light curves (Fig.~\ref{fig_chan_lc_SiS}). It could be interpreted by the instability of strong masers during the NIR maximum time and the dying out of the strongest maser spots near the NIR minimum light.

The variation of the HCN $(01^{1f}0)$\,J=3-2 line profile shows extreme asymmetry. The lack of variation in the red wing of the HCN line profile can be explained with at least two possible mechanisms: (1) the HCN maser region could be much smaller than that of the SiS maser so that the red part has been totally obscured by the star; (2) the HCN maser is unsaturated and the observed blue shifted masers are the magnified stellar radiation. The much broader velocity range of varying velocity channels (in the blue line wing) than the SiS line indicates that the HCN maser transition is inverted at almost all radii within the maximum radius of $\sim 8.1$\,$R_*$. This salient difference indicates some differences in the maser pumping processes of the two molecules.

\subsubsection{Comparison with other SiS masers in IRC\,+10216} \label{sec:discuss_SiS_maser}

Millimeter maser lines of SiS in IRC\,+10216 have been reported in previous works, e.g., the v=0 J=1-0 line by \citet{Henk1983}, the J=5-4 and 15-14 lines by \citet{Saha1984}, and more higher rotational maser transitions interferometrically mapped by \citet{Fonf2006} and \citet{Fonf2014}. These masers are possibly pumped through line overlap \citep{Fonf2006}. Our new results in this work confirm the variability of the SiS J=14-13 line and its maser nature in IRC\,+10216. Below we briefly compare the velocities of SiS maser lines found in this star by other authors  with the varying velocity channels identified from this work.

\begin{figure}
 \centering
 \includegraphics[angle=270,scale=0.34]{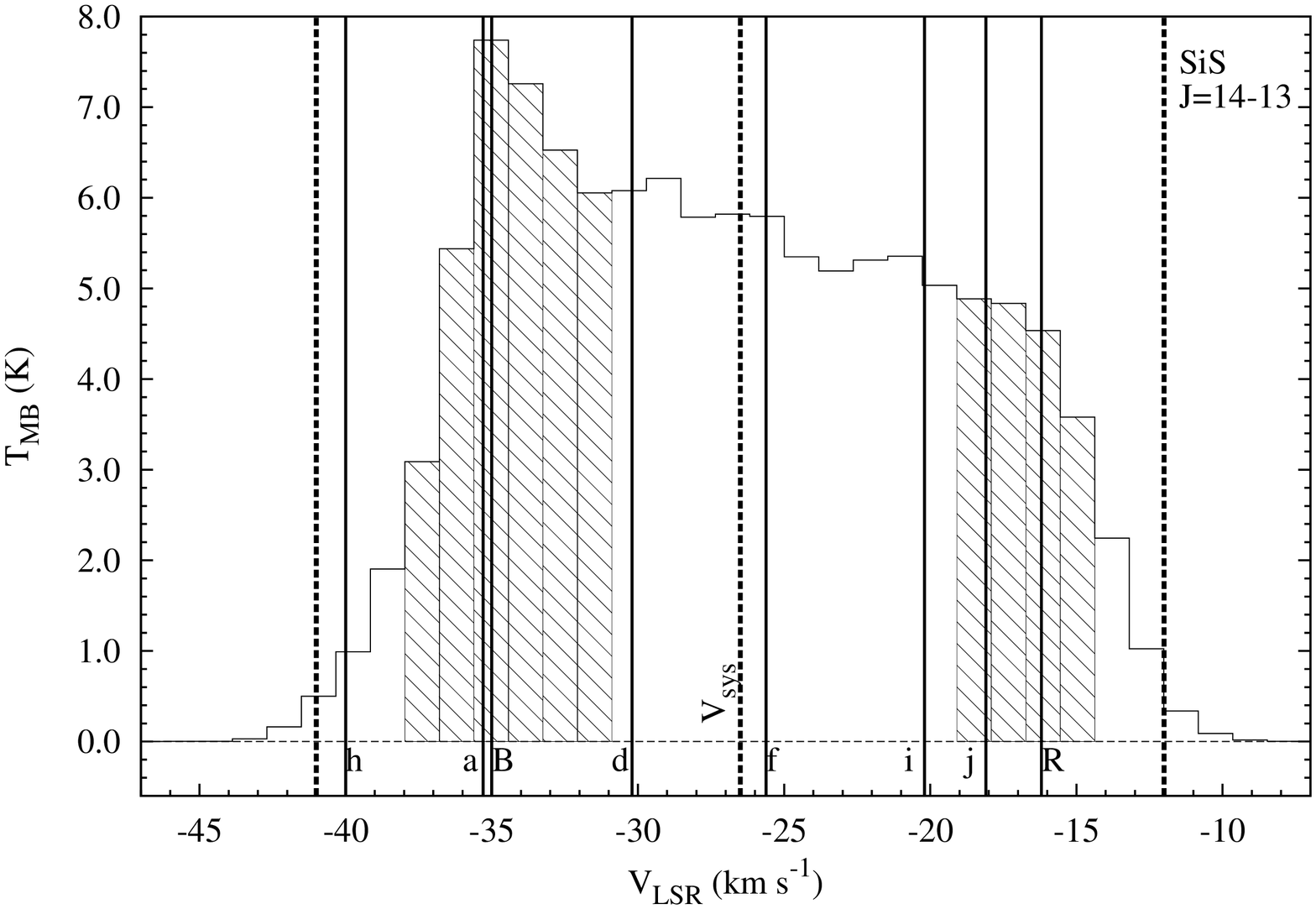}
 \includegraphics[angle=270,scale=0.34]{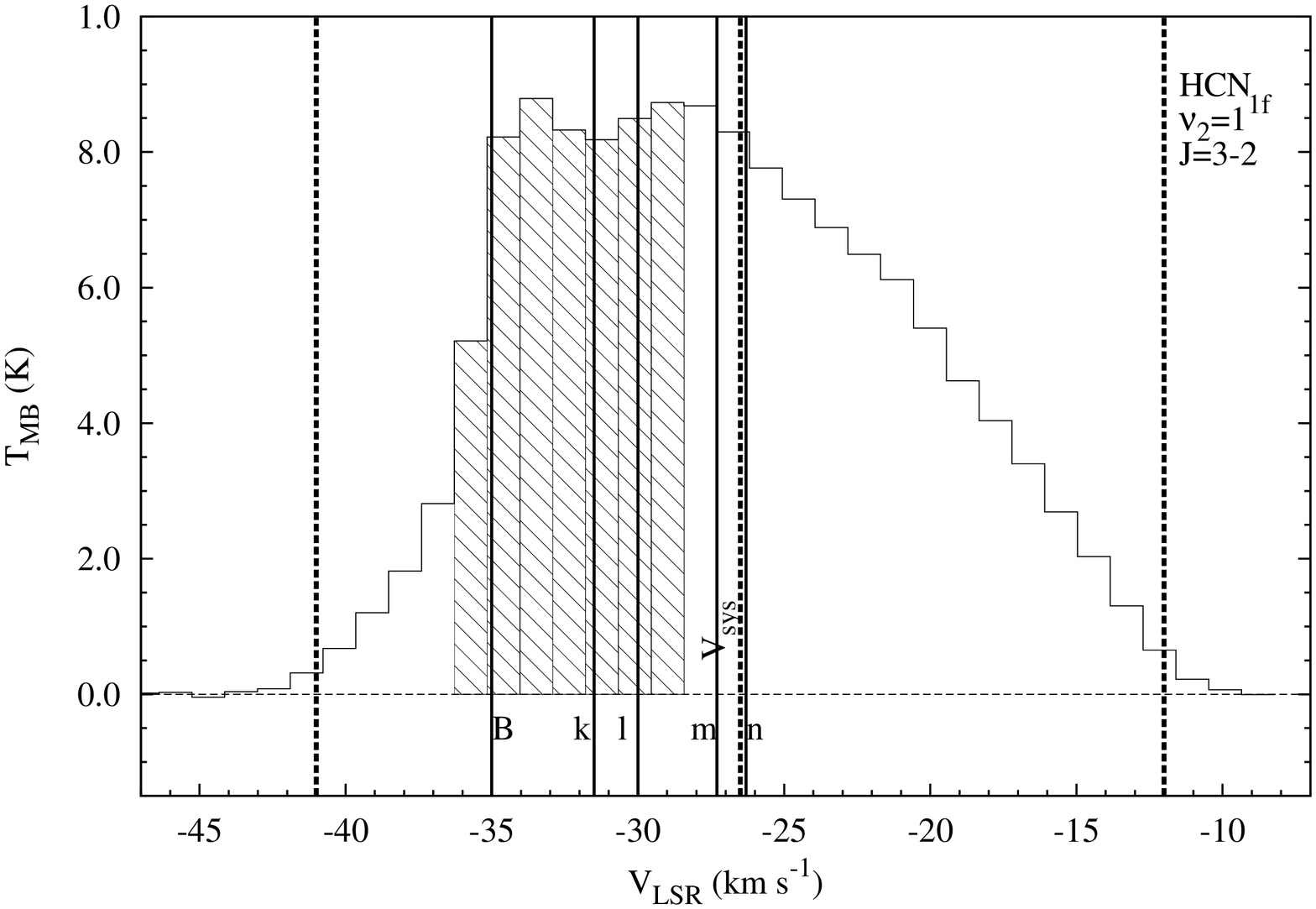}
 \caption{Compare the other maser feature velocities of the same molecules in IRC\,+10216 from literature with our average line profiles of the two maser lines SiS\,J=14-13 (top) and HCN\,$\nu_2=1^{1f}$\,J=3-2 (bottom). The varying velocity channels found in this work are hatched. The thick dashed vertical lines mark the systemic velocity (V$_{\rm sys}=-26.5$\,km\,s$^{-1}$) of the object and the two CSE expansion terminal velocities ($\pm14.5$\,km\,s$^{-1}$), the vertical full lines indicate the velocity channels with peak variation amplitudes from this work (the peak-channels; B for the blue one and R for the red one) and the maser feature velocities from literature (marked with lower case letters; see their definitions in Sects.~\ref{sec:discuss_SiS_maser} and \ref{sec:discuss_HCN_maser}).}
 \label{fig_maser_profiles}
\end{figure}

\citet{Fonf2006} found multiple candidate maser features in the line profiles of three ground state transitions of SiS: J=11-10, 14-13 and 15-14. We adopt the five major maser features from them and plot them with our average line profile in Fig.~\ref{fig_maser_profiles}: 
`a' (-35.3\,km\,s$^{-1}$), 
`d' (-30.2\,km\,s$^{-1}$), 
`f'  (-25.62\,km\,s$^{-1}$), 
`i' (-20.2\,km\,s$^{-1}$), 
`j'  (-18.1\,km\,s$^{-1}$). It agrees to our findings that maser features appear in both blue and red wings of this line. However, only their SiS\,J=14-13 and 15-14 maser components `a' and `j' are in the varying velocity channels discovered in our data. The other small line features could be stable local density structures related to the complex non-spherical structures of the inner CSE. The SiS\,J=1-0 maser of \citet{Henk1983} is also shown as `h' (-40\,km\,s$^{-1}$). It is outside of our varying velocity ranges and appears near the blue edge of the line. Perhaps this maser is pumped in the steady wind region by IR light \citep{Gong2017}.

\subsubsection{Comparison with other HCN masers in IRC\,+10216} \label{sec:discuss_HCN_maser}

HCN maser lines in IRC\,+10216 have also been found decades ago, e.g., the $(02^00)$ J=1-0 line by \citet{Guil1987}, \citet{Luca1988} and \citet{Luca1992}, $(01^{1e}0)$\,J=2-1 (originally assigned as $01^{1c}0$) by \citet{Luca1989}, $(04^00)$\,J=9-8 by \citet{Schi2000}, and even in the cross vibration ladder transition $(11^10)-(04^00)$\,J=10-9 by \citet{Schi2003}. Various maser pumping mechanisms such as IR light pumping plus line overlap \citep{Dinh2000,Luca1989,Shin2009} and chemical pumping \citep{Schi2000,Schi2003} have been discussed. Our line variability data offer several new insights to the HCN maser properties.

The velocities of other known HCN maser lines in IRC\,+10216 are compared with our velocity range of line variability in the bottom panel of Fig.~\ref{fig_maser_profiles}: HCN\,$(02^00)$\,J=1-0 at -30\,km\,s$^{-1}$ \citep[`l',][]{Guil1987,Luca1988},
$(01^{1e}0)$\,J=2-1 at -31.5\,km\,s$^{-1}$ 
\citep[`k',][]{Luca1989}, 
$(04^00)$\,J=9-8 at -27.3\,km\,s$^{-1}$ \citep[`m',][]{Schi2000}, and 
$(11^10)$-$(04^00)$\,J=10-9 at -26.3\,km\,s$^{-1}$ \citep[`n',][]{Schi2003}. 
Their velocities agree to our findings that maser lines only appear in the blue line wing. The two HCN maser lines with smaller vibrational quantum numbers (`l' and `k') are within the varying velocity channels, while the other two HCN maser lines that involve hotter vibrational levels (`m' and `n') are not in the varying channels but are very close to $V_{\rm sys}$, indicating they are inverted very close to the central stars, perhaps in the hottest pulsation shock regions.

\section{Summary} \label{sec:summary}

We have monitored the SiS\,J=14-13 and HCN\,$\nu_2=1^{1f}$\,J=3-2 maser lines together with several other millimeter lines toward IRC\,+10216 during a time span of 523 days. Relative light curves of line intensity with respect to specified reference lines are obtained for eight molecular line features (single lines or blended line groups); line shape light curves are also obtained for the two strongest lines.  

The relative line intensity light curves show regular long term variations on time scales of several hundred days with amplitudes of 5\%-30\%. The amplitudes are smaller than that of IR light and bolometric luminosity variations. Variation periods and phases can be either similar to or different from the NIR light. They possibly originate from the inner dense part of the CSE, particularly the region within or under the wind acceleration zone. Therefore, these varying lines have the potential to be developed into new probes of the wind launching processes in AGB stars.

The SiS\,J=14-13 and HCN\,$\nu_2=1^{1f}$\,J=3-2 lines show line shape variability mainly in two narrow velocity ranges and roughly follow the variation of the NIR light. However, the HCN line shape variation periods are generally longer by 100-200 days than the NIR light period (630 day). There exist blue-red asymmetries: the variation amplitude is larger in the blue line wing than in red line wing for both lines. There is also a velocity dependence of the periods and phases for both lines. The varying components are the main causes of the asymmetrical average line profiles of the two lines. Their velocities and amplitudes support them to be masers in the zone where the wind is still accelerating, with brightness temperatures higher than $10^5-10^6$\,K. Particularly, the HCN maser may be inverted all the way down to radii close to the photosphere, as indicated by the velocities of its varying channels. 

The line profile variation of the two maser lines is also analyzed by two approaches: line profile normalization and light curve decomposition. The results from the two methods are very similar, which supports that the remaining flux uncertainties after the in-band calibration are not very large. The decomposition approach provides new insights: the co-varying component in the systemic velocity channel has smaller amplitude than the differential variation components in the line wings; the amplitudes of both varying components are smaller than the strength of the constant components; the periods and maximum phases are also quite different between the co-varying and differential variation components.

In the strongest blue shifted velocity channels of the SiS\,J=14-13 line profile, additional variation with shorter periods of about 2-4 weeks near the maximum time and flat light curves around the minimum time are also found. They could be interpreted by the instability of strong masers during the NIR maximum light and the dying out of the strongest maser spots near the NIR minimum light.

\section*{Acknowledgments}

We thank the referee for the comments that greatly helped to improve our manuscript. JH is grateful to Dr. Tomasz Kaminski for carefully reading the manuscript and for many thoughtful critics and suggestions that have helped improve this work much. JH also thanks Dr. Holger S. P. M{\"u}ller for some discussions upon the labeling format of vibrational states of some of the observed lines. The financial support to Dinh-V-Trung from Vietnam's National Foundation for Science and Technology (NAFOSTED) under contract 103.99-2014.82 is greatly acknowledged. This work is also partly sponsored by the Chinese Academy of Sciences (CAS), through a grant to the CAS South America Center for Astronomy (CASSACA) in Santiago, Chile.

\vspace{5mm}
\facilities{ARO SMT}
\software{GILDAS/CLASS}

\appendix

\section{Line identification results} \label{sec:lineiden_detail}

\begin{figure*}
 \centering
\includegraphics[angle=0,scale=.475]{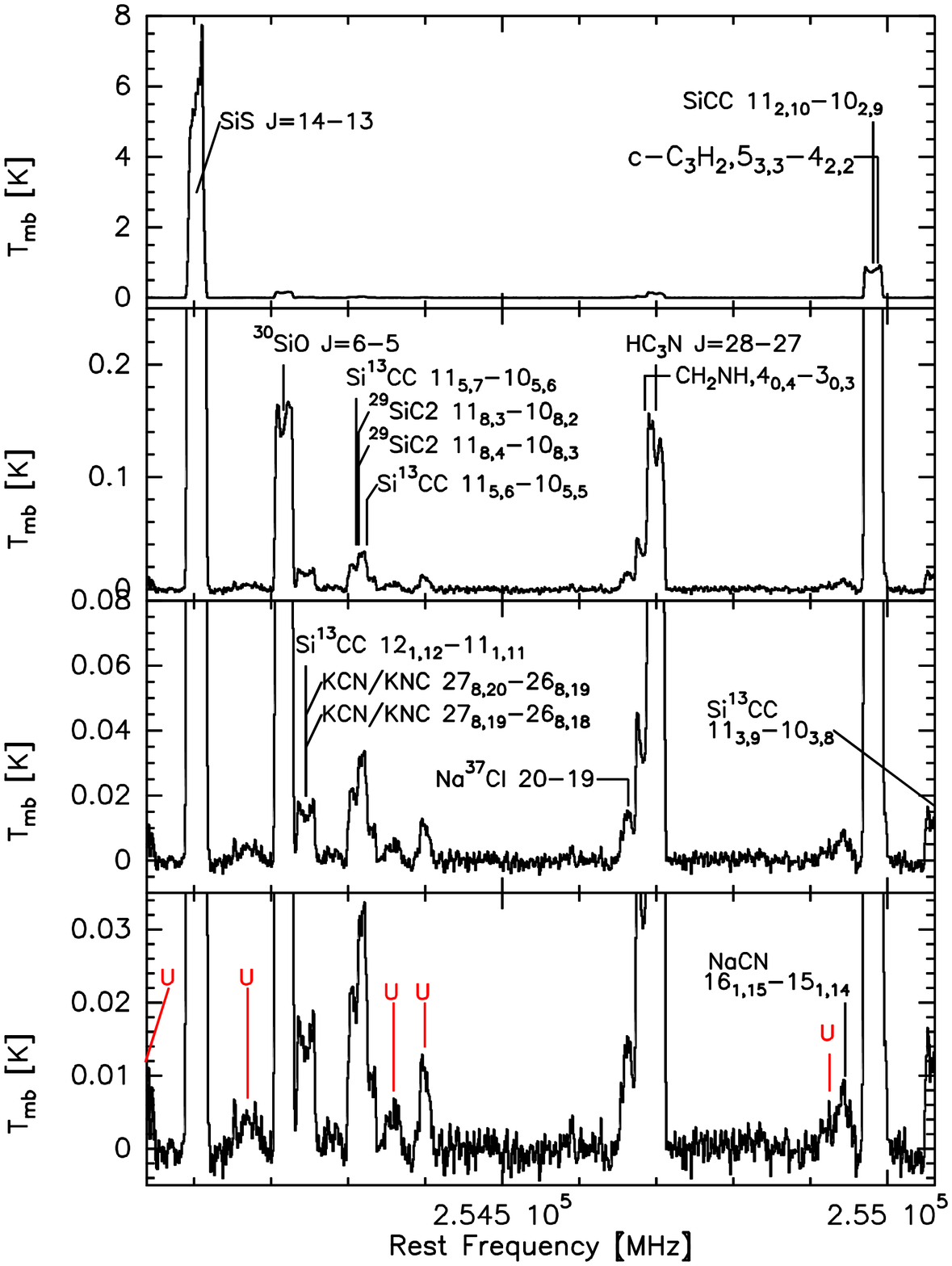}
\includegraphics[angle=0,scale=.475]{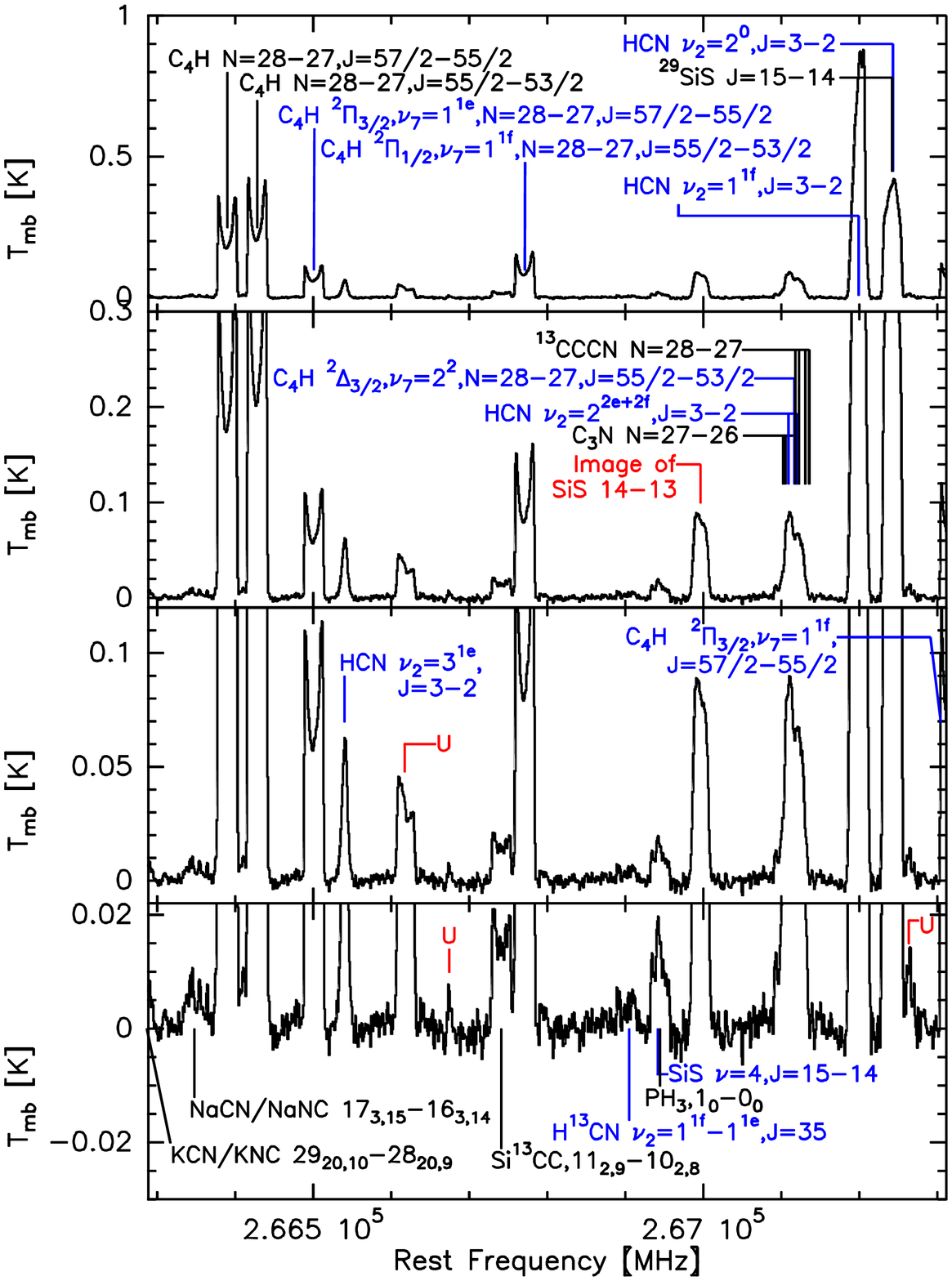}
 \caption{Average spectra showing lines detected in the LSB and USB and their carriers. Rotational transitions in ground vibration state are labeled in black, those in excited vibration states are in blue, unidentified features or side band image leaking are in red. See the discussions in Appendix.~\ref{sec:lineiden_detail}. ({\em A colorful version is available on-line only.})}
 \label{fig_SBplots}
\end{figure*}

The detailed list of line carriers of each detected line feature is given in Table~\ref{tab_lines} while the line profile averaged over all epochs are plotted in Fig.~\ref{fig_SBplots}. Concerning the format of quantum numbers for the vibrational states, there is some divergence in the literature. For example, the works of \citet{Ziur86} and \citet{Tene10b} used letters `d' and `c' to differentiate the two parity cases of the $\lambda$-doubling states of linear molecules, while others \citep[e.g.,][]{Zeli03} used `e' and `f' that were recommended by \citet{Brow75}. In this work, we will follow the convention of Brown et al and the $\lambda$-doublets of HCN is adopted from \citet{Zeli03} and that of C$_4$H from \citet{Yama87}. Here are some comments about the line carriers:
\begin{enumerate}{
\item{{\it CH$_2$NH\,4$_{0,4}-3_{0,3}$}\\
The CH$_2$NH\,4$_{0,4}-3_{0,3}$ line at 254685.20\,MHz shows a double peaked line profile partially blended with the strong HC$_3$N\,28-27 and weaker Na$^{37}$Cl\,20-19 lines. It was first detected in IRC\,+10216 by \citet{He08} who took it as an unidentified line. Its identity was first recognized in this star by \citet{Tene10a}.
}
\item{{\it Hot C$_4$H lines in bending states}\\
At least three hot C$_4$H lines in the bending state ($\nu_7=1^1$) are detected to be free of line blending. All of them show double peak line profiles, indicating that their emission regions are extended compared to the SMT beam size of $29\arcsec$ and are not very optically thick. Although the C$_4$H\,$^2\Pi_{3/2}\,\nu_7=1^{1f}$\,N=28-27 line at 267316.33\,MHz appears at the edge of the USB, the double peak shape of its partial line profile is still clear. \citet{Guel87} and \citet{Cook15} have shown that the rotational transitions of C$_4$H in the $\nu_7=1^1$ state mainly arise from a ring like structure of $15\arcsec$ radius, together with a much weaker component near the central star. The C$_4$H\,$^2\Delta_{3/2}\,\nu_7=2^2$\,N=28-27 line at 267116.98\,MHz is blended with several other lines of C$_3$N, $^{13}$CCCN and HCN. Thus its detection is only tentative.
}
\item{{\it Narrow lines}\\
Beside the three unidentified narrow lines that will be discussed in the next paragraphs, there are another two identified narrow lines: HCN\,$\nu_2=3^{1e}$\,J=3-2 at 266540\,MHz and SiS\,$\nu=4$\,J=15-14 at 266941.75\,MHz. The former has a ${\rm FWHM}=4.2\pm 0.1$\,km\,s$^{-1}$. The latter is a narrow spike on top of the broad PH$_3\,1_0-0_0$ line. Both narrow lines should originate in regions within or under the wind acceleration zone.
}
\item{{\it Unidentified lines}\\
All the eight U-lines have S/N ratios in terms of integrated line intensity higher than about 5. However, most of the eight U-lines have peak $T_{\rm mb}<10$\,mK, except U266618 that has a peak $T_{\rm mb}=\sim 40$\,mK. 
U266618 only appeared at six epochs and is found to be anti-correlated with the light curve of the leaked image of the strong line SiS\,14-13 from the image side band, indicating that it might be a false feature related to the side band leaking of SiS\,14-13. 

Three U-lines are definitely narrower than the majority of thermal lines: U254359 has a $V_{\rm exp}=12.3\pm 1$\,km\,s$^{-1}$, U254400 has a $V_{\rm exp}=5.9\pm0.5$\,km\,s$^{-1}$ and U266674 has a $V_{\rm exp}=1.9\pm0.6$\,km\,s$^{-1}$, all being much narrower than the terminal envelop expansion velocity 15\,km\,s$^{-1}$. Thus they are either lines from regions inside or under the wind acceleration zone or weak maser lines. U267263 also looks like a narrow line, but because of the line blending, we can not tell if it is the blue horn of a blended double peak line profile. 
}}
\end{enumerate}
\begin{table}
 \centering
 \begin{minipage}{170mm}  
  \caption{Detected lines.}
  \label{tab_lines}
  \begin{tabular}{l@{ }l@{ }l|l@{ }l@{ }l}
  \hline
  \hline
Freq(MHz) & trans                               & Notes        & Freq(MHz) & trans                                         & Notes       \\
\hline
254037.00 & U254037                             & incomplete   & 266283.63 & NaCN/NaNC\,$17_{4,13}-16_{4,12}$              & incomplete  \\
254103.20 & SiS\,14-13                          &              & 266347.60 & NaCN/NaNC\,$17_{3,15}-16_{3,14}$              &             \\
254169.59 & U254170                             & no-T2010     & 266389.92 & C$_4$H\,N=28-27\,J=57/2-55/2                  &             \\
254216.14 & $^{30}$SiO\,6-5                     &              & 266428.18 & C$_4$H\,N=28-27\,J=55/2-53/2                  &             \\
254244.97 & Si$^{13}$CC\,$12_{1,12}-11_{1,11}$  & blended      & 266500.88 & C$_4$H\,$^2\Pi_{3/2}\,\nu_7=1^{1e}$\,N=28-27  &             \\
254245.40 & KCN/KNC\,$27_{8,20}-26_{8,19}$      & blended      & 266540.00 & HCN\,$\nu_2=3^{1e}$\,J=3-2                    &             \\
254245.40 & KCN/KNC\,$27_{8,19}-26_{8,18}$      & blended      & 266617.79 & U266618                                       &             \\
254310.37 & Si$^{13}$CC\,$11_{5,7}-10_{5,6}$    & blended      & 266674.40 & U266674                                       &             \\
254313.37 & $^{29}$SiCC\,$11_{8,3}-10_{8,2}$    & blended      & 266740.96 & Si$^{13}$CC\,$11_{2,9}-10_{2,8}$              &             \\
254313.37 & $^{29}$SiCC\,$11_{8,4}-10_{8,3}$    & blended      & 266771.19 & C$_4$H\,$^2\Pi_{1/2}\,\nu_7=1^{1f}$\,N=28-27  &             \\
254324.42 & Si$^{13}$CC\,$11_{5,6}-10_{5,5}$    & blended      & 266904.99 & H$^{13}$CN\,$\nu_2=1^{1f}-1^{1e}$\,J=35       &             \\
254358.87 & U254359                             & no-T2010     & 266941.75 & SiS\,$\nu=4$\,J=15-14                         & blended     \\
254399.56 & U254400                             & no-T2010     & 266944.51 & PH$_3\,1_0-0_0$                               & blended     \\
254663.46 & Na$^{37}$Cl\,20-19                  & blended      & 267103.31 & CCCN\,N=27-26\,J=55/2-53/2                    & blended     \\
254685.20 & CH$_2$NH\,4$_{0,4}-3_{0,3}$         & blended      & 267109.14 & HCN\,$\nu_2=2^{2f}$\,J=3-2                    & blended     \\
254699.50 & HC$_3$N\,28-27                      & blended      & 267116.98 & C$_4$H\,$^2\Delta_{3/2}\,\nu_7=2^2$\,N=28-27  & blended     \\
254924.96 & U254925                             & no-T2010     & 267117.89 & $^{13}$CCCN\,N=28-27\,J=57/2-55/2\,F1=29-28   & blended     \\
254945.21 & NaCN\,$16_{1,15}-15_{1,14}$         &              & 267120.10 & HCN\,$\nu_2=2^{2e}$\,J=3-2                    & blended     \\
254981.49 & SiCC\,$11_{2,10}-10_{2,9}$          & blended      & 267121.23 & CCCN\,N=27-26\,J=53/2-51/2                    & blended     \\
254987.64 & c-C$_3$H$_2$\,$5_{3,3}- 4_{2,2}$    & blended      & 267122.89 & $^{13}$CCCN\,N=28-27\,J=57/2-55/2\,F1=28-27   & blended     \\
255060.77 & Si$^{13}$CC\,$11_{3,9}-10_{3,8}$    & incomplete   & 267130.89 & $^{13}$CCCN\,N=28-27\,J=55/2-53/2\,F1=28-27   & blended     \\
          &                                     &              & 267135.89 & $^{13}$CCCN\,N=28-27\,J=55/2-53/2\,F1=27-26   & blended     \\
          &                                     &              & 267199.28 & HCN\,$\nu_2=1^{1f}$\,J=3-2                    &             \\
          &                                     &              & 267242.14 & $^{29}$SiS\,J=15-14                           &             \\
          &                                     &              & 267243.20 & HCN\,$\nu_2=2^0$\,J=3-2                       &             \\
          &                                     &              & 267263.48 & U267263                                       &             \\
          &                                     &              & 267316.33 & C$_4$H\,$^2\Pi_{3/2}\,\nu_7=1^{1f}$\,N=28-27  & incomplete  \\
\hline
\end{tabular} Notes:\\
no-T2010: the unidentified line was observed but not detected by \citet{Tene10b}. \\
\end{minipage}
\end{table}

\section{Notes on line profile variation of weaker line features} \label{sec:App_subvrange}

Line shape variation of four relatively strong line features other than the SiS and HCN maser lines are discussed in broader velocity ranges in this section. Roughly velocity ranges of about 10\,km\,s$^{-1}$ wide are adopted so that a typical line from IRC\,+10216 ($\sim$29\,km\,s$^{-1}$ wide) can be divided into three parts: line center part and blue and red shifted line wings. The central velocity range is chosen as the reference to calibrate the other velocity ranges in the line wings to reveal their relative variabilities. For those blended line features, we carefully arrange the velocity ranges so that each range is dominated by the blue- or red-shifted or central component of the blended lines. The central velocity range of the possibly strongest blended line is selected as the reference for normalization (the detailed division is given for individual line features below). 

The normalized light curves of the four line features are shown in Figs.~\ref{fig_29SiS_HCN} and \ref{fig_C4H55}. The main features are:
\begin{figure}
 \centering
 \includegraphics[angle=0,scale=0.4]{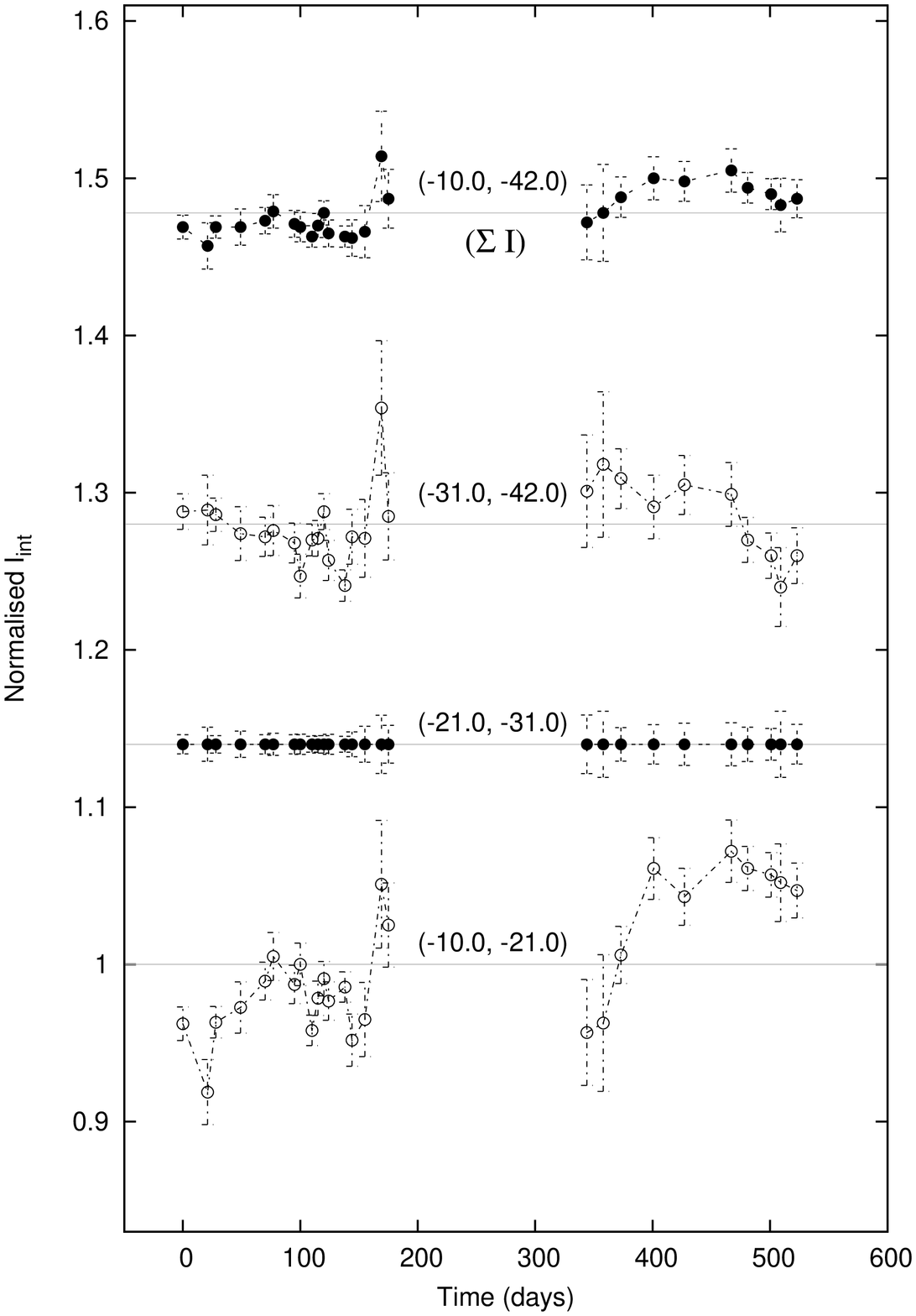}
 \includegraphics[angle=0,scale=0.4]{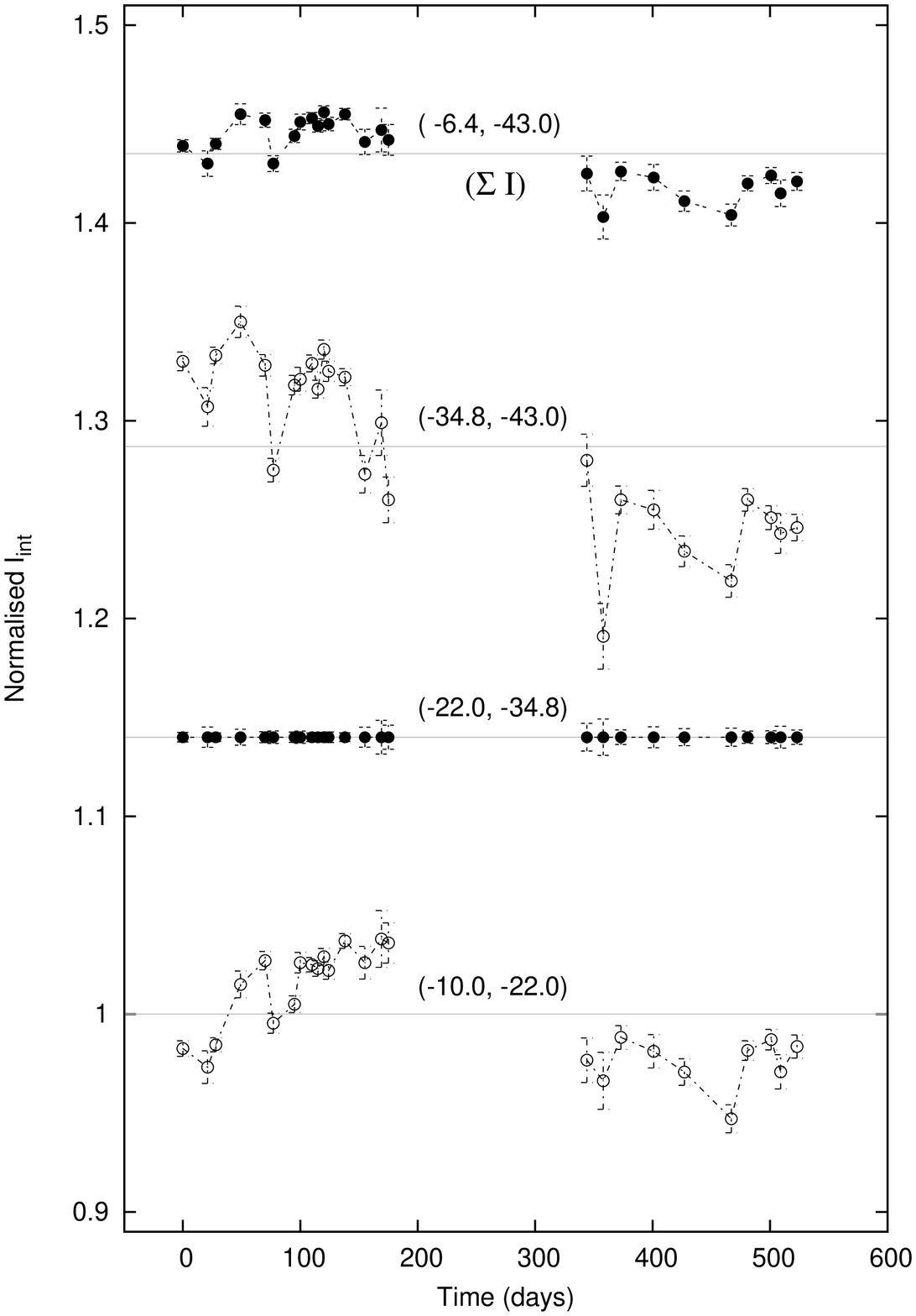}
\caption{The normalized light curves of integrated line intensity in velocity ranges (marked with numbers in unit of km\,s$^{-1}$) for the blended line feature $^{29}$SiS\,J=15-14 \& HCN\,$\nu_2=2^0$\,J=3-2 {\em (left)} and the LSB calibrator line SiCC\,$11_{2,10}-10_{2,9}$ {\em (right)}. Alternative open and filled circles are used to differentiate neighboring light curves. The velocity range used as the reference for normalization can be recognised as the straight light in each plot. The top curve (marked with `$\rm \Sigma I$') is for the integrated line strength of the whole line feature that is also calibrated by the reference velocity range. The different light curves are shifted for clarity.}
 \label{fig_29SiS_HCN}
\end{figure}
\begin{figure}
 \centering
 \includegraphics[angle=0,scale=0.4]{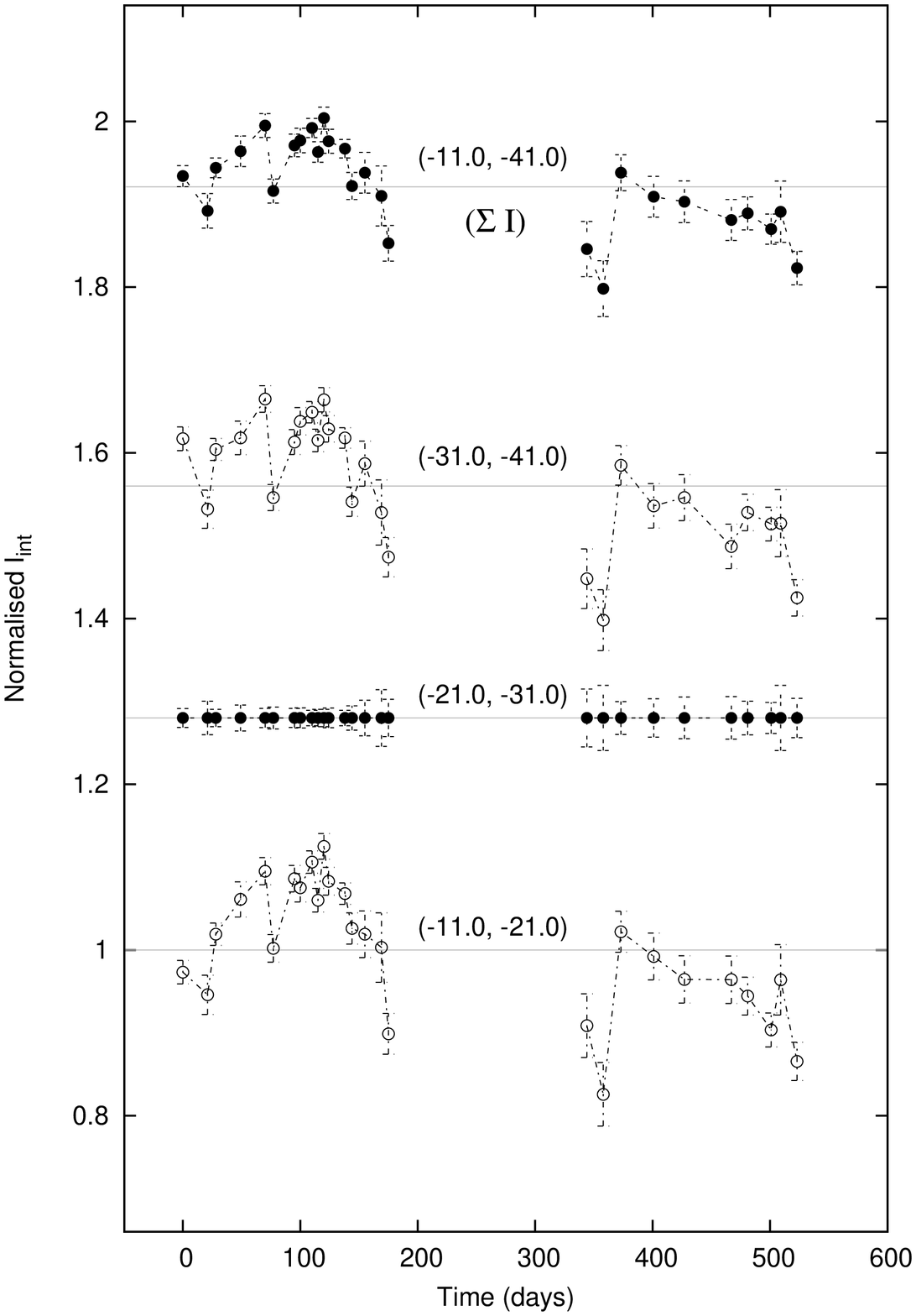}
 \includegraphics[angle=0,scale=0.4]{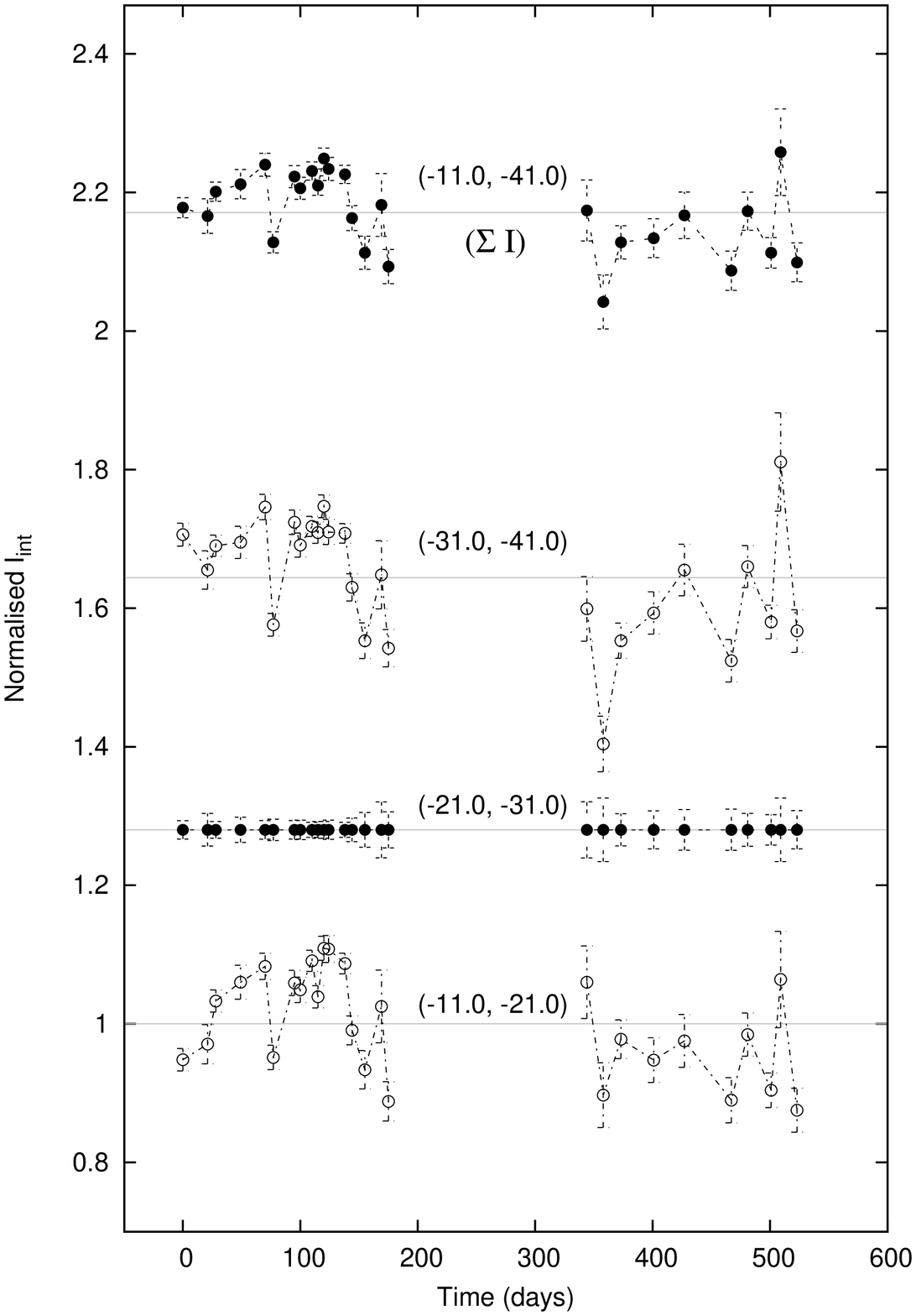}
 \caption{The same as Fig.~\ref{fig_29SiS_HCN} but for the USB calibrator lines C$_4$H\,N=28-27\,J=55/2-53/2 (left) and J=57/2-55/2 (right).}
 \label{fig_C4H55}
\end{figure}
\begin{itemize}
\item{$^{29}$SiS \& HCN\,$\nu_2=2^0$ line feature (Fig.~\ref{fig_29SiS_HCN}, left panel)\\
The two blended lines only have a small velocity shift from each other of about 1\,km\,s$^{-1}$. Thus we define three velocity ranges for the line feature, with the central reference velocity range centered around their average frequency. The bundaries of the three velocity ranges are (-42, -31, -21, -10)\,km\,s$^{-1}$.

This line feature is weak and only show marginal variability in both blue and red parts of the line profile. The variation amplitude is no larger than 20\% (with large error bars). The blue and red wings have a weak trend to opposite variations so that the total line intensity in the top curve in the figure shows little variation.
}
\item{SiCC\,$11_{2,10}-10_{2,9}$ (Fig.~\ref{fig_29SiS_HCN}, right panel)\\
This line is blended with c-C$_3$H$_2$\,$5_{3,3}-4_{2,2}$. We divide the line profile into four velocity ranges, with boundaries at (-43, -34.76, -22, -10, -3.76)\,km\,s$^{-1}$, so that the blue, central and red parts of each of the two lines can be sampled as evenly as possible. As a result, the three bluest ranges cover 21\%, 42\% and 37\% (from blue to red) of the 29\,km\,s$^{-1}$ wide full velocity range of the SiCC line respectively, while the three reddest ranges cover 39\%, 40\% and 21\% (from blue to red) of the full velocity range of the c-C$_3$H$_2$ line respectively. The reddest range turns out to be very weak and thus is omitted from Fig.~\ref{fig_29SiS_HCN} (it also indicates that the c-C$_3$H$_2$ line is very weak). We select the middle range (-34.76, -22)\,km\,s$^{-1}$ of the rest three velocity ranges (also the middle part of the SiCC line) as the reference for normalization.

Although this line is the calibrator line in the LSB for the in-band calibration, its line profile is found to be varying with time. The largest variation is about 8\%. The blue and red parts of the line profile vary differently. The light curve of the blue part looks more or less similar to that of the USB calibrator lines in Fig.~\ref{fig_C4H55}, while the light curve of the red part looks more differently.
}
\item{C$_4$H\,N=28-27\,J=55/2-53/2 \& J=57/2-55/2 (Fig.~\ref{fig_C4H55})\\
The two lines do not suffer from blending. We define three symmetrical velocity ranges with boundaries at (-41, -31, -21, -11)\,km\,s$^{-1}$. The central range is taken as the reference for normalization.

Although the two lines are taken as calibrator lines in the USB, their line profiles are found to be varying with time in a similar manner. The largest variation is about 17\%. Both the blue and red wings of the line profiles show irregular variation patterns. 
}
\end{itemize}

The line shape variation in the broad velocity ranges in Figs.~\ref{fig_29SiS_HCN}-\ref{fig_C4H55} can hardly be interpreted with physical processes such as IR light or shock driven variation or masers. The three flux calibrator lines (of SiCC and C$_4$H) have strong contribution from the extended cold part of the spherical CSE. Usually no line profile variation is expected with such a spherical emission region. Even if the inner hot and dense part of the CSE contributes a certain amount of variation to the central velocity range of the three lines, the resulting line profile variation in Figs.~\ref{fig_29SiS_HCN}-\ref{fig_C4H55} should be regular, blue-red symmetrical and periodical, which is contrary to the irregular behaviors seen in the figures. The $^{29}$SiS \& HCN\,$\nu_2=2^0$ blended line feature could have a considerable contribution from the inner hot CSE that is liable to variation. However, the contribution of the varying emission should be also blue-red symmetrical, which is not the case in Fig.~\ref{fig_29SiS_HCN}. We have no reasonable interpretation for the $^{29}$SiS \& HCN\,$\nu_2=2^0$ line feature, but will give a possible interpretation for the line shape variation in the three flux calibrator lines: random telescope pointing offsets.
 
It has been known that some rotational lines of both C$_4$H and SiCC have a prominent extended ring-like component (a spherical structure in three dimension) with a diameter of about $30\arcsec$ in their emission maps \citep[e.g.,][]{Guel1993,Daya1993,Taka1992,Gens1992,Luca1995,Cook15}. The ring size is comparable to the telescope beam size in this work. This is consistent with the double peak line profiles of all the three calibrator lines in our Fig.~\ref{fig_SBplots}. The SiCC\,$4_{2,3}-2_{2,2}$ line map from \citet{Luca1995} and the maps of C$_4$H lines in both ground and vibration states from \citet{Cook15} also shows a central emission component, though. Therefore, the roughly blue-red symmetrical but irregular variation of the line shapes of the three lines can be self-consistently explained as follows: the contribution from the ring like emission component varies sensitively to the random telescope pointing offsets and produce variation in the central velocity channels of the line profile; the central point like component (perhaps also an smooth extended component) is not sensitive the pointing fluctuation and thus contributes some unaffected emission to all velocity channels. In this case, the apparent variation of the double peaks with respect to the central velocity channels of the three lines is actually due to the variation of the central velocity channels themselves (the reference velocity ranges).

If the pointing fluctuation interpretation is true, the total line intensity of the three calibrator lines should be also affected to some extent. The impact of this effect to the in-band calibration of line intensity in Sect.~\ref{sec:var-Iint} 
can be roughly estimated by assuming that only the central 1/3 portion of the line profiles is subjected to such variation with pointing fluctuation. For the two C$_4$H calibrator lines in the USB, because of the sharp double peak profiles, the central velocity channels only contribute to about 1/5 of the integrated line intensity (2/5 by the blue and 2/5 by the red part). The apparent profile variation amplitude of the blue and red parts (actually reflecting the variation of the central part) is at most 17\% in Fig.~\ref{fig_C4H55}, corresponding to an integrated line intensity calibration error of at most $17\%\times 1/5 \approx 3.4\%$ for lines in USB. For the SiCC calibrator line in the LSB, the double peaks are not very sharp, thus the varying central part of the line profile may contribute to about 1/3 of the integrated line intensity. The maximum variation amplitude of the blue and red parts of its line profile amounts to about 8\% in Fig.~\ref{fig_29SiS_HCN}. This can be transformed into a line intensity calibration error of at most $8\% \times 1/3 \approx 2.7\%$ for lines in the LSB. This kind of calibration error should randomly vary from epoch to epoch. These errors are relatively small and we do not try to make correction for them. But a lesson can be learned is that the extended ring like emission component can incur large errors through its sensitive response to the pointing fluctuation of a telescope beam comparable to the ring size.

\bibliography{irc10216-mm-var}

\begin{thebibliography}{}
\expandafter\ifx\csname natexlab\endcsname\relax\def\natexlab#1{#1}\fi
\providecommand{\url}[1]{\href{#1}{#1}}

\bibitem[{{Bessell} {et~al.}(1996){Bessell}, {Scholz}, \& {Wood}}]{Bess1996}
{Bessell}, M.~S., {Scholz}, M., \& {Wood}, P.~R. 1996, \aap, 307, 481

\bibitem[{{Brown} {et~al.}(1975){Brown}, {Hougen}, {Huber}, {Johns}, {Kopp},
  {Lefebvre-Brion}, {Merer}, {Ramsay}, {Rostas}, \& {Zare}}]{Brow75}
{Brown}, J.~M., {Hougen}, J.~T., {Huber}, K.-P., {et~al.} 1975, Journal of
  Molecular Spectroscopy, 55, 500

\bibitem[{{Carlstrom} {et~al.}(1990){Carlstrom}, {Olofsson}, {Johansson},
  {Nguyen-Q-Rieu}, \& {Sahai}}]{Carl1990}
{Carlstrom}, U., {Olofsson}, H., {Johansson}, L.~E.~B., {Nguyen-Q-Rieu}, \&
  {Sahai}, R. 1990, in From Miras to Planetary Nebulae: Which Path for Stellar
  Evolution?, ed. M.~O. {Mennessier} \& A.~{Omont}, 170

\bibitem[{{Cernicharo} {et~al.}(2000){Cernicharo}, {Gu{\'e}lin}, \&
  {Kahane}}]{Cern2000}
{Cernicharo}, J., {Gu{\'e}lin}, M., \& {Kahane}, C. 2000, \aaps, 142, 181

\bibitem[{{Cernicharo} {et~al.}(2015){Cernicharo}, {Marcelino}, {Ag{\'u}ndez},
  \& {Gu{\'e}lin}}]{Cern2015}
{Cernicharo}, J., {Marcelino}, N., {Ag{\'u}ndez}, M., \& {Gu{\'e}lin}, M. 2015,
  \aap, 575, A91

\bibitem[{{Cernicharo} {et~al.}(2010){Cernicharo}, {Waters}, {Decin},
  {Encrenaz}, {Tielens}, {Ag{\'u}ndez}, {De Beck}, {M{\"u}ller}, {Goicoechea},
  {Barlow}, {Benz}, {Crimier}, {Daniel}, \& {et al{.}}}]{Cern2010}
{Cernicharo}, J., {Waters}, L.~B.~F.~M., {Decin}, L., {et~al.} 2010, \aap, 521,
  L8

\bibitem[{{Cernicharo} {et~al.}(2014){Cernicharo}, {Teyssier},
  {Quintana-Lacaci}, {Daniel}, {Ag{\'u}ndez}, {Velilla-Prieto}, {Decin},
  {Gu{\'e}lin}, {Encrenaz}, {Garcia-Lario}, {de Beck}, {Barlow}, {Groenewegen},
  {Neufeld}, \& {Pearson}}]{Cern2014}
{Cernicharo}, J., {Teyssier}, D., {Quintana-Lacaci}, G., {et~al.} 2014, \apjl,
  796, L21

\bibitem[{{Cherchneff}(2011)}]{Cher2011}
{Cherchneff}, I. 2011, \aap, 526, L11

\bibitem[{{Cherchneff}(2012)}]{Cher2012}
---. 2012, \aap, 545, A12

\bibitem[{{Cooksy} {et~al.}(2015){Cooksy}, {Gottlieb}, {Killian}, {Thaddeus},
  {Patel}, {Young}, \& {McCarthy}}]{Cook15}
{Cooksy}, A.~L., {Gottlieb}, C.~A., {Killian}, T.~C., {et~al.} 2015, \apjs,
  216, 30

\bibitem[{{Dayal} \& {Bieging}(1993)}]{Daya1993}
{Dayal}, A., \& {Bieging}, J.~H. 1993, \apjl, 407, L37

\bibitem[{{Decin} {et~al.}(2015){Decin}, {Richards}, {Neufeld}, {Steffen},
  {Melnick}, \& {Lombaert}}]{Deci2015}
{Decin}, L., {Richards}, A.~M.~S., {Neufeld}, D., {et~al.} 2015, \aap, 574, A5

\bibitem[{{Dinh-V-Trung} \& {Rieu}(2000)}]{Dinh2000}
{Dinh-V-Trung}, \& {Rieu}, N.-Q. 2000, \aap, 361, 601

\bibitem[{{Elitzur}(1992)}]{Elit1992}
{Elitzur}, M., ed. 1992, Astrophysics and Space Science Library, Vol. 170,
  {Astronomical masers}

\bibitem[{{Fleischer} {et~al.}(1995){Fleischer}, {Gauger}, \&
  {Sedlmayr}}]{Flei1995}
{Fleischer}, A.~J., {Gauger}, A., \& {Sedlmayr}, E. 1995, \aap, 297, 543

\bibitem[{{Fonfr{\'{\i}}a} {et~al.}(2014){Fonfr{\'{\i}}a},
  {Fern{\'a}ndez-L{\'o}pez}, {Ag{\'u}ndez}, {S{\'a}nchez-Contreras}, {Curiel},
  \& {Cernicharo}}]{Fonf2014}
{Fonfr{\'{\i}}a}, J.~P., {Fern{\'a}ndez-L{\'o}pez}, M., {Ag{\'u}ndez}, M.,
  {et~al.} 2014, \mnras, 445, 3289

\bibitem[{{Fonfr{\'{\i}}a Exp{\'o}sito} {et~al.}(2006){Fonfr{\'{\i}}a
  Exp{\'o}sito}, {Ag{\'u}ndez}, {Tercero}, {Pardo}, \& {Cernicharo}}]{Fonf2006}
{Fonfr{\'{\i}}a Exp{\'o}sito}, J.~P., {Ag{\'u}ndez}, M., {Tercero}, B.,
  {Pardo}, J.~R., \& {Cernicharo}, J. 2006, \apjl, 646, L127

\bibitem[{{Ford} {et~al.}(2003){Ford}, {Neufeld}, {Goldsmith}, \&
  {Melnick}}]{Ford2003}
{Ford}, K.~E.~S., {Neufeld}, D.~A., {Goldsmith}, P.~F., \& {Melnick}, G.~J.
  2003, \apj, 589, 430

\bibitem[{{Gautschy-Loidl} {et~al.}(2004){Gautschy-Loidl}, {H{\"o}fner},
  {J{\o}rgensen}, \& {Hron}}]{Gaut2004}
{Gautschy-Loidl}, R., {H{\"o}fner}, S., {J{\o}rgensen}, U.~G., \& {Hron}, J.
  2004, \aap, 422, 289

\bibitem[{{Gensheimer} {et~al.}(1992){Gensheimer}, {Likkel}, \&
  {Snyder}}]{Gens1992}
{Gensheimer}, P.~D., {Likkel}, L., \& {Snyder}, L.~E. 1992, \apjl, 388, L31

\bibitem[{{Gobrecht} {et~al.}(2016){Gobrecht}, {Cherchneff}, {Sarangi},
  {Plane}, \& {Bromley}}]{Gobr2016}
{Gobrecht}, D., {Cherchneff}, I., {Sarangi}, A., {Plane}, J.~M.~C., \&
  {Bromley}, S.~T. 2016, \aap, 585, A6

\bibitem[{{Gong} {et~al.}(2015){Gong}, {Henkel}, {Spezzano}, {Thorwirth},
  {Menten}, {Wyrowski}, {Mao}, \& {Klein}}]{Gong2015}
{Gong}, Y., {Henkel}, C., {Spezzano}, S., {et~al.} 2015, \aap, 574, A56

\bibitem[{{Gong} {et~al.}(2017){Gong}, {Henkel}, {Ott}, {Menten}, {Morris},
  {Keller}, {Claussen}, {Grasshoff}, \& {Mao}}]{Gong2017}
{Gong}, Y., {Henkel}, C., {Ott}, J., {et~al.} 2017, ArXiv e-prints,
  arXiv:1706.02446

\bibitem[{{Groenewegen} \& {Ludwig}(1998)}]{Groe1998}
{Groenewegen}, M.~A.~T., \& {Ludwig}, H.-G. 1998, \aap, 339, 489

\bibitem[{{Groenewegen} {et~al.}(2012){Groenewegen}, {Barlow}, {Blommaert},
  {Cernicharo}, {Decin}, {Gomez}, {Hargrave}, {Kerschbaum}, {Ladjal}, {Lim},
  {Matsuura}, {Olofsson}, {Sibthorpe}, {Swinyard}, {Ueta}, \&
  {Yates}}]{Groe2012}
{Groenewegen}, M.~A.~T., {Barlow}, M.~J., {Blommaert}, J.~A.~D.~L., {et~al.}
  2012, \aap, 543, L8

\bibitem[{{Guelin} {et~al.}(1987){Guelin}, {Cernicharo}, {Navarro}, {Woodward},
  {Gottlieb}, \& {Thaddeus}}]{Guel87}
{Guelin}, M., {Cernicharo}, J., {Navarro}, S., {et~al.} 1987, \aap, 182, L37

\bibitem[{{Guelin} {et~al.}(1993){Guelin}, {Lucas}, \& {Cernicharo}}]{Guel1993}
{Guelin}, M., {Lucas}, R., \& {Cernicharo}, J. 1993, \aap, 280, L19

\bibitem[{{Guilloteau} {et~al.}(1987){Guilloteau}, {Omont}, \&
  {Lucas}}]{Guil1987}
{Guilloteau}, S., {Omont}, A., \& {Lucas}, R. 1987, \aap, 176, L24

\bibitem[{{Han} {et~al.}(1998){Han}, {Mao}, {Lu}, {Wu}, {Sun}, {Wang}, {Pei},
  {Fan}, {Tang}, \& {Ji}}]{Han1998}
{Han}, F., {Mao}, R.~Q., {Lu}, J., {et~al.} 1998, \aaps, 127, 181

\bibitem[{{He} {et~al.}(2008){He}, {Dinh-V-Trung}, {Kwok}, {M{\"u}ller},
  {Zhang}, {Hasegawa}, {Peng}, \& {Huang}}]{He08}
{He}, J.~H., {Dinh-V-Trung}, {Kwok}, S., {et~al.} 2008, \apjs, 177, 275

\bibitem[{{Henkel} {et~al.}(1983){Henkel}, {Matthews}, \& {Morris}}]{Henk1983}
{Henkel}, C., {Matthews}, H.~E., \& {Morris}, M. 1983, \apj, 267, 184

\bibitem[{{Highberger} {et~al.}(2000){Highberger}, {Apponi}, {Bieging},
  {Ziurys}, \& {Mangum}}]{High2000}
{Highberger}, J.~L., {Apponi}, A.~J., {Bieging}, J.~H., {Ziurys}, L.~M., \&
  {Mangum}, J.~G. 2000, \apj, 544, 881

\bibitem[{{Hoefner} {et~al.}(1998){Hoefner}, {Jorgensen}, {Loidl}, \&
  {Aringer}}]{Hoef1998}
{Hoefner}, S., {Jorgensen}, U.~G., {Loidl}, R., \& {Aringer}, B. 1998, \aap,
  340, 497

\bibitem[{{H{\"o}fner} {et~al.}(2003){H{\"o}fner}, {Gautschy-Loidl}, {Aringer},
  \& {J{\o}rgensen}}]{Hoef2003}
{H{\"o}fner}, S., {Gautschy-Loidl}, R., {Aringer}, B., \& {J{\o}rgensen}, U.~G.
  2003, \aap, 399, 589

\bibitem[{{Huggins} \& {Healy}(1986)}]{Hugg1986}
{Huggins}, P.~J., \& {Healy}, A.~P. 1986, \apj, 304, 418

\bibitem[{{Jeffers} {et~al.}(2014){Jeffers}, {Min}, {Waters}, {Canovas},
  {Pols}, {Rodenhuis}, {de Juan Ovelar}, {Keller}, \& {Decin}}]{Jeff2014}
{Jeffers}, S.~V., {Min}, M., {Waters}, L.~B.~F.~M., {et~al.} 2014, \aap, 572,
  A3

\bibitem[{{Kim} {et~al.}(2015){Kim}, {Lee}, {Mauron}, \& {Chu}}]{Kim2015}
{Kim}, H., {Lee}, H.-G., {Mauron}, N., \& {Chu}, Y.-H. 2015, \apjl, 804, L10

\bibitem[{{Le{\~a}o} {et~al.}(2006){Le{\~a}o}, {de Laverny}, {M{\'e}karnia},
  {de Medeiros}, \& {Vandame}}]{Leao2006}
{Le{\~a}o}, I.~C., {de Laverny}, P., {M{\'e}karnia}, D., {de Medeiros}, J.~R.,
  \& {Vandame}, B. 2006, \aap, 455, 187

\bibitem[{{Loidl} {et~al.}(1999){Loidl}, {H{\"o}fner}, {J{\o}rgensen}, \&
  {Aringer}}]{Loid1999}
{Loidl}, R., {H{\"o}fner}, S., {J{\o}rgensen}, U.~G., \& {Aringer}, B. 1999,
  \aap, 342, 531

\bibitem[{{Lucas} \& {Cernicharo}(1989)}]{Luca1989}
{Lucas}, R., \& {Cernicharo}, J. 1989, \aap, 218, L20

\bibitem[{{Lucas} {et~al.}(1995){Lucas}, {Gu{\'e}lin}, {Kahane}, {Audinos}, \&
  {Cernicharo}}]{Luca1995}
{Lucas}, R., {Gu{\'e}lin}, M., {Kahane}, C., {Audinos}, P., \& {Cernicharo}, J.
  1995, \apss, 224, 293

\bibitem[{{Lucas} \& {Guilloteau}(1992)}]{Luca1992}
{Lucas}, R., \& {Guilloteau}, S. 1992, \aap, 259, L23

\bibitem[{{Lucas} {et~al.}(1988){Lucas}, {Omont}, \& {Guilloteau}}]{Luca1988}
{Lucas}, R., {Omont}, A., \& {Guilloteau}, S. 1988, \aap, 194, 230

\bibitem[{{Marigo} {et~al.}(2016){Marigo}, {Ripamonti}, {Nanni}, {Bressan}, \&
  {Girardi}}]{Mari2016}
{Marigo}, P., {Ripamonti}, E., {Nanni}, A., {Bressan}, A., \& {Girardi}, L.
  2016, \mnras, 456, 23

\bibitem[{{Men'shchikov} {et~al.}(2001){Men'shchikov}, {Balega}, {Bl{\"o}cker},
  {Osterbart}, \& {Weigelt}}]{Mens01}
{Men'shchikov}, A.~B., {Balega}, Y., {Bl{\"o}cker}, T., {Osterbart}, R., \&
  {Weigelt}, G. 2001, \aap, 368, 497

\bibitem[{{Menten} {et~al.}(2012){Menten}, {Reid}, {Kami{\'n}ski}, \&
  {Claussen}}]{Ment12}
{Menten}, K.~M., {Reid}, M.~J., {Kami{\'n}ski}, T., \& {Claussen}, M.~J. 2012,
  \aap, 543, A73

\bibitem[{{Menut} {et~al.}(2007){Menut}, {Gendron}, {Schartmann}, {Tuthill},
  {Lopez}, {Danchi}, {Wolf}, {Lagrange}, {Flament}, {Rouan}, {Cl{\'e}net}, \&
  {Berruyer}}]{Menu2007}
{Menut}, J.-L., {Gendron}, E., {Schartmann}, M., {et~al.} 2007, \mnras, 376, L6

\bibitem[{{M{\"u}ller} {et~al.}(2005){M{\"u}ller}, {Schl{\"o}der}, {Stutzki},
  \& {Winnewisser}}]{Mull05}
{M{\"u}ller}, H.~S.~P., {Schl{\"o}der}, F., {Stutzki}, J., \& {Winnewisser}, G.
  2005, Journal of Molecular Structure, 742, 215

\bibitem[{{M{\"u}ller} {et~al.}(2001){M{\"u}ller}, {Thorwirth}, {Roth}, \&
  {Winnewisser}}]{Mull01}
{M{\"u}ller}, H.~S.~P., {Thorwirth}, S., {Roth}, D.~A., \& {Winnewisser}, G.
  2001, \aap, 370, L49

\bibitem[{{Murakawa} {et~al.}(2005){Murakawa}, {Suto}, {Oya}, {Yates}, {Ueta},
  \& {Meixner}}]{Mura2005}
{Murakawa}, K., {Suto}, H., {Oya}, S., {et~al.} 2005, \aap, 436, 601

\bibitem[{{Neufeld} {et~al.}(2011){Neufeld}, {Gonz{\'a}lez-Alfonso}, {Melnick},
  {Szczerba}, {Schmidt}, {Decin}, {de Koter}, {Sch{\"o}ier}, \&
  {Cernicharo}}]{Neuf2011}
{Neufeld}, D.~A., {Gonz{\'a}lez-Alfonso}, E., {Melnick}, G.~J., {et~al.} 2011,
  \apjl, 727, L28

\bibitem[{{Nguyen-Q-Rieu} {et~al.}(1984){Nguyen-Q-Rieu}, {Bujarrabal},
  {Olofsson}, {Johansson}, \& {Turner}}]{Nguy1984}
{Nguyen-Q-Rieu}, {Bujarrabal}, V., {Olofsson}, H., {Johansson}, L.~E.~B., \&
  {Turner}, B.~E. 1984, \apj, 286, 276

\bibitem[{{Nowotny} {et~al.}(2005){Nowotny}, {Aringer}, {H{\"o}fner},
  {Gautschy-Loidl}, \& {Windsteig}}]{Nowo2005}
{Nowotny}, W., {Aringer}, B., {H{\"o}fner}, S., {Gautschy-Loidl}, R., \&
  {Windsteig}, W. 2005, \aap, 437, 273

\bibitem[{{Nowotny} {et~al.}(2010){Nowotny}, {H{\"o}fner}, \&
  {Aringer}}]{Nowo2010}
{Nowotny}, W., {H{\"o}fner}, S., \& {Aringer}, B. 2010, \aap, 514, A35

\bibitem[{{Patel} {et~al.}(2011){Patel}, {Young}, {Gottlieb}, {Thaddeus},
  {Wilson}, {Menten}, {Reid}, {McCarthy}, {Cernicharo}, {He}, {Br{\"u}nken},
  {Trung}, \& {Keto}}]{Pate2011}
{Patel}, N.~A., {Young}, K.~H., {Gottlieb}, C.~A., {et~al.} 2011, \apjs, 193,
  17

\bibitem[{{Quintana-Lacaci} {et~al.}(2016){Quintana-Lacaci}, {Cernicharo},
  {Ag{\'u}ndez}, {Velilla Prieto}, {Castro-Carrizo}, {Marcelino}, {Cabezas},
  {Pe{\~n}a}, {Alonso}, {Z{\'u}{\~n}iga}, {Requena}, {Bastida}, {Kalugina},
  {Lique}, \& {Gu{\'e}lin}}]{Quin2016}
{Quintana-Lacaci}, G., {Cernicharo}, J., {Ag{\'u}ndez}, M., {et~al.} 2016,
  \apj, 818, 192

\bibitem[{{Sahai} {et~al.}(1984){Sahai}, {Wootten}, \& {Clegg}}]{Saha1984}
{Sahai}, R., {Wootten}, A., \& {Clegg}, R.~E.~S. 1984, \apj, 284, 144

\bibitem[{{Schilke} {et~al.}(2000){Schilke}, {Mehringer}, \&
  {Menten}}]{Schi2000}
{Schilke}, P., {Mehringer}, D.~M., \& {Menten}, K.~M. 2000, \apjl, 528, L37

\bibitem[{{Schilke} \& {Menten}(2003)}]{Schi2003}
{Schilke}, P., \& {Menten}, K.~M. 2003, \apj, 583, 446

\bibitem[{{Shenavrin} {et~al.}(2011){Shenavrin}, {Taranova}, \&
  {Nadzhip}}]{Shen11}
{Shenavrin}, V.~I., {Taranova}, O.~G., \& {Nadzhip}, A.~E. 2011, Astronomy
  Reports, 55, 31

\bibitem[{{Shinnaga} {et~al.}(2009){Shinnaga}, {Young}, {Tilanus},
  {Chamberlin}, {Gurwell}, {Wilner}, {Hughes}, {Yoshida}, {Peng}, {Force},
  {Friberg}, {Bottinelli}, {Van Dishoeck}, \& {Phillips}}]{Shin2009}
{Shinnaga}, H., {Young}, K.~H., {Tilanus}, R.~P.~J., {et~al.} 2009, \apj, 698,
  1924

\bibitem[{{Takano} {et~al.}(1992){Takano}, {Saito}, \& {Tsuji}}]{Taka1992}
{Takano}, S., {Saito}, S., \& {Tsuji}, T. 1992, \pasj, 44, 469

\bibitem[{{Tenenbaum} {et~al.}(2010{\natexlab{a}}){Tenenbaum}, {Dodd}, {Milam},
  {Woolf}, \& {Ziurys}}]{Tene10a}
{Tenenbaum}, E.~D., {Dodd}, J.~L., {Milam}, S.~N., {Woolf}, N.~J., \& {Ziurys},
  L.~M. 2010{\natexlab{a}}, \apjl, 720, L102

\bibitem[{{Tenenbaum} {et~al.}(2010{\natexlab{b}}){Tenenbaum}, {Dodd}, {Milam},
  {Woolf}, \& {Ziurys}}]{Tene10b}
---. 2010{\natexlab{b}}, \apjs, 190, 348

\bibitem[{{Turner}(1987)}]{Turn1987}
{Turner}, B.~E. 1987, \aap, 183, L23

\bibitem[{{Tuthill} {et~al.}(2005){Tuthill}, {Monnier}, \& {Danchi}}]{Tuth2005}
{Tuthill}, P.~G., {Monnier}, J.~D., \& {Danchi}, W.~C. 2005, \apj, 624, 352

\bibitem[{{Velilla Prieto} {et~al.}(2015){Velilla Prieto}, {Cernicharo},
  {Quintana-Lacaci}, {Ag{\'u}ndez}, {Castro-Carrizo}, {Fonfr{\'{\i}}a},
  {Marcelino}, {Z{\'u}{\~n}iga}, {Requena}, {Bastida}, {Lique}, \&
  {Gu{\'e}lin}}]{Veli2015}
{Velilla Prieto}, L., {Cernicharo}, J., {Quintana-Lacaci}, G., {et~al.} 2015,
  \apjl, 805, L13

\bibitem[{{Yamamoto} {et~al.}(1987){Yamamoto}, {Saito}, {Guelin}, {Cernicharo},
  {Suzuki}, \& {Ohishi}}]{Yama87}
{Yamamoto}, S., {Saito}, S., {Guelin}, M., {et~al.} 1987, \apjl, 323, L149

\bibitem[{{Zelinger} {et~al.}(2003){Zelinger}, {Amano}, {Ahrens},
  {Br{\"u}nken}, {Lewen}, {M{\"u}ller}, \& {Winnewisser}}]{Zeli03}
{Zelinger}, Z., {Amano}, T., {Ahrens}, V., {et~al.} 2003, Journal of Molecular
  Spectroscopy, 220, 223

\bibitem[{{Ziurys} \& {Turner}(1986)}]{Ziur86}
{Ziurys}, L.~M., \& {Turner}, B.~E. 1986, \apjl, 300, L19

\end{thebibliography}
\bibliographystyle{aasjournal.bst}

\end{document}